\documentclass[aps,prc,twocolumn,showpacs,superscriptaddress,floatfix]{revtex4-2}
\usepackage{graphicx}% Include figure files
\usepackage{dcolumn}% Align table columns on the decimal point
\usepackage{bm}% bold math
\usepackage{hyperref}% add hypertext capabilities
\usepackage[utf8]{inputenc}
\hypersetup{
  colorlinks=true,        % false: boxed links; true: colored links
  linkcolor=blue,         % color of internal links
  citecolor=cyan,         % color of links to bibliography
  urlcolor=cyan            % color of external links
}

\usepackage[utf8]{inputenc}
\usepackage{colortbl}
\usepackage{amsmath}
\usepackage{times}
\usepackage{multirow}
\usepackage[normalem]{ulem}
\usepackage{rotating}
\usepackage{amssymb}

\newcommand{\be}{\begin{equation}}
\newcommand{\ee}{\end{equation}}
\newcommand{\bea}{\begin{eqnarray}}
\newcommand{\eea}{\end{eqnarray}}

\usepackage{xcolor}

\definecolor{color1}{HTML}{440154}
\definecolor{color2}{HTML}{481568}
\definecolor{color3}{HTML}{482677}

\definecolor{color18}{HTML}{B8DE29}
\definecolor{color19}{HTML}{DCE318}
\definecolor{color20}{HTML}{FDE725}
\begin{document}

\title{Neutron star  equation of state: identifying hadronic matter characteristics}

\author{Constan\c ca Provid\^encia}
 \email{cp@uc.pt}
 \affiliation{CFisUC, Department of Physics, University of Coimbra, 3004-516 Coimbra, Portugal.}
\author{Tuhin Malik}
 \email{tm@uc.pt}
 \affiliation{CFisUC, Department of Physics, University of Coimbra, 3004-516 Coimbra, Portugal.}
 \author{Milena Bastos Albino}
 \email{milena.albino@usp.br}
 \affiliation{CFisUC, Department of Physics, University of Coimbra, 3004-516 Coimbra, Portugal.}
 \author{Márcio Ferreira}
 \email{marcio.ferreira@uc.pt}
 \affiliation{CFisUC, Department of Physics, University of Coimbra, 3004-516 Coimbra, Portugal.}

\date{\today}

\begin{abstract} 
The general behavior of the nuclear equation of state (EOS), relevant for the description of neutron stars (NS), is studied within a  relativistic mean field description of nuclear matter. Different formulations, both with density dependent couplings and with non-linear mesonic terms, are considered and their predictions compared and discussed. A special attention is drawn to the effect on the neutron star properties of the inclusion of exotic degrees of freedom as hyperons. Properties such as the speed of sound, the trace anomaly, the proton fraction and  the onset of direct Urca processes inside neutron stars are discussed. The knowledge of the general behavior of the hadronic equation of state and the implication it has on the neutron star properties will allow to identify signatures of a deconfinement phase transition discussed in other studies.
\end{abstract}

\maketitle

\section{Introduction} \label{sec:intro}
Neutron stars (NS) are objects with several extreme properties which make them   a true laboratory for dense baryonic matter. Under the extreme conditions existing in their interior it is expected, for instance, that quark deconfinement may occur in the center of NS. These are also the ideal objects to study very asymmetric nuclear matter which cannot be tested in the laboratory. In the present multi-messenger era, astrophysical observations are starting to impose some stringent constraints on the equation of state (EOS) of the high density baryonic matter. These constraints come from the gravitational wave detection by the LIGO Virgo collaboration as the detection of the binary neutron star merger GW170817 \cite{TheLIGOScientific:2017qsa} or the GW190425 \cite{LIGOScientific:2020aai} , from  radio data  \cite{Demorest2010,Antoniadis2013,NANOGrav:2019jur,Fonseca:2021wxt,Romani:2021xmb} or the recent x-ray observations of NICER allowing a prediction of both the NS mass and radius \cite{Riley:2019yda,Miller:2019cac,Riley:2021pdl,Miller:2021qha,Raaijmakers:2021uju}.

The nuclear matter EOS at low densities is constrained  not only by well know nuclear matter properties as the binding energy, saturation density and incompressibility  \cite{Margueron2018a}, but also from {\it ab-initio} calculations of pure neutron matter based on a chiral effective field  theoretical description \cite{Hebeler2013,Drischler:2017wtt,Drischler:2020yad}. At very high densities, $\sim 40\rho_0$ where $\rho_0$ is the nuclear saturation density, perturbative QCD calculations have been performed and they also impose strong constraints \cite{Kurkela2009,Kurkela:2014vha}. Although the pQCD EOS is determined at densities that are not attained inside neutron stars,  it was shown recently that these constraints may affect in a non-trivial way the EOS inside neutron stars \cite{Gorda:2022jvk}.

From the measurement of mass and radius of several NS, one expects to be able to recover the EOS. The integration of the Tolman-Oppenheimer-Volkoff (TOV) equations establishes a one-to-one relation between the mass-radius curve and the pressures-energy density function \cite{Lindblom:1992}. Several methods have been used to obtain the EoS from the known mass-radius curve such as Baseyan inference \cite{Steiner:2010fz,Ozel:2010fw,Steiner:2012xt,Raithel:2016bux} or neural network techniques \cite{Fujimoto_2018,Fujimoto_2020,fujimoto2021extensive,Ferreira:2019bny,morawski2020neural,Carvalho:2023ele}, see \cite{Zhou:2023pti} for a recent review on the application of machine learning techniques to learn about QCD matter under extreme conditions.  Another problem is the determination of the NS composition from the knowledge of the EOS. Several studies have been carried out with the objective of extracting the proton fraction. Starting from a Taylor expansion  representation of the EOS in the parabolic approximation for the asymmetry, it was shown that the proton fraction could not be recovered from the $\beta$-equilibrium EOS  \cite{Mondal2021,Imam:2021dbe,Tovar2021}. In \cite{Mondal2021}, the authors attribute the failure to the existence of  multiple solutions. In \cite{Imam:2021dbe}, the reason was assigned to the occurrence of correlations between the nuclear matter parameters.

Many studies have been performed with the objective of determining the EOS of strongly interacting matter constrained by observations and  well accepted {\it ab-initio} calculations as the ones reported above. In order, to span the whole phase space that joins the low density to the high density constraints different interpolation schemes have been undertaken  based in agnostic descriptions of the EOS. Among these we can point out the use of a piecewise polytropic interpolation  \cite{Kurkela:2014vha,Annala:2021gom}, a spectral  interpolation \cite{Lindblom:2010bb}, a  speed-of-sound interpolation \cite{Annala2019,Altiparmak:2022bke,Somasundaram:2022ztm}, meta-models based on Taylor expansions \cite{Margueron:2017eqc,Margueron:2017lup,Ferreira:2019bgy,Xie:2019sqb,Xie:2020tdo,Ferreira:2021pni,Thi:2021jhz} 
or a nonparametric inference of the EOS \cite{Landry:2018prl,Essick:2019ldf, Gorda:2022jvk,Zhou:2023cfs}. These studies have been used to infer signatures of the presence of deconfined matter inside neutron stars, for instance, by analyzing the behavior of the speed of sound with density \cite{Altiparmak:2022bke,Somasundaram:2022ztm} or the trace anomaly which may signal the restoration of conformal symmetry \cite{Fujimoto:2022ohj,Annala:2023cwx}. However, the above approaches are not able determine the composition of neutron stars. %\red{Marcio: poderia completar com outros trabalhos que estarão em falta?}

% discuss composition

The present chapter reviews recent work developed within the framework of a relativistic mean-field (RMF)  description of hadronic matter at zero temperature having as main objective the determination of the region in the neutron star mass radius diagram, and corresponding EOS, in conformity with present observations and {\it ab-initio} constraints. A Bayesian inference  will be applied in the search for the parameters of the models. In comparison with the agnostic approaches described above, our perspective has an underlying microscopic model, which allows us to discuss composition, including proton fraction or hyperon content. We consider this information completes the one obtained from the agnostic descriptions of the EOS, and  may bring extra clues into the interpretation of the results obtained. In the following chapters we will review the methodology and results obtained in the works \cite{Malik:2022ilb,Malik:2022jqc,Malik:2022zol,Malik:2023mnx}.  In particular, we will  compare outputs obtained considering the different microscopic models in order to assess the generality and the specificity of the conclusions. The microscopic models based in a Lagrangian formulation used in these works may be divided in two classes: i)  the Lagrangian density is formulated in terms of  constant parameters and include non-linear mesonic terms as proposed in \cite{Boguta1977,Mueller1996}. These models are designated by NL; ii) the Lagrangian density contains only quadratic mesonic terms and is expressed in terms of couplings that have an explicit density dependence  as explored in \cite{Typel1999,Typel2009}. In this class, we consider two different parametrizations of the couplings, the one proposed in \cite{Typel1999} which we designate by DDH and the one used in \cite{Malik:2022ilb} designated as DDB. We will also discuss the limitations of this second class of models concerning the high density behavior of the coupling to the $\varrho$-meson, which defines the density dependence of the symmetry energy, and we will propose a generalization that overcomes the limitation \cite{Malik:2022zol}. Lastly, and considering recent interest in identifying signatures of deconfinement and of imposing high density pQCD constraints, we will discuss some of the physical quantities examined, including the speed of sound, polytropic index and trace anomaly and discuss the limitations enforced by pQCD.

Some other works have been developed in the last years using a Bayesian inference approach to constraint the parameters of RMF models including, \cite{Traversi:2020aaa} where a simpler version of the NL description was considered,  \cite{Sun:2022yor}  which has restricted the  $\Lambda-\omega$ couplings in hyperonic stellar matter  imposing as constraints the  GW and NICER observations, \cite{Beznogov:2022rri} where the authors have studied how the pure neutron matter  pressure and energy per particle constrains the isovector behavior of nuclear matter, and studied several correlations between nuclear matter properties  (NMP) and NS properties, 
\cite{Huang:2023grj} where the authors have constrained the NL model from the present available NS observations and tested how constraining might be the future observations programmed for eXTP \cite{extp_watts} and STROBE-X \cite{strobex}.  

In the present chapter, we will first present the microscopic models used to perform the study, the Bayesian inference methodology, together with the priors, the  data chosen to fit the models. We next compare the behavior of the different data sets generated, including the nuclear matter properties (NMP) and the neutron star properties, including the speed of sound and the proton fraction. The inclusion of hyperons will be discussed as well as the onset of the nucleon direct Urca processes. We will also refer to some properties that are directly connected to QCD, in particular the trace anomaly and the constraints imposed by pQCD on the generated data sets of EOS.

\section{Formalism}

In this section,  we briefly summarize the frameworks that will be applied to describe 
the nuclear or hadronic matter  EOS.  We will start by introducing the models through the definition of the Lagrangian density. As referred in the Introduction, two different classes are  considered. They define  the density dependence of the EOS through completely different approaches: i)  density dependent couplings are introduced (DDH and DDB models); ii) non-linear mesonic terms are included (NL).

\subsection{The model\label{sec:model}}
The equation of state of nuclear matter is determined from the Lagrangian density that describes the nuclear system.  The degrees of freedom include the nucleons of mass $m$ described by Dirac spinors ${\Psi}$, and the meson fields, the scalar isoscalar $\sigma$ field, the vector isoscalar $\omega$ field, and the vector isovector $\varrho$ field,  with masses $m_i,\, i=\sigma, \, \omega,\, \varrho$, which describe the nuclear interaction.  The parameters $\Gamma_i$ or $g_i$, $i=\sigma, \, \omega,\, \varrho$ designate the couplings of the mesons to the nucleons.
The Lagrangian density  is given by
\begin{equation}
\begin{aligned}
\mathcal{L}=& \bar{\Psi}\Big[\gamma^{\mu}\left(i \partial_{\mu}-\Gamma_{\omega} A_{\mu}^{(\omega)}-%\frac{1}{2}
\Gamma_{\varrho} {\boldsymbol{t}} \cdot \boldsymbol{A}_{\mu}^{(\varrho)}\right) %\\&
-\left(m-\Gamma_{\sigma} \phi\right)\Big] \Psi  \\&
+ \frac{1}{2}\left(\partial_{\mu} \phi \partial^{\mu} \phi-m_{\sigma}^{2} \phi^{2} \right) \\
&-\frac{1}{4} F_{\mu \nu}^{(\omega)} F^{(\omega) \mu \nu} 
+\frac{1}{2}m_{\omega}^{2} A_{\mu}^{(\omega)} A^{(\omega) \mu} \\&
-\frac{1}{4} \boldsymbol{F}_{\mu \nu}^{(\varrho)} \cdot \boldsymbol{F}^{(\varrho) \mu \nu} 
+ \frac{1}{2} m_{\varrho}^{2} \boldsymbol{A}_{\mu}^{(\varrho)} \cdot \boldsymbol{A}^{(\varrho) \mu}+ \mathcal{L}_{NL},
\end{aligned}
\label{lagrangian}
\end{equation}
where the last term $\mathcal{L}_{NL}$ is null if density dependent couplings $\Gamma_i$ are chosen, or includes self-interacting and mixed meson terms if the meson-nucleon couplings  are taken as constant parameters. In order to distinguish, we will designate the constant couplings by the lower case  letter $g_i$ in the NL formulation. In the above expression  $\gamma^\mu $ and $\boldsymbol{t}$ designate, respectively, the Dirac matrices and the isospin operator. The  vector meson field strength tensors are defined as  $F^{(\omega, \varrho)\mu \nu} = \partial^ \mu A^{(\omega, \varrho)\nu} -\partial^ \nu A^{(\omega, \varrho) \mu}$.  

\subsubsection{Density dependent description}
The density dependent models include meson-nucleon couplings  $\Gamma_i$, that depend on the total nucleonic density $\rho$, and is defined as
\begin{equation}
  \Gamma_{i}(\rho) =\Gamma_{i,0} ~ h_i(x)~,\quad x = \rho/\rho_0~, \,i=\sigma, \omega, \varrho, 
\end{equation}
with  $\Gamma_{i,0}$ the couplings at saturation density $\rho_0$. For the isoscalar mesons,  $\sigma$ and $\omega$,  two  parametrizations $h_i$ are considered:
\begin{equation}
h_i(x) = \exp[-(x^{a_i}-1)]
\label{hm1}
\end{equation}
as in \cite{Malik:2022zol}, giving origin to the DDB sets, and
\begin{eqnarray}
h_i(x) =a_M\frac{1+b_i(x+d_i)^2}{1+c_i(x+d_i)^2} \, ,
\label{hm1a}
\end{eqnarray}
as in \cite{Typel1999,Typel2009},  and originating the DDH data sets.
The $\varrho$-meson nucleon coupling is defined as in \cite{Typel1999}
\begin{equation}
h_\varrho(x) = \exp[-a_\varrho (x-1)] ~.
\label{hm2}
\end{equation}

\subsubsection{Non-linear meson terms}
The model introduced in \cite{Mueller:1996pm} is defined with constant couplings, which we designate by $g_i,\, i=\sigma,\, \omega,\, \varrho$, and, instead, includes non-linear meson terms in the Lagrangian density, which are defined by
\begin{eqnarray}
  \mathcal{L}_{NL}=&-\frac{1}{3} b g_\sigma^3 (\sigma)^{3}-\frac{1}{4} c g_\sigma^4 (\sigma)^{4}+\frac{\xi}{4!}(g_{\omega}^2\omega_{\mu}\omega^{\mu})^{2} \nonumber \\ &+\Lambda_{\omega}g_{\varrho}^{2}\boldsymbol{\varrho}_{\mu} \cdot \boldsymbol{\varrho}^{\mu} g_{\omega}^2\omega_{\mu}\omega^{\mu}.
  \label{lagrangian}
\end{eqnarray}
 The parameters multiplying each one of these terms  $b,\, c,$ $\xi$, $\Lambda_\omega$ will be fixed together with the meson-nucleon couplings $g_i$   	by imposing nuclear matter and NS observational constraints.

 The parameters $b,\, c,$ in front of the $\sigma$ self interacting terms control the nuclear matter incompressibility at saturation \cite{Boguta1977}. The $\xi$ term  was introduced in \cite{Sugahara:1993wz} to modulate the high density dependence of the EoS, the larger $\xi$ the softer the EOS. The non-linear $\omega-\varrho$ term influences the density dependence of the symmetry energy \cite{Cavagnoli:2011ft}.
 
The equations of motion for the meson fields are given by
		\begin{eqnarray}
			{\sigma}&=& \frac{g_{\sigma}}{m_{\sigma,{\rm eff}}^{2}}\sum_{i} \rho^s_i\label{sigma}\\
			{\omega} &=&\frac{g_{\omega}}{m_{\omega,{\rm eff}}^{2}} \sum_{i} \rho_i \label{omega}\\
			{\varrho} &=&\frac{g_{\varrho}}{m_{\varrho,{\rm eff}}^{2}}\sum_{i} t_{3i} \rho_i, \label{rho}
		\end{eqnarray}
 where $\rho^s_i$ and $\rho_i$ are, respectively, the scalar density and the number density of nucleon $i$, and the effective meson masses are defined as
 \begin{eqnarray}
   m_{\sigma,{\rm eff}}^{2}&=& m_{\sigma}^{2}+{ b g_\sigma^3}{\sigma}+{c g_\sigma^4}{\sigma}^{2} \label{ms} \\ 
    m_{\omega,{\rm eff}}^{2}&=& m_{\omega}^{2}+ \frac{\xi}{3!}g_{\omega}^{4}{\omega}^{2} +2\Lambda_{\omega}g_{\varrho}^{2}g_{\omega}^{2}{\varrho}^{2}\label{mw}\\
    m_{\varrho,{\rm eff}}^{2}&=&m_{\varrho}^{2}+2\Lambda_{\omega}g_{\omega}^{2}g_{\varrho}^{2}{\omega}^{2}. \label{mr}
 \end{eqnarray}
 The non-linear meson terms define effective meson masses that depend on the density:
i)  $m_{\omega,{\rm eff}}$ increases with the $\omega$-field and, as a consequence, the $\omega$ field is not  proportional to the density for a non zero $\xi$, but increases with a power of $\rho$ smaller than one; ii) $m_{\varrho,{\rm eff}}$ increases with the density $\rho$, and, therefore, as the density increases the $\varrho$ field becomes weaker, resulting in a softer symmetry energy. The magnitude of the softening depends on $\xi$: the larger $\xi$ the smaller the softening.

Notice that the  meson equations, i.e. Eqs. (\ref{sigma}), (\ref{omega}) and (\ref{rho}), are also valid for the DDB and DDH models with the replacement $m_{i,eff}\to m_i$, since in the last two descriptions non-linear terms are not present.

\subsection{Bayesian inference procedure}
\label{sec:bayesian_inf}
The model parameters are determined within a Bayesian inference procedure, i.e.  applying  Bayes' theorem \cite{Gelman2013}, based on observed or experimental data, designated by fit data. The Bayesian parameter optimization system is determined from four different inputs that must be given: the prior, the likelihood function, the fit data, and the sampler.

{\it The Prior:-} The prior domain in our Bayesian setup is determined from a  Latin hypercube sampling, allowing the parameters of the underlying RMF model to vary so that a broad range of nuclear matter saturation properties is spanned. For each of  the different RMP models considered a uniform prior is defined.

{\it The Fit Data:-} As fit data we have considered for the three RMF models (see  Table \ref{tab1}):
 the nuclear saturation density $\rho_0$, the binding energy per nucleon $\epsilon_0$, the incompressibility coefficient $K_0$, and the symmetry energy $J_{\rm sym,0}$, all evaluated at $\rho_0$. We also include the pressure of pure neutron matter (PNM) at densities of 0.08, 0.12, and 0.16 fm$^{-3}$ from N$^3$LO calculations in chiral effective field theory (chEFT) \cite{Hebeler2013}, considering  2$\times$ N$^3$LO data uncertainty. Finally, it is also required that the maximum NS mass  is at least 2$M_\odot$. This requirement is introduced in the likelihood with uniform probability.
 
{\it The Log-Likelihood:-} A log-likelihood function is optimized as a cost function for the  fit data defined in Table \ref{tab1}. It is defined by the equation below, Eq.  \ref{loglik}, taking into account the uncertainties $\sigma_j$ associated with each data point $j$,
\begin{equation}
\label{loglik}
    Log (\mathcal{L}) = -0.5 \times \sum_j  \left\{ \left(\frac{d_j-m_j(\boldsymbol{\bm{\theta}})}{\sigma_j}\right)^2 + Log(2 \pi \sigma_j^2)\right\}.
\end{equation} 
The maximum NS mass  is treated differently, using a step function probability.

To populate the multi-dimensional posterior, we employ the nested sampling algorithm \cite{Skilling2004}, specifically the PyMultinest sampler \cite{Buchner:2014nha,buchner2021nested}, which is well-suited for low-dimensional problems. The EoS data set for subsequent analyses will be generated using the full posterior, which contains 25287 EoS. The posterior obtained for the three data sets is given in Table \ref{tab:para}, in Appendix A.

\begin{table}[!ht]
\centering
 \caption{ The constraints used as fit data in the Baseyian inference are: binding energy per nucleon $\epsilon_0$, incompressibility $K_0$, symmetry energy $J_{\rm sym,0}$ at the nuclear saturation density $\rho_0$, each with a 1$\sigma$ uncertainty, the pressure of pure neutron matter (PNM) at densities of 0.08, 0.12, and 0.16 fm$^{-3}$, obtained from a chEFT calculation \cite{Hebeler2013}, considering a 2$\times$ N$^3$LO uncertainty for the PNM pressure and  the maximum mass of neutron stars must exceed 2$M_\odot$.}
  \label{tab1}
 \setlength{\tabcolsep}{1.5pt}
      \renewcommand{\arraystretch}{1.1}
\begin{tabular}{cccc}
\hline 
\hline 
\multicolumn{4}{c}{Constraints}                                                        \\
\multicolumn{2}{c}{Quantity}                     & Value/Band  & Ref     \\ \hline
\multirow{3}{*}{\shortstack{NMP \\  {[}MeV{]} }} 
& $\rho_0$ & $0.153\pm0.005$ & \cite{Typel1999}    \\
& $\epsilon_0$ & $-16.1\pm0.2$ & \cite{Dutra:2014qga}   \\
                               & $K_0$           & $230\pm40$   & \cite{Shlomo2006,Todd-Rutel2005}    \\
                              & $J_{\rm sym, 0}$           & $32.5\pm1.8$  & \cite{Essick:2021ezp}   \\
                              
                               &                 &                &                                                   \\
  \shortstack{PNM \\ {[}MeV fm$^{-3}${]}}                  & $P(\rho)$       & $2\times$ N$^{3}$LO    & \cite{Hebeler2013}   \\
  &$dP/d\rho$&$>0$&\\
%                               &                 &                &                           &                           \\
\shortstack{NS mass \\ {[}$M_\odot${]}}        & $M_{\rm max}$   & $>2.0$     &  \cite{Fonseca:2021wxt}      \\ 

\hline 
\end{tabular}
\end{table}

\begin{figure}
    \centering
    \includegraphics[width=0.95\linewidth]{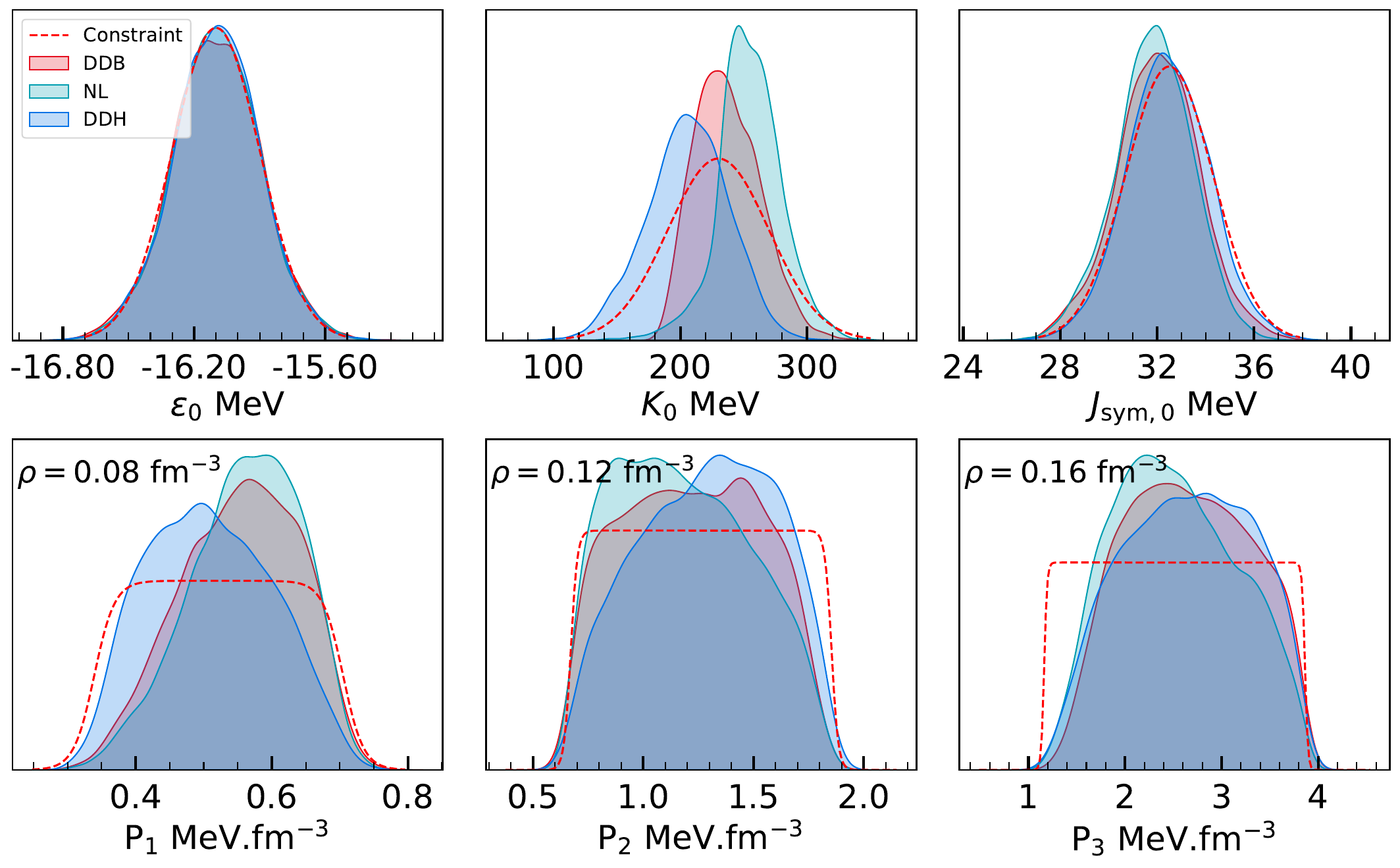}
    \caption{Fit data considered to constrain all EOS data set built for the present study, both for nucleonic and for hyperonic matter.}
    \label{fig:target}
\end{figure}

\section{Results}
In the following subsections, we compare the performance of the three different microscopic models used to generate the  data sets within  Bayesian inference calculations that consider as fit data the ones presented in Table \ref{tab1}. Both NS  and NMP will be compared. We will also discuss the effect of including hyperons, as well as the proton fraction and the onset of the nucleonic direct Urca processes.  Finally, the behavior of the speed of sound and trace anomaly with the baryonic density will be discussed and the compatibility with pQCD constraints will be commented.

\subsection{NL, DDB and DDH: a comparison \label{sec:comp}}

A comparison of  the performance of the three frameworks concerning the reproduction of the fit data is summarized in Fig. \ref{fig:target}. The chosen fit data were the same for the three frameworks and are given in Table \ref{tab1}. All models reproduce the binding energy $\epsilon_0$ and symmetry energy $J_{sym,0}$ at saturation in a similar way. The largest differences concern the incompressibility with DDH preferring smaller values and NL preferring larger ones. DDB peaks at the maximum of the fit data but with a much smaller width. Concerning the pure neutron matter (PNM) constraints the three frameworks satisfy the constraint imposed at the larger density in a similar way, but there are differences at the lowest and intermediate densities with DDH concentrating at lower pressure values for the lowest density. These behaviors will be reflected in the NMP and NS properties.

\begin{figure*}
    \centering
    \includegraphics[width=0.85\linewidth]{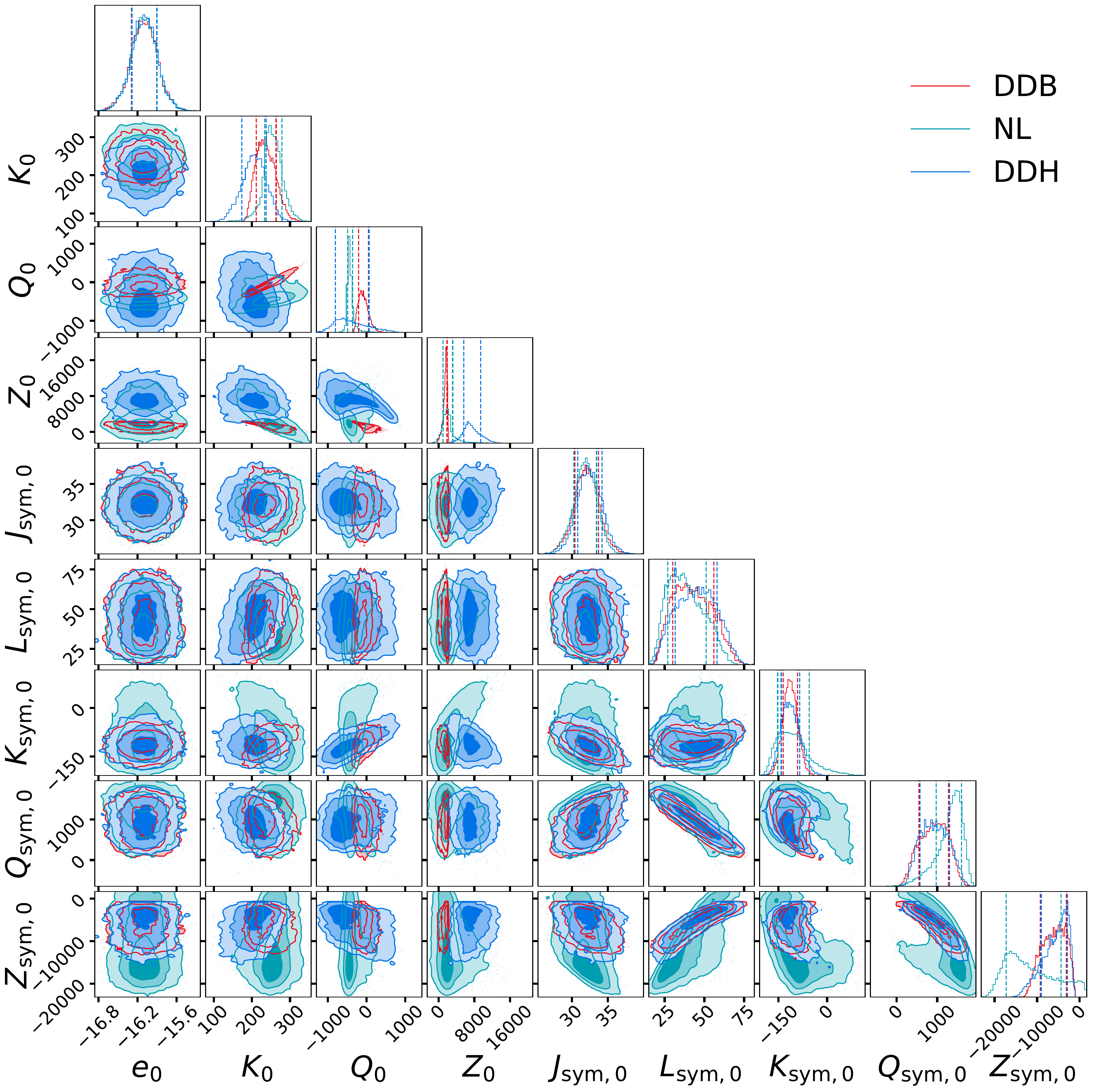}
    \caption{Corner plot comparing the nuclear matter properties of the three nucleonic data sets DDB. NL and DDH, in particular, the symmetric nuclear matter properties at saturation density $\rho_0$ defined by Eq. (\ref{x0}): binding energy    $\epsilon_0$, incompressibility $K_0$, skewness $Q_0$ and kurtosis $Z_0$; the symmetry energy properties at saturation  defined by Eq. (\ref{xsym}): symmetry energy $J_{\rm sym,0}$, slope $L_{\rm sym,0}$, incompressibility $K_{\rm sym,0}$, skewness $Q_{\rm sym,0}$ and kurtosis $Z_{\rm sym,0}$.}
    \label{fig:nmp}
\end{figure*}

Having verified that the three frameworks reproduce the fit data, we analyze next the NMP at saturation. This is summarized in the corner plot shown in  Fig. \ref{fig:nmp}  and in Table \ref{tab:nmp}, where, considering the parabolic approximation for the energy of nuclear matter per particle with the isospin asymmetry $\delta=(\rho_p-\rho_n)/\rho$ at nuclear density $\rho$,
\bea
\epsilon(\rho,\delta)\simeq  \epsilon(\rho,0)+S(\rho)\delta^2,
 \label{eq:eden}
\eea 
the parameters corresponding to the symmetric nuclear matter energy per particle $ \epsilon(\rho,0)$ and the symmetry energy $S(\rho)$ expansion around saturation density  till fourth order $n$ are given by: (i) for the symmetric nuclear matter, the energy per nucleon $\epsilon_0=\epsilon(\rho_0,0)$ ($n=0$), the incompressibility coefficient $K_0$ ($n=2$), the
skewness  $Q_0$ ($n=3$),  and  the kurtosis $Z_0$ ($n=4$), respectively, defined by
\begin{equation}
X_0^{(n)}=3^n \rho_0^n \left (\frac{\partial^n \epsilon(\rho, 0)}{\partial \rho^n}\right)_{\rho_0}, \, n=2,3,4;
\label{x0}
\end{equation}
(ii) for the symmetry energy,  the symmetry energy at saturation 
 $J_{\rm sym,0}$ ($n=0$), 
\begin{equation}
J_{\rm sym,0}= S(\rho_0)=\frac{1}{2} \left (\frac{\partial^2 \epsilon(\rho,\delta)}{\partial\delta^2}\right)_{\delta=0},
\end{equation}
the slope $L_{\rm sym,0}$ ($n=1$),  the curvature $K_{\rm sym,0}$ ($n=2$),  the skewness $Q_{\rm sym,0}$ ($n=3$), and  the kurtosis $Z_{\rm sym,0}$ ($n=4$), 
respectively, defined as
\begin{equation}
X_{\rm sym,0}^{(n)}=3^n \rho_0^n \left (\frac{\partial^n S(\rho)}{\partial \rho^n}\right )_{\rho_0},\, n=1,2,3,4.
\label{xsym}
\end{equation}

\begin{table*}[]
\caption{Nuclear matter properties at saturation density, median values and 90\% CI, of the three data sets, DDB, NL and DDH.  Symmetric nuclear matter properties at saturation density $\rho_0$ defined by Eq. (\ref{x0}): binding energy    $\epsilon_0$, incompressibility $K_0$, skewness $Q_0$ and kurtosis $Z_0$. Symmetry energy properties at saturation  defined by Eq. (\ref{xsym}): symmetry energy $J_{\rm sym,0}$, slope $L_{\rm sym,0}$, incompressibility $K_{\rm sym,0}$, skewness $Q_{\rm sym,0}$ and kurtosis $Z_{\rm sym,0}$. }
\label{tab:nmp}
 \setlength{\tabcolsep}{8pt}
      \renewcommand{\arraystretch}{1.2}
\begin{tabular}{ccccccccccccc}
\hline \hline 
\multicolumn{3}{c}{\multirow{2}{*}{Model}}              & $\rho_0$  & $\epsilon_0$ & $K_0$ & $Q_0$ & $Z_0$ & $J_{\rm sym,0}$ & $L_{\rm sym,0}$ & $K_{\rm sym,0}$ & $Q_{\rm sym,0}$ & $Z_{\rm sym,0}$ \\ \cline{4-13} 
\multicolumn{3}{c}{}                                    & fm$^{-3}$ & \multicolumn{9}{c}{MeV}                                                                                                        \\ \hline
\multirow{3}{*}{DDB}  & \multicolumn{2}{c}{median}      & 0.152     & -16.10       & 235   & -90   & 1585  & 32.05           & 42              & -114            & 935             & -5941           \\
                      & \multirow{2}{*}{90 \% CI} & min & 0.142     & -16.43       & 199   & -262  & 486   & 29.15           & 25              & -149            & 364             & -10751          \\
                      &                           & max & 0.164     & -15.76       & 282   & 162   & 2043  & 34.81           & 63              & -76             & 1434            & -2128           \\
                      &                           &     &           &              &       &       &       &                 &                 &                 &                 &                 \\
\multirow{3}{*}{NL}   & \multicolumn{2}{c}{median}      & 0.152     & -16.10       & 254   & -440  & 1952  & 31.89           & 37              & -109            & 1367            & -12613          \\
                      & \multirow{2}{*}{90 \% CI} & min & 0.145     & -16.43       & 213   & -516  & 243   & 29.08           & 23              & -171            & 629             & -19118          \\
                      &                           & max & 0.160     & -15.77       & 297   & -247  & 5295  & 34.41           & 58              & -3              & 1710            & -394            \\
                      &                           &     &           &              &       &       &       &                 &                 &                 &                 &                 \\
\multirow{3}{*}{DDH} & \multicolumn{2}{c}{median}      & 0.156     & -16.10       & 206   & -460  & 7189  & 32.44           & 45              & -114            & 930             & -5215           \\
                      & \multirow{2}{*}{90 \% CI} & min & 0.144     & -16.43       & 150   & -978  & 4459  & 29.68           & 25              & -157            & 412             & -11529          \\
                      &                           & max & 0.167     & -15.78       & 257   & 395   & 10908 & 35.24           & 65              & -64             & 1491            & -2078           \\ \hline
\end{tabular}
\end{table*}

% \usepackage{multirow}

\begin{table}
\caption{The Neutron star properties, median values and 90\% CI, of the three nucleon data sets, DDB, NL and DDH. The following properties are given: the maximum mass $M_{\rm max}$ and respective baryon mass $M_{\rm B,max}$, radius $R_{\rm max}$, speed of the sound squared at the center $c_s^2$ and central baryonic density $\rho_c$, and the radius and tidal deformability of the 1.4$M_\odot$ star, $R_{\rm 1.4}$ and $\Lambda_{\rm 1.4}$.}
\label{tab:ns}
\setlength{\tabcolsep}{1.pt}
      \renewcommand{\arraystretch}{1.2}
      \centering
\begin{tabular}{cccccccccc}
\hline \hline 
\multicolumn{3}{c}{\multirow{2}{*}{Model}}             & $M_{\rm max}$ & $M_{\rm B,max}$ & $R_{\rm max}$ & $R_{\rm 1.4}$ & $\Lambda_{\rm 1.4}$ & $C_s^2$ & $\rho_c$  \\ \cline{4-10} 
\multicolumn{3}{c}{}                                   & \multicolumn{2}{c}{$M_\odot$}   & \multicolumn{2}{c}{km}        & …                   & $c^2$   & fm$^{-3}$ \\ \hline
\multirow{3}{*}{DDB} & \multicolumn{2}{c}{median}      & 2.148         & 2.567           & 11.13         & 12.66         & 466                 & 0.649   & 1.002     \\
                     & \multirow{2}{*}{90 \% CI} & min & 2.022         & 2.396           & 10.54         & 12.04         & 334                 & 0.520   & 0.865     \\
                     &                           & max & 2.366         & 2.857           & 11.85         & 13.28         & 648                 & 0.718   & 1.121     \\
                     &                           &     &               &                 &               &               &                     &         &           \\
\multirow{3}{*}{NL}  & \multicolumn{2}{c}{median}      & 2.062         & 2.446           & 10.92         & 12.44         & 423                 & 0.576   & 1.051     \\
                     & \multirow{2}{*}{90 \% CI} & min & 2.006         & 2.370           & 10.52         & 12.08         & 347                 & 0.446   & 0.904     \\
                     &                           & max & 2.260         & 2.715           & 11.70         & 13.03         & 582                 & 0.685   & 1.127     \\
                     &                           &     &               &                 &               &               &                     &         &           \\
\multirow{3}{*}{DDH} & \multicolumn{2}{c}{median}      & 2.242         & 2.712           & 10.97         & 12.21         & 423                 & 0.750   & 0.986     \\
                     & \multirow{2}{*}{90 \% CI} & min & 2.037         & 2.439           & 10.15         & 11.46         & 273                 & 0.727   & 0.887     \\
                     &                           & max & 2.380         & 2.898           & 11.52         & 12.75         & 546                 & 0.763   & 1.170     \\ \hline
\end{tabular}
\end{table}

\begin{figure*}
    \centering
    \includegraphics[width=0.8\linewidth]{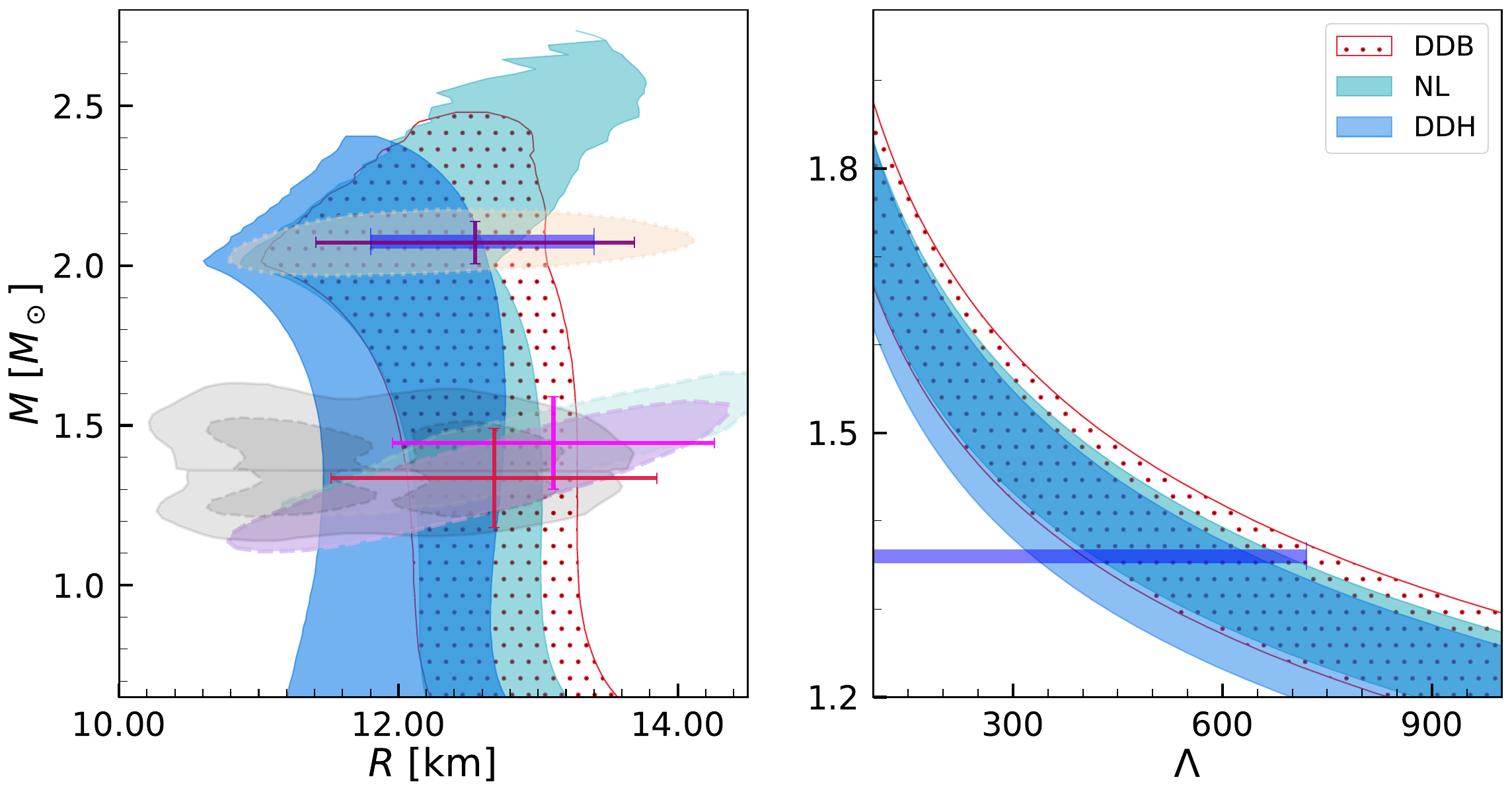}
    \caption{ NS mass-radius (left) and tidal deformability-mass regions obtained from the 90\% CI for the conditional probabilities $P(R|M)$ and $P(\Lambda|M)$ for DDB (dotted red), DDH  (blue) and NL (green) frameworks. The blue horizontal bar on the left panel indicates the 90\% CI radius for the pulsar PSR J0740+6620 with $M=$2.08$M_\odot$   obtained combining observational data from GW170817 and NICER as well as  nuclear saturation properties \cite{Miller:2021qha}. The gray shaded regions indicate the 90\% (solid) and 50\% (dashed) CI of the LIGO/Virgo analysis for high mass (top) and low mass (bottom)  components of the NS binary that originated the GW170817 event \cite{LIGOScientific:2018hze}. The NICER $1\sigma$ (68\%) credible zone of the 2-D mass-radii posterior distribution for the  PSR J0030+0451 (lilas and light green) \cite{Riley:2019yda,Miller:2019cac}, and the  PSR J0740 + 6620 (light orange) \cite{Riley:2021pdl,Miller:2021qha} are also included. The horizontal (radius) and vertical (mass) error bars reflect the $1\sigma$ credible interval derived for  NICER data's 1-D marginalized posterior distribution.}
    \label{fig:mrl}
\end{figure*}

Some comments are in order: i) as discussed before the incompressibility of DDH models peaks at a lower values than the other two, which present a similar behavior, and spreads over a larger range of values;  ii) concerning the skewness and kurtosis, which define the high density behavior of the EOS, DDH presents a very wide spread for the skewness from low negative to high positive values, and the kurtosis takes the largest values, to compensate the low incompressibility values it may take. This is necessary for the model to satisfy the 2$M_\odot$ constraint imposed. Concerning the other two models, DDB presents the most restricted distribution which we can identify as a subset of the one NL defines, that is disjoint from the set defined by DDH  for the kurtosis; it is interesting to verify that NL may take  small and even negative values of the kurtosis; %\red{can we see if this is related  with the speed of sound? }; 
iv) concerning the symmetry energy the distribution presented by the three models for the symmetry energy and slope at saturation is similar.  However, there  are differences  in the higher order parameters, in particular, $K_{sym,0}$ and $Z_{sym,0}$: the two models with density dependent coupling, DDH and DDB, behave in a similar way but NL spreads along a wider range of values and $K_{sym,0}$ may take positive values and $Z_{sym,0}$ takes very large negative values. The differences encountered are in part due to the fact that for DDH and DDB models the coupling of the $\varrho$-meson to the nucleons tends to zero at sufficiently large densities. Generalizing the parametrization of the  $\varrho$-meson coupling  will allow to go beyond this limitation. We will come back to this problem in one of the following sections.

NS properties, as the mass and radius, are determined from the integration of the Tolmann-Oppenheimer-Volkoff equations for spherical stars in statiscal equilibrium \cite{TOV1,TOV2}, see \cite{book.Glendenning1996} for a review. The tidal deformabilities $\Lambda$, quantities that are obtained from the detection of gravitational waves \cite{LIGOScientific:2017ync},  are obtained integrating the equations obtained in \cite{Hinderer2008}.

The radius and tidal deformability for NS with a given mass have been calculated within the three frameworks and the results are plotted in Fig. \ref{fig:mrl}, on the left side the radius-mass  and on the right side the tidal deformability-mass. Results of several observations, in particular, from the LIGO Virgo Collaboration for the  GW170817 \cite{LIGOScientific:2018cki} and from NICER \cite{Riley:2019yda,Miller:2019cac,Riley:2021pdl,Miller:2021qha} for the pulsars PSR J0030+0451 and  PSR J0740 + 6620, have been included.  The three data sets show different properties which reflect the different NMP the different sets have as discussed before.
The main conclusions that can be drawn are: i) NL data set is the most restricted for low mass stars presenting intermediate radii mostly between 12 and 13 km. DDH, the data set with smallest $K_0$ values, presents the smallest radii for low mass stars, $\sim 11.5-12.5$~km, while the DDB set predicts the largest radii, $\sim 12-13.5$~km. The density dependence of the EOS at high densities is strongly influenced by the non-linear terms in the NL data set and the function that defines the density dependence of the meson couplings in the other two sets, DDB and DDH. DDH  data set is soft at low densities so that low mass stars have a quite small radius, but at large densities becomes stiff  to allow maximum mass stars with almost 2.5$M_\odot$. The DDB data set allows for larger maximum masses than DDH, however, NL data set  attains the largest masses, close to 2.75$M_\odot$. All data sets agree with the presently  available  NS observations.
In the right panel, the tidal deformabilities are plotted as a function of the mass for the three data sets. Their behavior follows the one obtained for the radii, with DDH having the smallest values and DDB the highest. Only some models of DDB are outside the 90\% CI obtained from GW170817 value for a 1.36$M_\odot$ star (see the blue horizontal bar).

\begin{figure}
    \centering
    \includegraphics[width=0.95\linewidth]{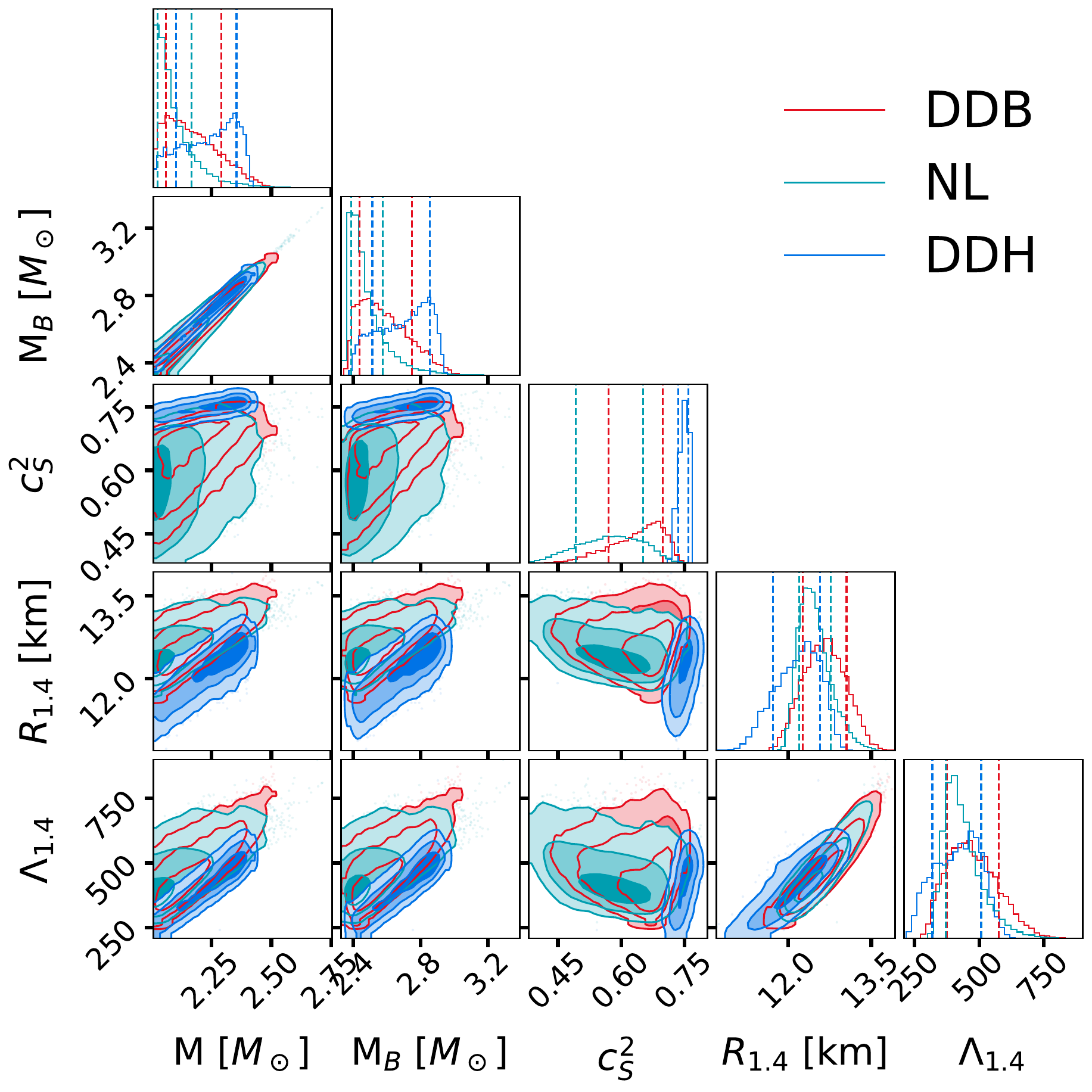}
    \caption{ Corner plot comparing the following NS properties obtained with the three data sets, DDB, NL and DDH: maximum gravitational mass $M$, maximum baryonic mass $M_B$ and respective central speed of sound  squared $c_s^2$, radius $R$ and tidal deformability  $\Lambda$  of a 1.4 $M_\odot$ star. }
    \label{fig:ns}
\end{figure}
The corner plot shown in  Fig. \ref{fig:ns} involving some NS properties allows some interesting conclusions: i) there is some correlation between the radius  and the tidal deformability of a 1.4$M_\odot$ stars and the  star maximum mass; ii) the central speed of sound squared of the maximum mass star is clearly model dependent: for DDH model,  $c_s^2\sim 0.7-0.8$ is pratically constant and quite high, while for the other two models $c_s^2$ can be as low as 0.45 or even lower and as high as 0.75; iii) it is precisely DDH with the largest speed of sound that predicts the smallest radii for 1.4$M_\odot$ stars.

\begin{table*}
\caption{Nuclear matter properties at saturation density, median values and 90\% CI, of the two data sets including hyperons, DDB-hyp and NL-hyp. Symmetric nuclear matter properties at saturation density $\rho_0$ defined by Eq. (\ref{x0}): binding energy    $\epsilon_0$, incompressibility $K_0$, skewness $Q_0$ and kurtosis $Z_0$. Symmetry energy properties at saturation  defined by Eq. (\ref{xsym}): symmetry energy $J_{\rm sym,0}$, slope $L_{\rm sym,0}$, incompressibility $K_{\rm sym,0}$, skewness $Q_{\rm sym,0}$ and kurtosis $Z_{\rm sym,0}$. }
\centering
\label{tab:nmp_hyp}
 \setlength{\tabcolsep}{7.5pt}
      \renewcommand{\arraystretch}{1.2}
\begin{tabular}{ccccccccccccc}
\hline \hline 
\multicolumn{3}{c}{\multirow{2}{*}{Model}}                & $\rho_0$  & $\epsilon_0$ & $K_0$ & $Q_0$ & $Z_0$ & $J_{\rm sym,0}$ & $L_{\rm sym,0}$ & $K_{\rm sym,0}$ & $Q_{\rm sym,0}$ & $Z_{\rm sym,0}$ \\ \cline{4-13} 
\multicolumn{3}{c}{}                                      & fm$^{-3}$ & \multicolumn{9}{c}{MeV}                                                                                                        \\ \hline
\multirow{3}{*}{DDB-hyp} & \multicolumn{2}{c}{median}     & 0.152     & -16.09       & 272   & 130   & 1425  & 32.15           & 43              & -98             & 966             & -6713           \\
                         & \multirow{2}{*}{90\% CI} & min & 0.147     & -16.39       & 247   & -16   & 803   & 29.48           & 26              & -127            & 354             & -12178          \\
                         &                          & max & 0.157     & -15.79       & 309   & 349   & 1680  & 34.83           & 65              & -59             & 1453            & -2723           \\ 
                         \\
\multirow{3}{*}{NL-hyp}  & \multicolumn{2}{c}{median}     & 0.150     & -16.09       & 296   & -117  & 2105  & 31.85           & 42              & -70             & 1312            & -13592          \\
                         & \multirow{2}{*}{90\% CI} & min & 0.144     & -16.41       & 270   & -246  & -405  & 29.16           & 31              & -127            & 895             & -18989          \\
                         &                          & max & 0.157     & -15.76       & 341   & 104   & 3078  & 34.44           & 57              & -12             & 1607            & -3893           \\ \\ 
\hline
\end{tabular}
\end{table*}

\subsection{Including hyperons \label{sec:hyp}}
In the inner core of a NS, non-nucleonic degrees of freedom may set in. In the present section we will discuss the onset of hyperons. As in \cite{Malik:2022zol}, we will introduce only two hyperons, the neutral $\Lambda$-hyperon and the negatively charged $\Xi^-$-hyperon. These two hyperons are the ones that appear in the largest fractions, either because of having the smallest hyperon mass as the $\Lambda$, or because of being negatively charged (the $\Xi^-$), and, therefore,  favorably replace the electrons and reduce the total pressure of the system. Both hyperons have an attractive potential in symmetric nuclear matter, and form hypernuclei. The binding energy of hyperons in hypernuclei has been used to fit the couplings of the hyperons to mesons in the RMF description of hadronic matter \cite{Fortin:2017cvt,Providencia:2018ywl}. Although the mass of the $\Sigma^-$-hyperon is smaller than the one of $\Xi^-$, it interacts repulsively with nuclear matter, as the non-existence of $\Sigma$-hypernuclei seems to show \cite{Gal:2016boi}. As a consequence, in NS matter its onset occurs at larger densities than the $\Xi^-$ onset \cite{Weissenborn:2011kb,Fortin2016,Fortin:2020qin,Stone:2019blq}.

The introduction of hyperons requires a generalization of the Dirac term of Eq. (\ref{lagrangian}) to include $\Lambda$ and $\Xi^-$ besides protons and neutrons,
\begin{eqnarray}
\mathcal{L}_D=&&\sum_{j=p,n,\Lambda,\Xi^-}\bar{\Psi}_j\Big[\gamma^{\mu}\left(i \partial_{\mu}-\Gamma_{\omega,j} A_{\mu}^{(\omega)}-%\frac{1}{2}
\Gamma_{\varrho,j} {\boldsymbol{t}}_j \cdot \boldsymbol{A}_{\mu}^{(\varrho)}\right) \Big.\nonumber\\&&
\Big.-\left(m-\Gamma_{\sigma,j} \phi\right)\Big] \Psi 
\end{eqnarray}
For the couplings of the hyperons to the vector-mesons we consider the SU(6) values for the  vector isoscalar mesons, $g_{\omega\Xi}=\frac{1}{3} g_{\omega N} = \frac{1}{2} g_{\omega\Lambda} $ and 
		%\label{eq:SU6-relation2}
$ g_{\phi\Xi} = 2 g_{\phi\Lambda} =- \frac{2\sqrt{2}}{3} g_{\omega N} $
%	\label{eq:SU6-relation2b}
and for the isovector $\varrho$-meson, $g_{\varrho\Xi} =  g_{\varrho N}$.
 In this last case the hyperon isospin also defines the strength of the coupling.  
 Having assumed these values for the couplings of the hyperons to the vector mesons, the coupling to the $\sigma$-meson is fitted to hypernuclei properties
\cite{Fortin:2017cvt,Fortin:2017dsj,Providencia:2018ywl}.  In general,  we express the couplings to the mesons as a fraction of the  nucleon couplings, $g_{m \,i}=x_{m\, i}\, g_{\sigma}$ with  $m=\sigma,\, \omega,\, \varrho$ and $i=\Lambda$ and $\Xi^-$. Considering  the results of the fits done in \cite{Fortin:2017cvt,Fortin:2017dsj,Providencia:2018ywl}, values between 0.609 and 0.622 were determined for the  fraction $x_{\sigma\Lambda}$, and will be adopted in the present study. For the fraction $x_{\sigma\Xi^-}$, the range  0.309 to 0.321 will be used, as determined from fits to  the binding energy of $\Xi^-$ in the  hypernuclei $^{15}_{\Xi}$C and $^{12}_{\Xi}$Be \cite{Fortin:2020qin}.

\begin{figure*}
    \centering
    \includegraphics[width=0.75\linewidth]{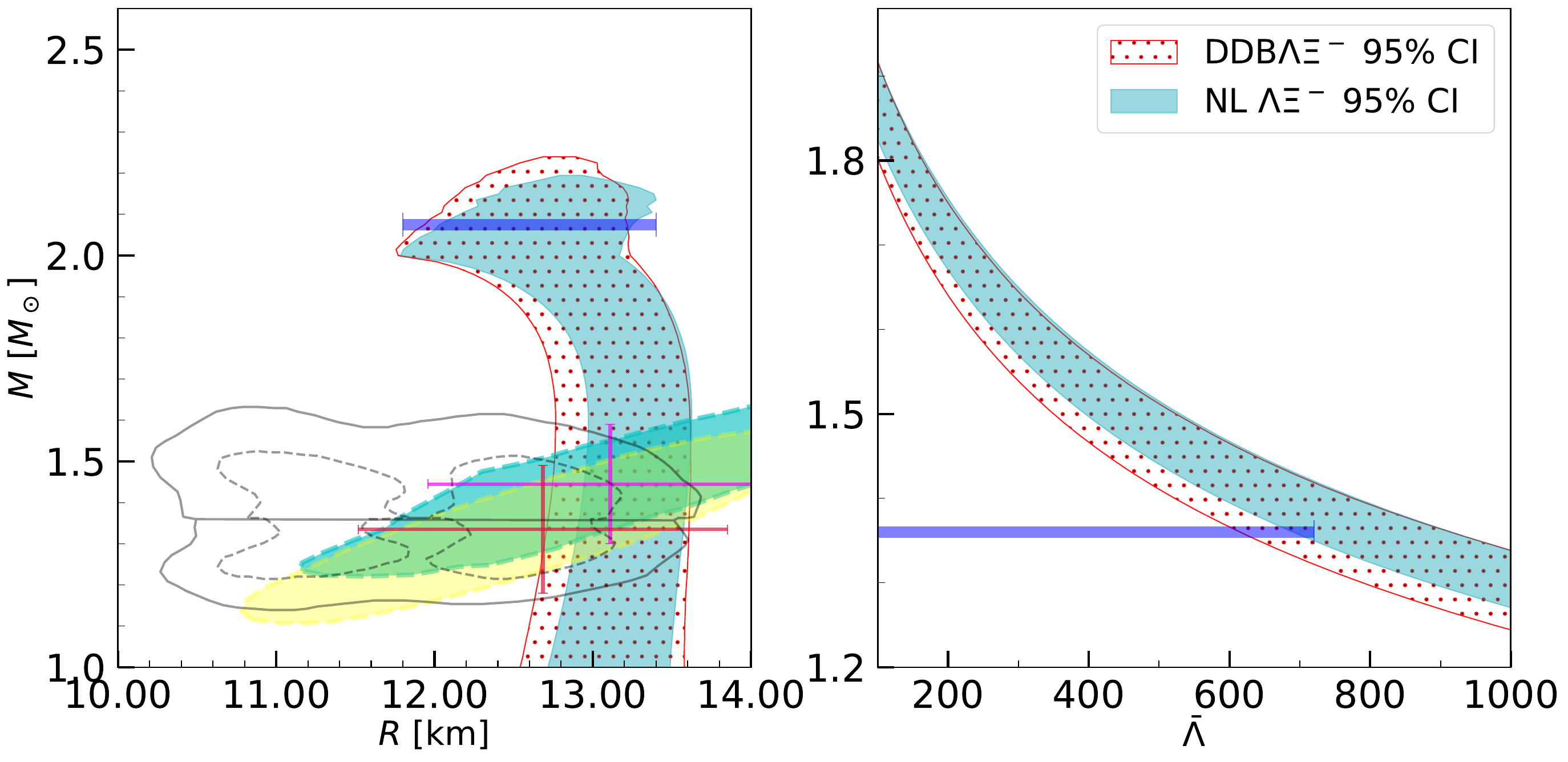}
    \caption{ The 90\% CI region for the hyperon data sets DDB-hyp (dotted red) and  NL-hyp (green)  derived using the conditional probabilities $P(R|M)$ (left) and $P(\Lambda|M)$ (right). The lines in the left panel indicate the 90\% (solid) and 50\% (dashed) CI for the binary components of the GW170817 event \cite{LIGOScientific:2018hze}. Also shown is the $1\sigma$ (68\%) credible 2-D posterior distribution in the  mass-radii domain from the millisecond pulsar PSR J0030+0451 (cyan and yellow) \cite{Riley:2019yda,Miller:2019cac} obtained from the NICER x-ray data. The horizontal (radius) and vertical (mass) error red bars reflect the $1\sigma$ credible interval derived for the same NICER data's 1-D marginalized posterior distribution. The blue bars represent  the radius of the PSR J0740+6620 at 2.08$M_\odot$ (left panel) and the tidal deformability at 1.36 $M_\odot$ (right panel) \cite{LIGOScientific:2018cki}. 
    }
    \label{fig:mrl_hyp}
\end{figure*}

We have performed calculations for hyperonic stars within two models, DDB and NL, imposing the same fit data that was considered to constrain the EOS data sets of nucleon matter, and which is summarized in Table \ref{fig:target}.  Chemical equilibrium dictates that:
\begin{eqnarray}
\mu_\Lambda&=& \sqrt{(m^*_\Lambda)^2+ k^2_{F\Lambda}}+x_{\omega\Lambda}\,g_\omega \omega=\mu_n    \label{muL}\\
\mu_\Xi&=&\sqrt{(m^*_\Xi)^2+ k^2_{F\Xi}}+x_{\omega\Xi}\,g_\omega \omega-\frac{1}{2} g_\varrho \varrho= \mu_n+\mu_e, \nonumber\\ \label{muX}
\end{eqnarray}
 where $m^*_i$ is the effective mass of hyperon $i$ and $k_{Fi}$ its Fermi momentum. Charge neutrality imposes that $\rho_p=\rho_\Xi+\rho_e+\rho_\mu.$

In Fig. \ref{fig:mrl_hyp}, predictions obtained  for the NS radius (left) and tidal deformability (right) for diferent NS masses  are plotted. In Table \ref{tab:nmp_hyp}, the median and the 90\% CI nuclear matter properties  of both data sets are summarized and in Table  \ref{tab:ns_hyp} some NS properties are given, in particular, the median and  the 90\%CI of the maximum mass, respective, baryonic mass, radius,  central speed of sound squared and central baryonic density, together with the radius and tidal deformability of a 1.4$M\odot$ star.

We first discuss the effect on the nuclear matter properties of including hyperons, comparing results of Tables \ref{tab:nmp} and \ref{tab:nmp_hyp}.  Isovector properties are essentially not affected for the DDB data set, and only slightly for the NL data set reflected in a small increase of the different properties. Isoscalar properties are the mostly affected: the incompressibility $K_0$ suffers an increase of 15\%-20\%, and the median skewness  becomes positive. The reason for this change is the fact that the presence of the onset of hyperons relieves the pressure inside the NS and the condition that 2$M_\odot$ stars must be described obliges the EOS to be  harder, mainly affecting the isoscalar channel of the EOS.

The implication of the hardening of the EOS is that larger NS radii are predicted (compare Fig. \ref{fig:mrl_hyp} left with \ref{fig:mrl} left). The median values of the radius of  $1.4\, M_\odot$ stars reflect clearly this effect: they increase from  12.66 (12.44)~km for DDB (NL) to 14.22 (13.11)~km, i.e. more than $\sim 0.5$~km or $\sim 5\%$. Measurements of the NS radius with an uncertainty smaller than $5\%$, such as the ones programmed with eXTP \cite{extp_watts} and STROBE-X \cite{strobex}, could distinguish between these two scenarios. Also the tidal deformability is strongly affected increasing its median value from 466 (423) to 650 (610), respectively, for DDB (NL), and the constraint imposed by GW170817 is essentially not satisfied (see Fig. \ref{fig:mrl_hyp} middle panel). Another important property that distinguishes both scenarios is the NS maximum mass
that decreases from a maximum value at 90\% CI of 2.37 (2.26) $M_\odot$ for DDB (NL) to 2.08 (2.13) $M_\odot$.  Concerning the NS properties in the center of the star it is pointed out the decrease of the speed of sound, its square decreasing essentially to values of the order of 0.5 or below in the presence of hyperons, when it takes values of the order of 0.6 going up to $\sim 0.7$ if only nucleon matter is considered.  In Sec. \ref{sec:speed} the speed of sound in matter with hyperons will be compared with the one obtained with nucleonic models. 

\begin{table}[]
\label{tab:ns_hyp}
\caption{NS properties, the median and the 90\% CI, of the data sets with hyperons, DDB-hyp and NL-hyp. The following properties are given: the maximum mass $M_{\rm max}$ and respective baryonic mass $M_{\rm B,max}$, radius $R_{\rm max}$, speed of the sound squared at the center $c_s^2$ and central baryonic density $\rho_c$, and the radius and tidal deformability of the 1.4$M_\odot$ star, $R_{\rm 1.4}$ and $\Lambda_{\rm 1.4}$.}
 \setlength{\tabcolsep}{1.5pt}
      \renewcommand{\arraystretch}{1.2}
\begin{tabular}{cccccccccc}
\hline \hline 
\multicolumn{3}{c}{\multirow{2}{*}{Model}}                & $M_{\rm max}$ & $M_{\rm B,max}$ & $R_{\rm max}$ & $R_{\rm 1.4}$ & $\Lambda_{\rm 1.4}$ & $C_s^2$ & $\rho_c$  \\ \cline{4-10} 
\multicolumn{3}{c}{}                                      & \multicolumn{2}{c}{$M_\odot$}   & \multicolumn{2}{c}{km}        & …                   & $c^2$   & fm$^{-3}$ \\ \hline
\multirow{3}{*}{NL-hyp} & \multicolumn{2}{c}{median}      & 2.024         & 2.357           & 11.82         & 13.22         & 659                 & 0.47    & 0.920     \\
                        & \multirow{2}{*}{90 \% CI} & min & 2.003         & 2.329           & 11.55         & 12.97         & 595                 & 0.41    & 0.860     \\
                        &                           & max & 2.083         & 2.433           & 12.20         & 13.52         & 758                 & 0.51    & 0.968     \\
                        &                           &     &               &                 &               &               &                     &         &           \\
\multirow{3}{*}{NL-hyp} & \multicolumn{2}{c}{median}      & 2.040         & 2.385           & 11.73         & 13.11         & 610                 & 0.48    & 0.932     \\
                        & \multirow{2}{*}{90 \% CI} & min & 1.992         & 2.322           & 11.48         & 12.76         & 526                 & 0.44    & 0.871     \\
                        &                           & max & 2.130         & 2.501           & 12.08         & 13.51         & 743                 & 0.50    & 0.964     \\ \hline
\end{tabular}
\end{table}

\begin{figure*}
    \centering
    \includegraphics[width=.9\linewidth]{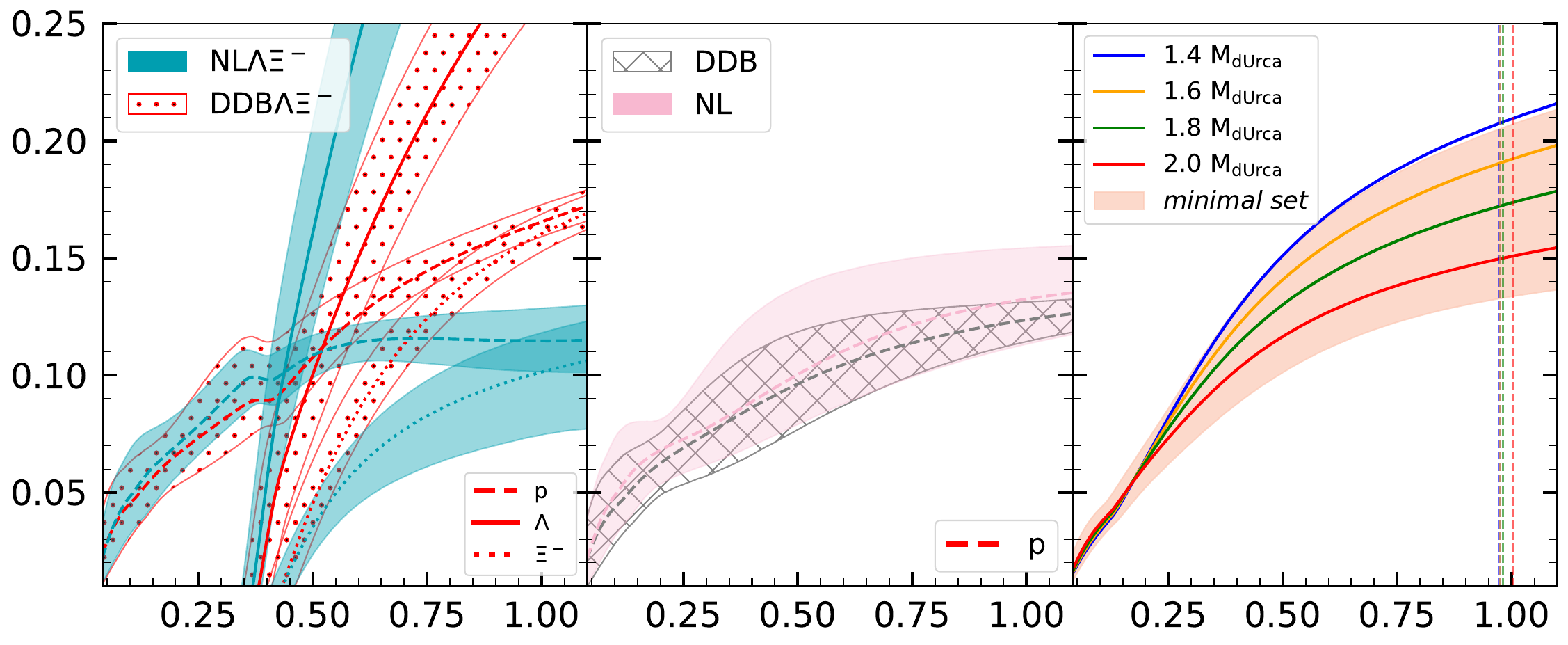}
    \caption{Proton and hyperons fractions for data sets DDB-hyp (red dotted) and NL-hyp (blue) (left panel), proton fractions for the data sets DDB (slashed region) and NL (pink region) (middle panel),  proton fraction  defining the minimal set compatible with chEFT PNM calculations at $2\sigma$  obtained varying the parameter  $y$ introduced in Eq. (\ref{hm2}) (pink region), or fixing $y$ to a given $M_{dUrca}$, i.e. 1.4, 1.6, 1.8, 2.0 $M_\odot$ (full lines). }
    \label{fig:proton}
\end{figure*}

\subsection{Onset of direct Urca}
In this subsection the dependence on the baryon density of the proton fraction of $\beta$-equilibrium matter  will be discussed. We consider two of the microscopic models discussed in Sec. \ref{sec:comp}, DDB and NL. Within the same models, the onset of hyperons and its influence on the proton fraction will also be commented using the data sets described in the previous section,  Sec. \ref{sec:hyp}.

The $\varrho$-meson coupling to the nucleon in  DDB and DDH data sets  decreases exponentially with the density. At 
 high densities it approaches zero, and,  as a consequence,  these models allow for very asymmetric matter at high densities since the symmetry energy is low. Therefore, the opening of nucleon direct Urca processes  \cite{Yakovlev:2000jp,Yakovlev:2004iq}
does not occur inside NS \cite{Fortin:2016hny,Fortin:2020qin,Fortin:2021umb}.  This is clearly seen from the middle panel in Fig. \ref{fig:proton}, where the median and 90\% CI bands of the proton fraction are plotted as a function of the baryonic density for the data sets DDB (slashed band) and for NL (pink band):  at large densities the proton fraction for DDB is smaller  and narrower than the NL proton fraction. The smaller width is also an indication that the $g_\varrho$ coupling of all models tends to the same value, zero, at high densities.  NL models, however, span a wider range of proton fractions, and in particular, the opening of direct Urca processes may occur in some models.

In the left panel of the same figure the fraction of protons is plotted together with the $\Lambda$ and $\Xi^-$ hyperon fractions for matter including hyperons. The $\Lambda$-hyperon is the first hyperon to set in just above twice saturation density, while the $\Xi^-$ sets in just below 3~$\rho_0$. The onset of the $\Lambda$-hyperon implies a decrease of the neutron fraction, decreasing the pressure caused by this species and, therefore, the system energy. As a consequence the proton fraction also decreases (see discussion in \cite{Malik:2022jqc}). As soon as the $\Xi^-$-hyperon sets in the proton fraction suffers an increase to compensate for the negatively charged hyperon. This behavior is well illustrated in the left panel of \ref{fig:proton}. The  effect if much stronger for the DDB-hyp data set because the $\varrho$-meson coupling is weaker and, therefore, the repulsive term that enters the $\Xi$ chemical potential is weaker, see Eq. (\ref{muX}). The coupling $g_\varrho$ ($\Gamma_\varrho$)  varies at  90\% CI  within the range $[9.55,14.60]$  ( $[6.97,8.78]$ ) for NL  (DDB) at saturation density. Including hyperons in the model these values change only slightly to $[9.79,14.31]$  for NL-hyp and  $[7.13,8.58]$  for  DDB-hyp at saturation density.

It was shown in Sec. \ref{sec:comp}, in particular, with the corner plot \ref{fig:nmp}, that while DDB and DDH data sets differ a lot when comparing the symmetric nuclear matter properties, the symmetry energy properties are very similar considering all orders of the Taylor expansion studied. In order to overcome the special feature of these models  of not allowing for nucleon direct Urca processes, in \cite{Malik:2022ilb} a generalization of the $\varrho$-meson coupling was proposed including a new parameter $y$. For the function $h_\varrho(x)$  that defines the
density dependence of the coupling  $\Gamma_\varrho$, see Eq. (\ref{hm2}),  we consider 
\begin{equation}
h_\varrho(x) = y ~ \exp[-a_\varrho (x-1)] + (y-1) ~, \quad 0<y\le 1~.
\label{hm2}
\end{equation}
\begin{figure}
    \centering
    \includegraphics[width=0.85\linewidth]{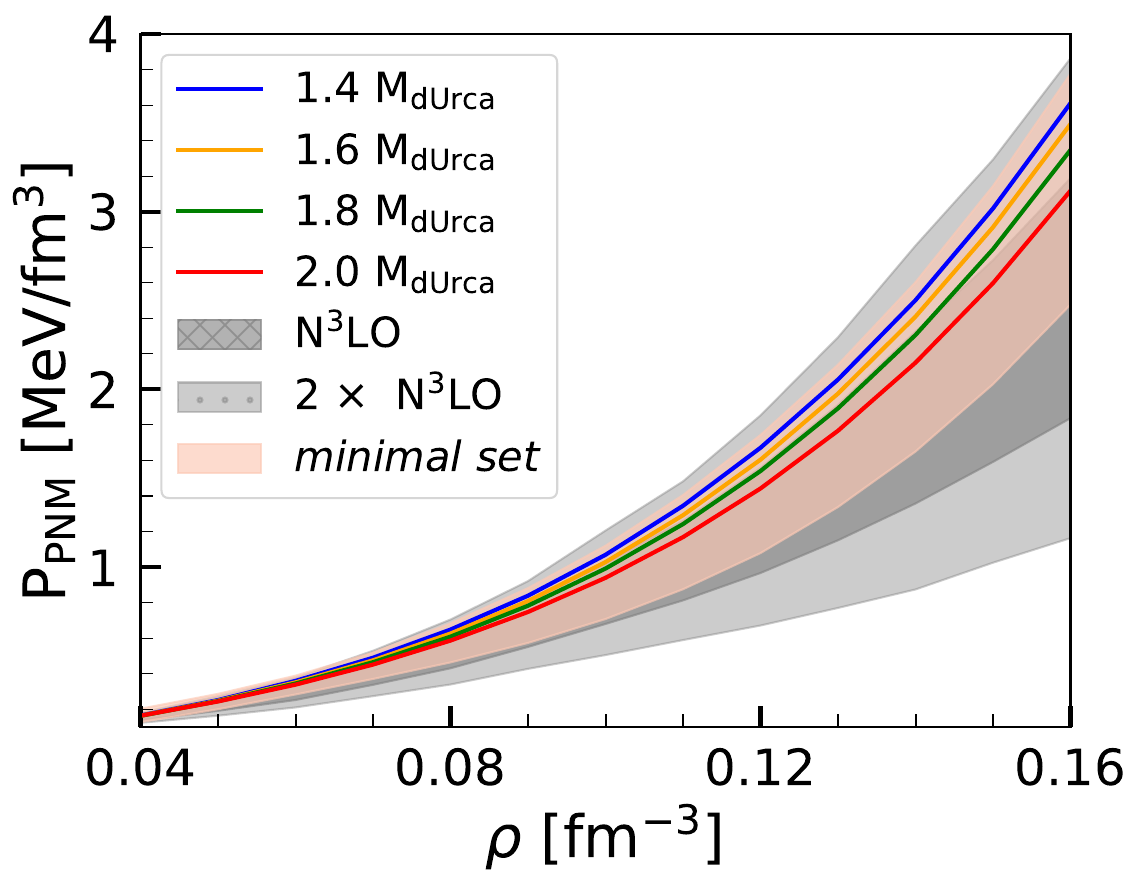}
    \caption{Pure neutron matter pressure as a function of the baryon density for the DDB model  using Eq. (\ref{hm2}) to define the $\varrho$-meson coupling:  
the minimal set compatible with chEFT PNM calculations at $2\sigma$  obtained varying the parameter  $y$  (pink region), or fixing $y$ to a given $M_{dUrca}$, i.e. 1.4, 1.6, 1.8, 2.0 $M_\odot$ (full lines). The dark (light) gray bands define the chEFT pressure at 1$\sigma (2\sigma)$ from \cite{Hebeler2013}.}
    \label{fig:pnm}
\end{figure}
Allowing $y$, imposing the constraints defined in Table \ref{tab1},  we have generated the PNM  pink band in Fig. \ref{fig:pnm}. In the same figure, the chEFT PNM EOS from \cite{Hebeler2013} is also included considering 1 $\sigma$ (dark gray) and 2 $\sigma$ (light gray), as well as the PNM EOS imposing  that  $M_{dUrca}$ is 1.4, 1.6, 1.8 and 2.0 $M_\odot$, where  we designate by $M_{dUrca}$, the mass of the star where nucleon  direct Urca processes set in its center.
In fact, the new parameter $y$ must be determined from NS properties that are sensitive to the high density behavior of the symmetry energy, such as the proton fraction. In particular, the onset of nucleon direct Urca  (dUrca) processes are an appropriate observation and were used in \cite{Malik:2022ilb} to constraint $y$.
In Fig. \ref{fig:proton} right panel, we show  the proton fraction (full lines) corresponding to different $M_{dUrca}$. This was possible by choosing the adequate $y$. The pink region spans the proton fraction compatible with PNM chEFT calculation at $2\sigma$, already defined in Fig. \ref{fig:pnm}.
 
These constraints derived from pure neutron matter exclude dUrca processes from stars with a mass $\lesssim$ 1.4 $M_\odot$ at $2\sigma$. If we restrict ourselves to 1$\sigma$, $M_{dUrca}$ rises to a value above 1.6 $M_\odot$.  These results are in agreement with the analysis performed in \cite{Beznogov:2015ewa}, where it is concluded that  NS cooling curves seem to indicate that  M$_{\rm dUrca}$ $\sim 1.6 - 1.8$ M$_\odot$.

\subsection{Speed of sound, trace anomaly, and pQCD constraints \label{sec:speed}}
Lately, some discussion has been concentrated on the behavior of the speed of sound with density. This quantity, which is directly related to the dependence of the pressure  on the energy density, is sensitive to the onset of new degrees of freedom and first-order phase transitions.  In particular, at high densities, it is expected that matter is deconfined and  exhibits conformal symmetry with the square of the speed of sound being equal to 1/3. One of the present great interests is to identify possible signatures of the presence of deconfined  quark matter inside NS. 

The general behavior of the speed of sound squared obtained from agnostic descriptions of the EOS of baryonic matter, that has been constrained by low-density pure neutron matter {\it ab-initio}  calculations \cite{Hebeler2013,Drischler:2017wtt,Drischler:2020yad} and the pQCD EOS at densities of the order $\gtrsim 40 \rho_0$, and by NS observations, includes a steep increase until an energy density of the order of $\sim 500$~MeV/fm is attained, followed by a decrease or flattening,  approaching 1/3 at high densities \cite{Annala2019,Altiparmak:2022bke,Somasundaram:2022ztm,Gorda:2022jvk,Kurkela:2022elj,Annala:2023cwx}, see also the discussion in \cite{Kojo:2020krb}.

\begin{figure*}
    \centering
    \includegraphics[width=0.32\linewidth]{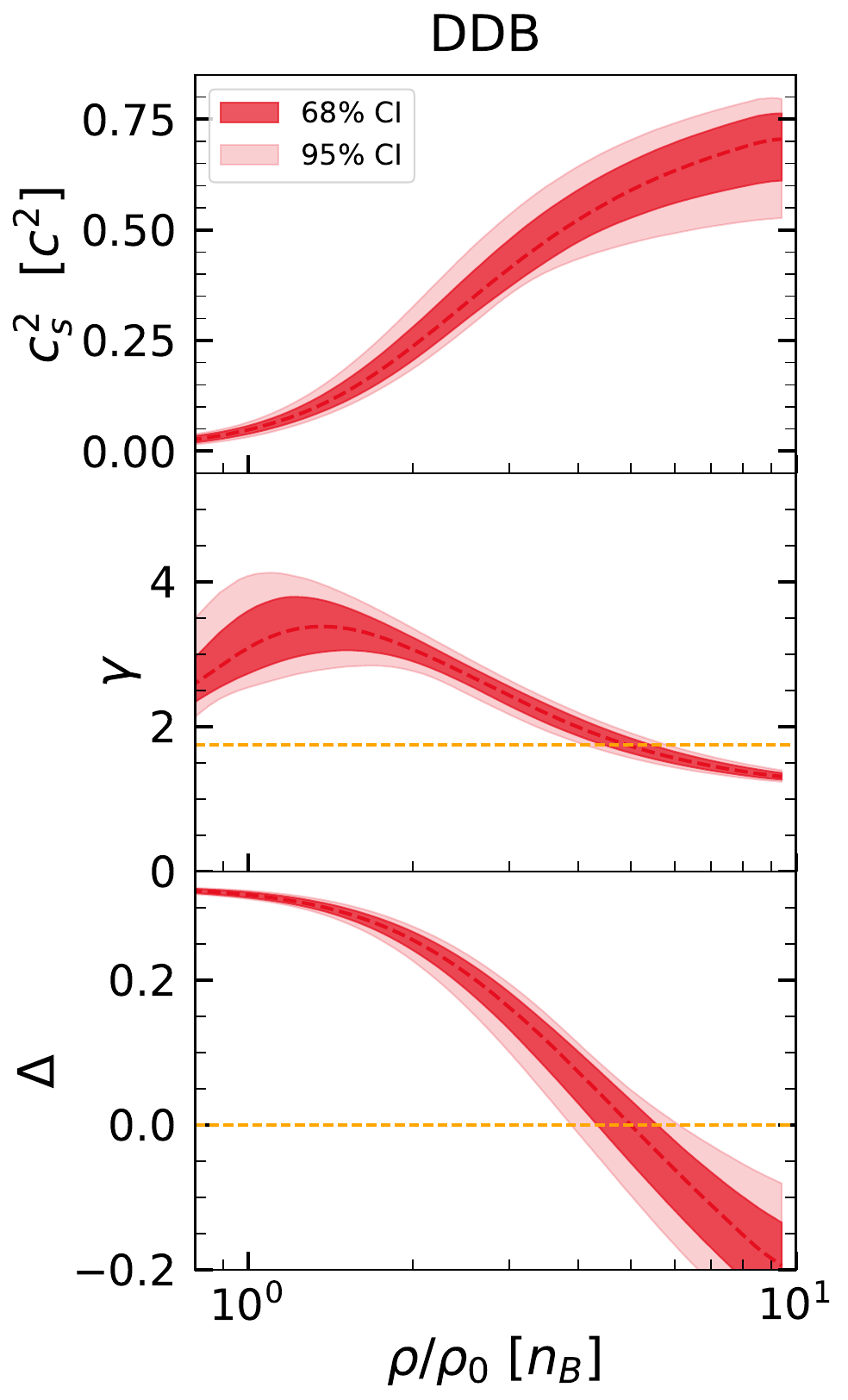}
    \includegraphics[width=0.32\linewidth]{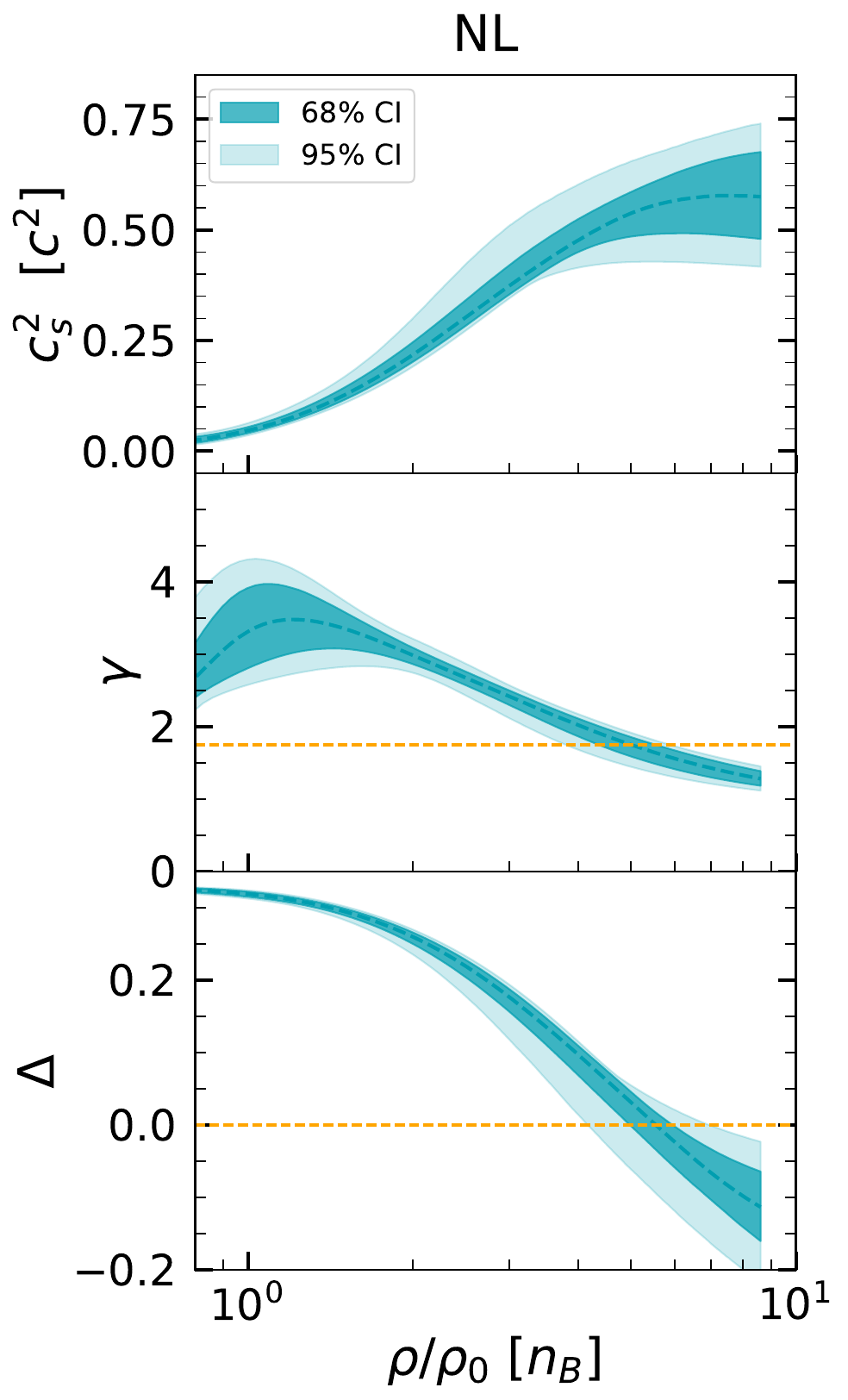}
    \includegraphics[width=0.32\linewidth]{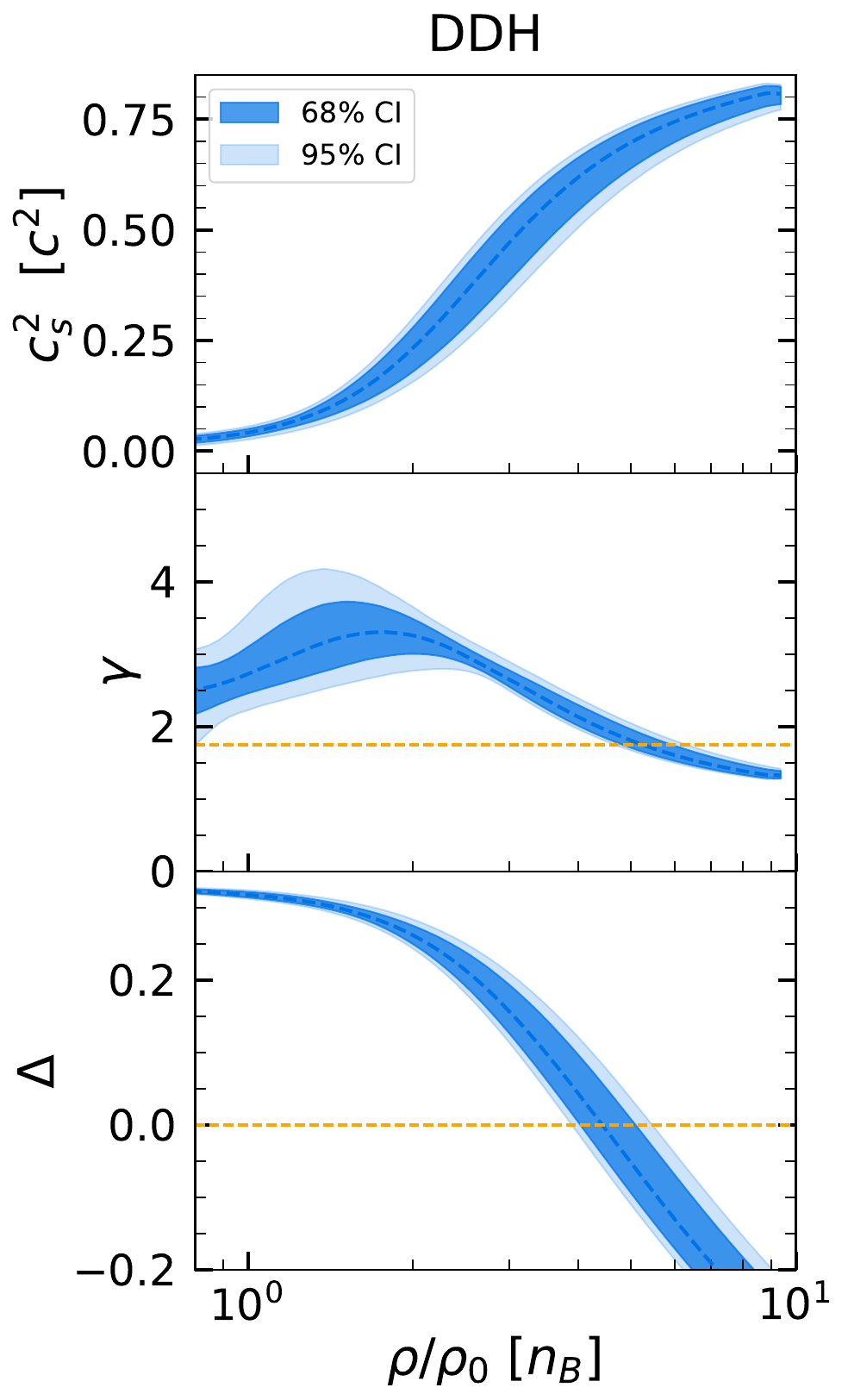}
    \caption{The speed of sound squared $c_s^2$, the polytropic index $\gamma=d \mbox{ln}P/d\mbox{ln}\epsilon$ and the trace anomaly $\Delta=1/3-P/\epsilon$ for the three data sets, DDB, NL and DDH. The horizontal lines in the $\gamma$ plots identifies the value 1.75.} %\red{I think it would be good to have a scale on the right side of the panels}. %\red{2- Could we se how csderiv behavings representing $c_s^2- (1/3-\Delta$, see eq. (6) of https://arxiv.org/pdf/2207.06753.pdf}}
    \label{fig:del}
\end{figure*}

In Fig. \ref{fig:del}, the top panels of the three columns show the behavior of the speed of sound squared for the three data sets DDB, NL and DDH, in particular, the 68\% and 95\% CI are shown. The different sets present a different behavior: for set DDH $c_s^2$ increases monotonically with a small dispersion, and attains values close to 0.8 for densities of the order of 1~fm$^{-3}$; set NL is on the other extreme, and above $\rho\sim 0.3$~fm$^{-3}$ shows a quite large dispersion including a flattening or slight decrease, never attaining values above 0.7 and presenting values that can go below 0.4; DDB shows an intermediate behavior, not so extreme as DDH, but also showing a monotonic increase. 

\begin{figure*}
    \centering
    \begin{tabular}{ccc}

  \includegraphics[width=0.33\linewidth]{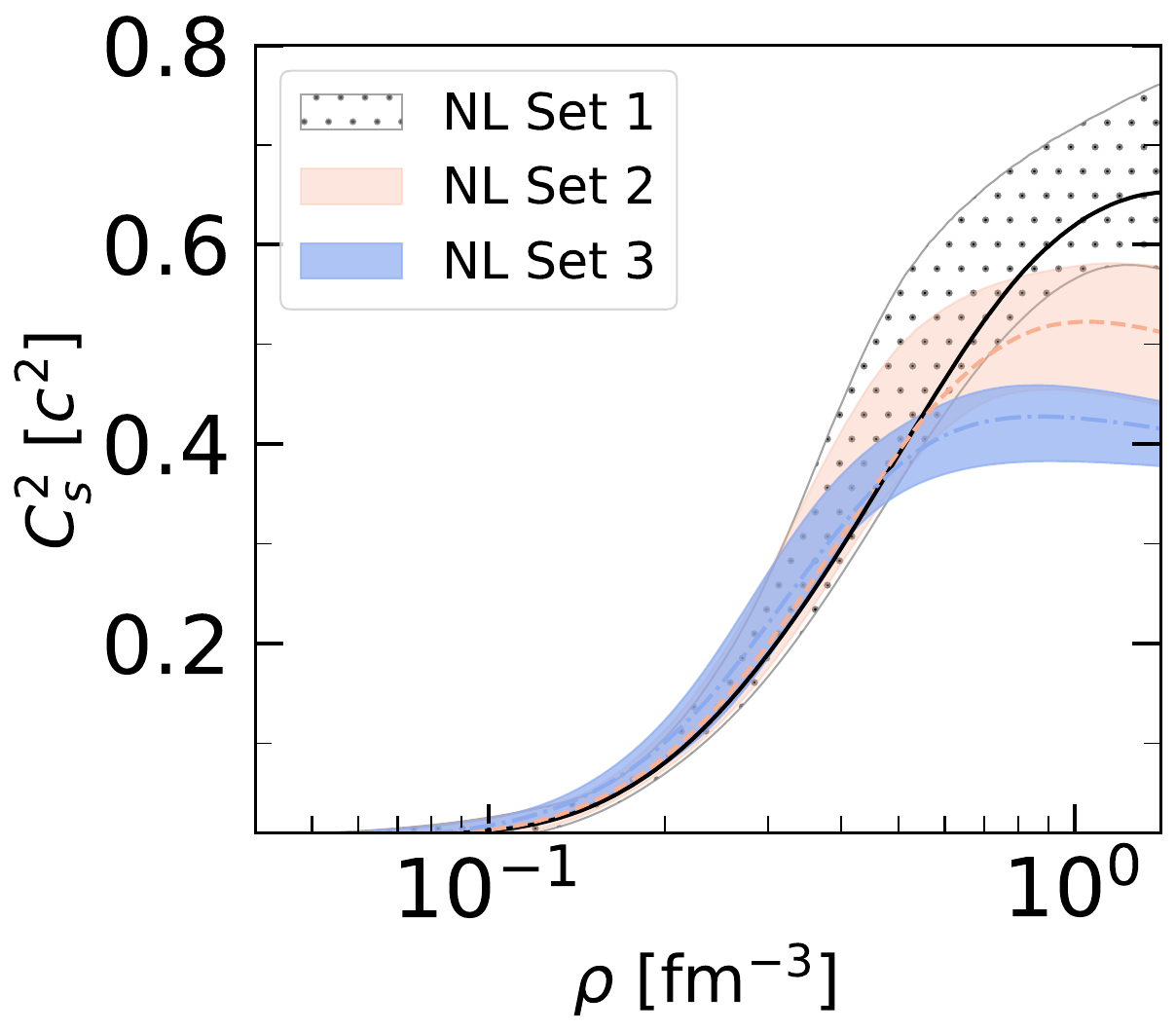}&
  \includegraphics[width=0.33\linewidth]{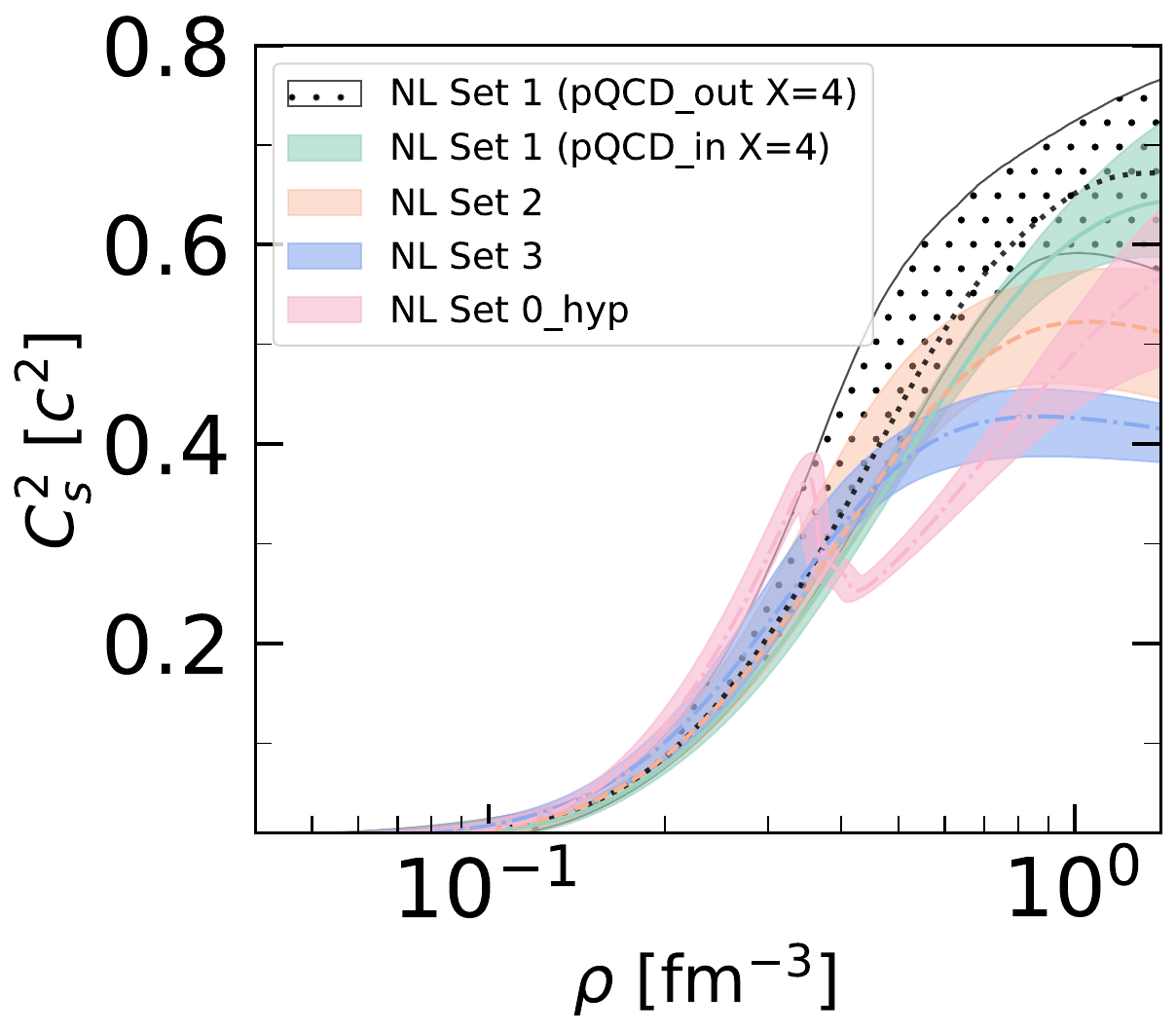} &
    \includegraphics[width=0.27\linewidth]{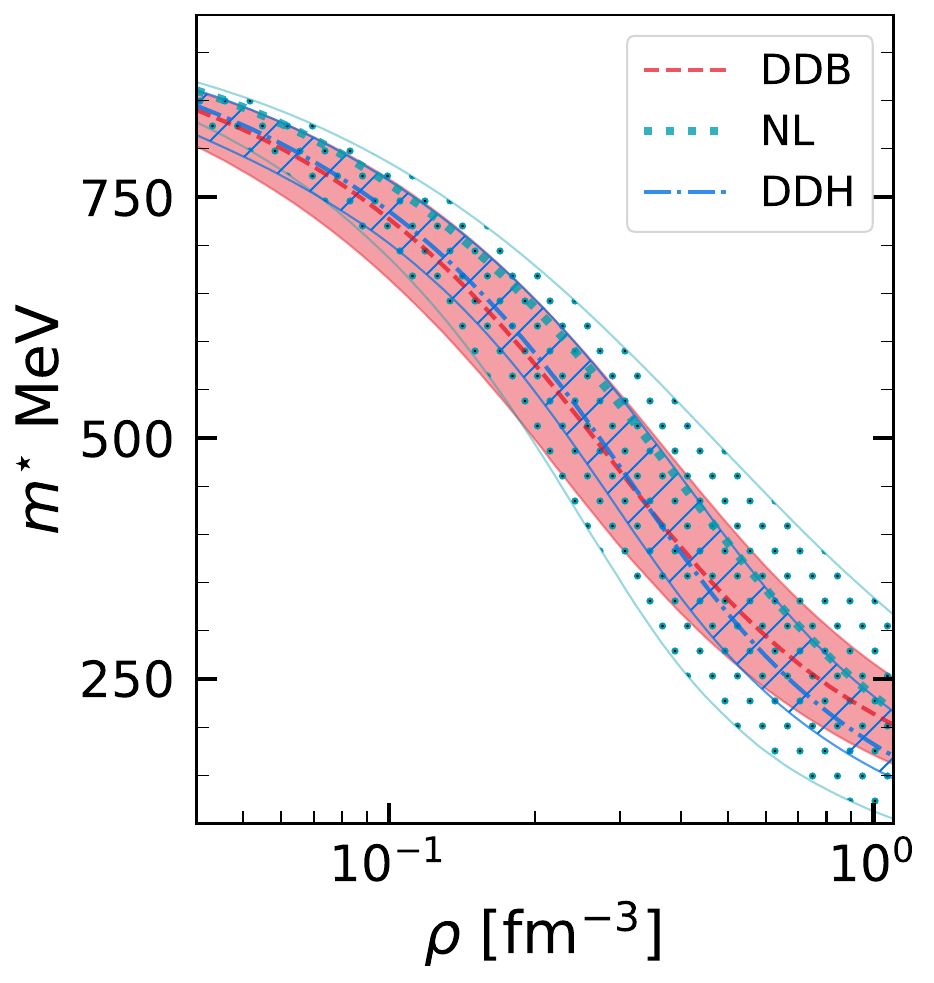}
   \end{tabular}
   \caption{The figure displays the median and 95\% credible interval of the 
    square of sound velocity ($c_s^2$) for the NL model  (left and middle panels)  and the  effective mass  (right) for DDB, NL, and DDH models as a function of baryon density. The left panel  highlights three distinct intervals of the parameter $\xi: \text{NL Set 1}, \text{NL Set 2},$ and $\text{NL Set 3}$. In the plot, NL Set 1 is represented by a black dotted region and corresponds to $\xi$ values within the interval [0, 0.004]. NL Set 2 is represented by an orange region and encompasses $\xi$ values within the range [0.004, 0.015]. NL Set 3 is depicted in blue and represents $\xi$ values within the interval [0.015, 0.04].  Each set of EOS contains a comparable number of samples, approximately 18,000 samples, providing a robust statistical basis for the displayed results. {In the middle panel, Set 1  was divided into two parts: green (black dotted)  EoS that satisfy  (do not satisfy)  pQCD constraints with X=4. In this panel the $c_s^2$ for the NL-hyp set is also shown (pink band).} 
    The right panel was obtained with the data sets presented in Sec. \ref{sec:comp}, which contain $\sim 15,000$ to 17,000 samples.}
    \label{fig:nl-cs2-allset}
\end{figure*}

The NL data set contains EOS with quite different behaviors at high densities, controlled by the  $\omega^4$ term. In the left panel of Fig. \ref{fig:nl-cs2-allset}, the speed of sound squared is plotted for different ranges of the parameter $\xi$,  for set 1 $\xi \in [0.0,0.004]$, for set 2 $\xi \in [0.004,0.015]$  and for set 3 $\xi \in [0.014,0.04]$. This parameter controls the contribution of the $\omega^4$ term in the Lagrangian density, and as discussed in \cite{Mueller:1996pm,Malik:2023mnx}, in the high density limit it makes  the speed of sound  squared go to 1/3. This indicates that a quite large range of values of the speed of sound squared are possible considering just nuclear degrees of freedom. In the middle panel, $c_s^2$ for the NL-hyp set has also been included. This set presents a peak above $2\rho_0$, when the hyperons set in, followed by a monotonous increase of the speed of sound, attaining  values $c_s^2\lesssim 0.6$ at 1~fm$^{-3}$.

Several quantities have been proposed as indicators of the presence of deconfined matter, including the polytropic index $\gamma=d \mbox{ln}P/d\mbox{ln}\epsilon$ \cite{Annala2019}, which takes the value 1 in conformal matter, the trace anomaly scaled by the energy density  introduced in \cite{Fujimoto:2022ohj} $\Delta=1/3-P/\epsilon$ which should approach zero in the conformal limit, and the derived quantity proposed in \cite{Annala:2023cwx} $d_c=\sqrt{\Delta^2+ {\Delta}^{'2}}$, where  $\Delta'= c_s^2 \, \left(1/\gamma-1\right)$ is the logarithmic derivative of $\Delta$ with respect to the energy density, which approaches zero in the conformal limit. In the following, we will discuss how  these quantities behave when we consider the different EOS data sets introduced in the present study. This will allow to identify properties that are still present in hadronic matter from properties that totally characterize deconfined matter.

In middle and bottom lines of Fig.  \ref{fig:del}, the  polytropic index $\gamma$ and  the trace anomaly $\Delta$ are plotted as a function of the baryonic density in units of the saturation density. The horizontal line in the $\gamma$ panels identifies the value 1.75 that has been proposed as indicating the transition to deconfined quark matter \cite{Annala2019}. For all models the polytropic index $\gamma$ increases until $\sim \rho_0$, followed by a monotonous decrease that goes below 1.75 at a density above $\sim 0.4$~fm$^{-3}$. The behavior of the three sets is quite consistent and it seems to indicate that a value of $\gamma<1.75$ is not enough to identify a phase transition to deconfined matter.  The normalized trace anomaly shows a behavior similar to the one discussed in \cite{Fujimoto:2022ohj}, where results from several studies \cite{Al-Mamun:2020vzu,Raaijmakers:2021uju,Gorda:2022jvk,Drischler:2021bup} have been compared, and it crosses the zero axis at densities of the order of 0.4-0.8~fm$^{-3}$, becoming afterwards negative. At sufficiently high densities this quantity should tend to the pQCD values that are slightly positive. Considering the models studied, for the NL data set (and even DDB) $\Delta$ shows a change of slope around $1$~fm$^{-3}$, which could match a  positive trace anomaly  in finite density QCD.

\begin{figure}
    \centering
    \includegraphics[width=0.49\linewidth]{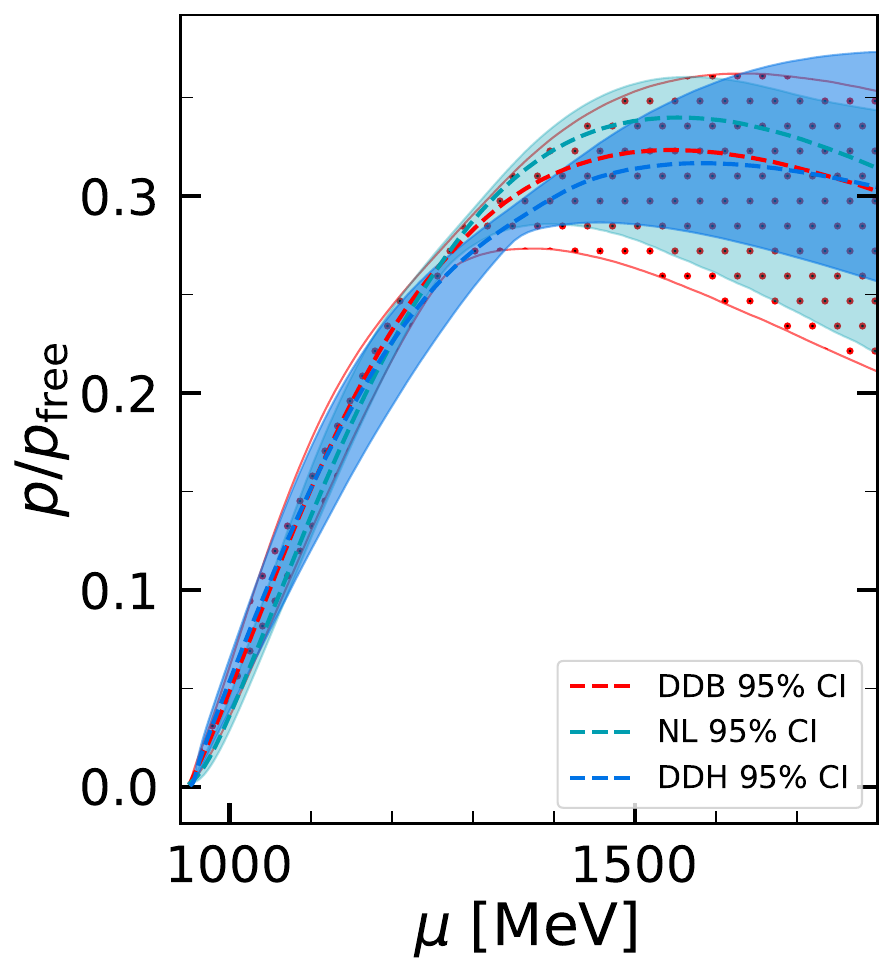}
    \includegraphics[width=0.49\linewidth]{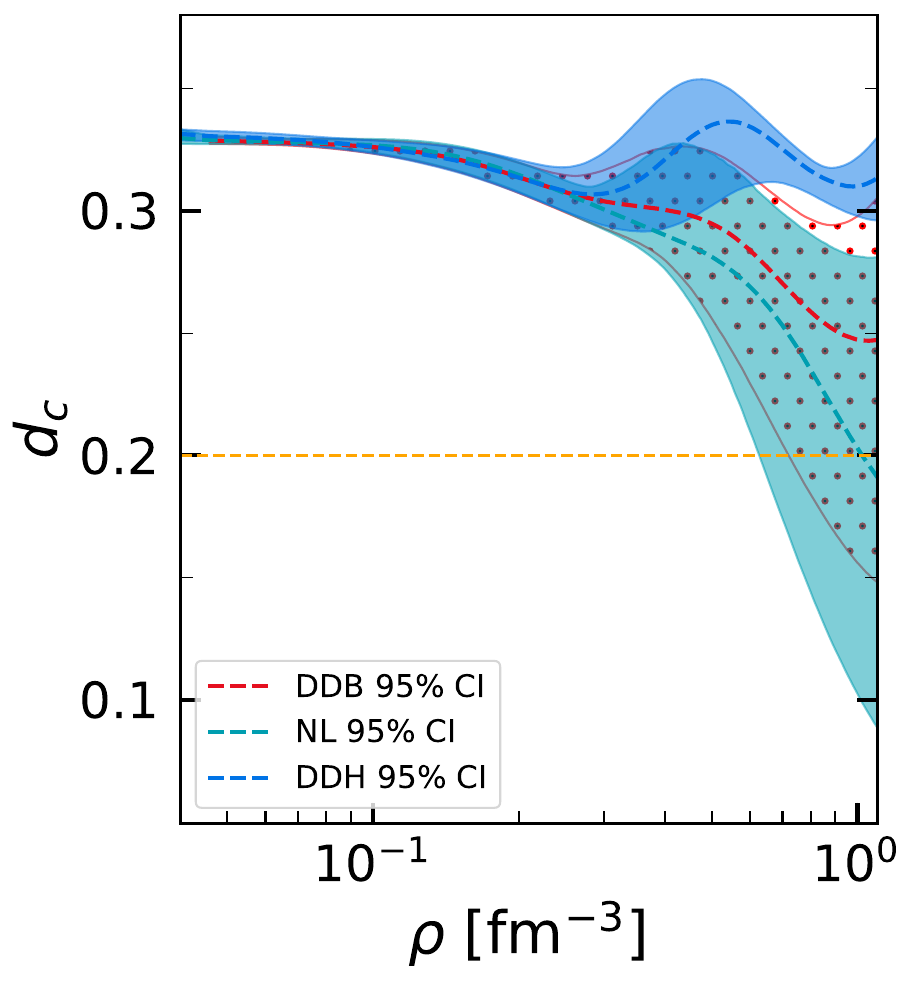}
     \caption{The figure illustrates the pressure normalized by the free pressure with respect to the baryon chemical potential $\mu$, and the relationship between $d_c$ and $\rho$ for three data sets: DDB, NL, and DDH, arranged from left to right. The median values are represented by lines, while the 95\% confidence interval regions are depicted as shaded bands.}
    \label{fig:pfrac}
\end{figure}

\begin{figure*}
    \centering
%    \begin{tabular}{cc}
    \includegraphics[width=0.9\linewidth]{figure/graph_pxeps_book.pdf}\\
    \includegraphics[width=0.3\linewidth]{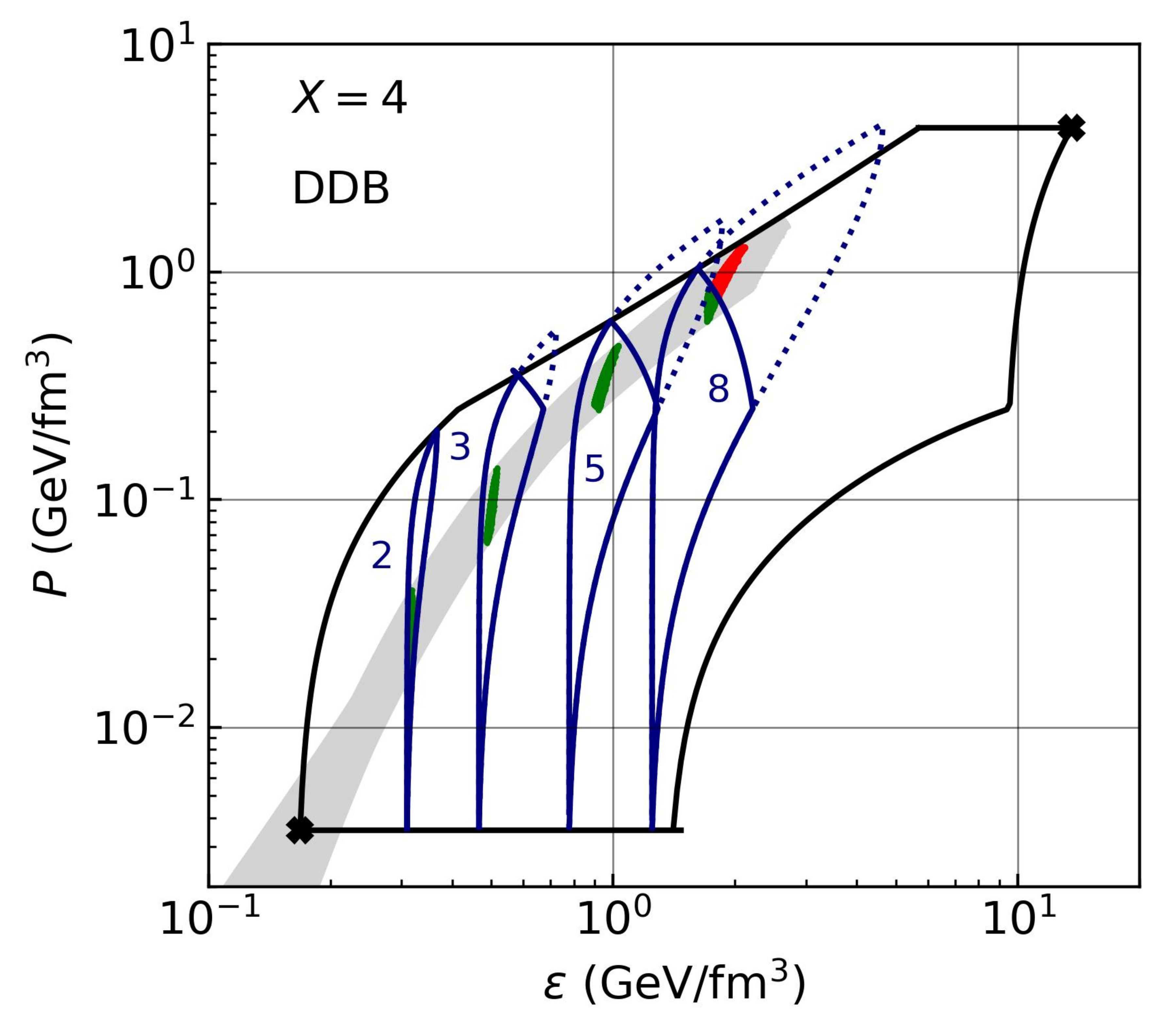}
%    \end{tabular}
     \caption{The pressure versus the energy density is shown for NL, NL-hyp and NL restricted to $\xi\in[0.015:0.04]$ (set 3) (top line, from left to right) and DDB (bottom line).
The constraints from Ref. \cite{Komoltsev:2021jzg} that ensure stability, causality, and thermodynamic consistency delimit the region inside the black solid line. The application of the constraints specifically to some baryon number densities,  $n = 2,$ 3, 5, and 8 $n_s$, defines the regions enclosed by the solid (dotted) blue lines that satisfy (do not satisfy) the pQCD constraints, respectively, from left to right (where $n_s$=0.16 fm$^{-3}$).
Results for the most constraining renormalization scale parameter, $X=4$, \cite{Kurkela2009} are given. The green and red dots represent, respectively, the models in our sets that satisfy and do not satisfy pQCD constraints. Notice that the central density of the maximum mass stars is $\rho_c\lesssim 7\, n_s.$}
    \label{fig:pQCD}
\end{figure*}

We have added to Fig. \ref{fig:nl-cs2-allset} the right panel where the effective nucleon mass is plotted as a function of the baryon density. It is seen that the mass decreases quite fast with density and at $\sim 1$ fm$^{-3}$ the effective masses are below 300 MeV, and may even reach $\sim 100$ MeV for some NL samples. This corresponds to an approximate chiral symmetry restoration and could be the explanation for a behavior similar to the one expected for deconfined matter.

We complete this discussion with Fig. \ref{fig:pfrac}, where the ratio of the pressure to the free particle pressure $p/p_{free}$ is plotted as a function of the baryon chemical potential and the quantity $d_c$ defined in \cite{Annala:2023cwx} as a function of the baryon density. For all sets the ratio $p/p_{free}$ saturates and even decreases for chemical potentials greater than 1300 MeV, after attaining a value of the order of 0.3.  Although  the dispersion is quite large, this ratio takes values approximately 0.1 smaller than the ones obtained in \cite{Annala:2023cwx}.  In \cite{Annala:2023cwx},  it is proposed that $d_c<0.2$ could identify the presence of deconfined matter. In fact, $d_c$ never goes below 0.2 for DDH, and for DDB the median stays above 0.2 although  values below 0.2 are compatible with the 95\% CI.  For the NL data set,  the median of $d_c$ may take values below 0.2 for densities above 1~fm$^{-3}$, and it will be interesting to understand the reason of this behavior.

As already referred before, several studies based in an agnostic description of the EOS constrain the generated EOS imposing at high densities the pQCD EOS. The baryon density for which pQCD EOS is defined, $\gtrsim 40\rho_0$, is outside the range of densities where the RMF models defined in Sec. \ref{sec:model} are valid. Using thermodynamic relations and causality   \cite{Komoltsev:2021jzg} showed that pQCD EOS impose constraints at densities existing in the interior of NS, in particular, for densities $\gtrsim2.2 \rho_0$. In the following, we study how these constraints affect our different data sets. In order to build the geometric construction proposed in \cite{Komoltsev:2021jzg},  we consider  the same constraints these authors chose although other choice could have been done, in particular, at low densities, since  our EOS may be considered well constrained until 2$\rho_0$, 
the density above which non-nucleon degrees of freedom may set in. Note also that the pQCD constraints depend on the renormalization scale \cite{Kurkela:2014vha}: we show results in Fig. \ref{fig:pQCD} for the scale imposing the strongest constraints, $X=4$, \cite{Malik:2023mnx}.   
The analysis was performed for sets NL, NL-hyp, NL  restricted to $\xi\in[0.015:0.04]$ (labelled as set 3) and DDB and the following conclusions may be drawn:
 a) the constraints are satisfied for $\xi>0.015$, i.e. set 3; b) only a few EOS from the set NL-hyp do not satisfy the constraints at 8$n_s$. Notice, however, that  at 90\% CI the largest density in the center of NS within this set is  $\sim \,6n\,_s$, and, therefore all hyperonic stars are compatible with pQCD.  In \cite{Malik:2023mnx}, it was shown that for smaller QCD renormalization scales the total set satisfies the pQCD constraints up to 8$n_s$; c) several EOS from  the set NL do not satisfy the pQCD constraints at densities $\sim8\,n_s$ or even $\sim 5n_s$ . These models have $\xi<0.004$, and the highest maximum masses. If these EOS are removed from the NL set, the absolute maximum mass  drops from $\sim2.75M_\odot$ to $\sim 2.5M_\odot$ for models that satisfy  pQCD with $X=1$, and  to $\sim 2.15M_\odot$ if $X=4$. In the middle panel of Fig. \ref{fig:nl-cs2-allset}, the speed of sound squared for the NL EOS with $\xi<0.004$, has been divided in two subsets, according to their capacity to satisfying (green band) or not  (black dotted band) the pQCD constraints. The EOS that satisfy these constraints present smaller values of $c_s^2$ at high densities. The bottom line of Fig. \ref{fig:pQCD}, we also show results for the DDB set, again taking the most constraining QCD scale: some EOS do not satisfy the 8~$n_s$ constraints, but at 90\% CI no star with a central density above 7$n_s$ was obtained.

\section{Conclusions}
It was an objective of the present work to analyze in a critical way the capacity that RMF models have to describe hadronic matter, and the overall implications when they are used to extract nuclear matter properties from NS observations. As RMF models, we have considered two of the main frameworks frequently used, RMF models including non-linear mesonic terms with constant coupling parameters, designated as NL \cite{Boguta1977,Sugahara:1993wz,Mueller:1996pm,Horowitz:2000xj,Malik:2023mnx}, and models with coupling parameters with an explicit dependence on the density, which do not include mesonic terms beyond quadratic terms \cite{Typel1999,Lalazissis2005,Typel2009,Malik:2022zol}. Other relativistic mean-field approaches have been left out, such as the chiral invariant nuclear model discussed in \cite{Koch:1987py} and developed later  in \cite{daProvidencia:2002pcg,Mishustin:2003wq,Pais:2016dng,Wei:2015aep}, or the inclusion of the isovector scalar meson as studied in \cite{Liu:2001iz}. They will be considered  in a  future work.

A set of fit-data has been imposed, constituted by some nuclear matter properties, the pure neutron matter pressure obtained within a chEFT, and a maximum star mass above 2$M_\odot$. A Bayesian inference formalism was applied to determine the coupling parameters probability distribution, and from these the NMP and NS properties were calculated.  We have shown the the mass-radius domain spanned by the  posterior of the three data sets  are not totally coincident, with the DDH framework predicting smaller radii, DDB larger radii and the NL larger maximum masses, all at 90\% confidence intervals. 

The inclusion of hyperons in these models was also discussed. It was shown that hyperons do not exclude 2$M_\odot$ stars, although maximum masses are much smaller than the ones attained with nucleonic models. However, the radius of canonical stars are larger if hyperonic degrees of freedom are introduced. This is due to the fact that in order to attain  a 2$M_\odot$ mass, and since the onset of hyperons softens de EOS, the nuclear matter parameters describing the symmetric nuclear matter  EOS have to be larger. This confirms similar conclusions drawn in \cite{Fortin:2014mya,Malik:2022jqc}.

The behavior of the proton fraction with density inside NS was also discussed. It was shown that the frameworks DDH and DDB have too small high density couplings to the $\varrho$-meson and as a consequence no nucleonic direct  Urca processes are predicted inside NS, as already discussed in \cite{Fortin:2016hny,Fortin:2020qin,Malik:2022ilb}. This limitation of the models with density dependent couplings was overcome with a generalization of the $\varrho$-meson coupling. The new parameter introduced may be constrained by observations on the cooling of NS. In fact,  in \cite{Malik:2022ilb} it was shown that above three times saturation density the symmetry energy is strongly correlated with the mass of NS characterized by the onset of nucleonic direct Urca processes at their center. {Constraints from chEFT  seem to rule out the direct Urca onset inside NS with a mass below 1.4~$M_\odot$}.

We have analised several EOS properties as the speed of sound, the trace anomaly, and the consistence of the RMF EOS with the predictions of pQCD. It was shown that within  DDH and DDB models the speed of sound are monotonically increasing functions of the density, while within the NL model, the speed of sound behavior is sensitive to the coupling of  $\omega^4$ term, and may present a maximum followed by a decreasing tendency with density. Its behavior may be confused with the onset of a non-nucleonic degree of freedom, as discussed in \cite{Annala2019,Annala:2021gom,Somasundaram:2022ztm}. The different behaviors of the three frameworks reflect the different functionals that define the EOS of each one and the lack of constraining high density observations or experimental data.  It was shown that the three models predict values of the polytropic index below 1.75 for densities above 0.4 to 0.7 fm$^{-3}$ and that the trace anomaly becomes negative for these densities. It was also discussed that the quantity related to the trace anomaly and its derivative introduced in \cite{Annala:2023cwx}, $d_c$, takes values generally above 0.2, a limit proposed in \cite{Annala:2023cwx} as defining a transition to deconfined quark matter, although within the models NL and DDB values below 0.2 are not excluded at densities above $\gtrsim 0.6$ fm$^{-3}$. Notice, however, while in \cite{Annala:2023cwx} this quantity may take values above 0.5, for the present three models it never takes values above 0.35. 

Using thermodynamic and causality  arguments together with  low density nuclear matter and high density pQCD constraints, a discussion similar to the one proposed in \cite{Komoltsev:2021jzg} was developed, and some of the models within RMF that do not satisfy the high density constraints have been identify. These are models with a very stiff high density EOS, although still causal.

\section*{ACKNOWLEDGMENTS} 
This work was partially supported by national funds from FCT (Fundação para a Ciência e a Tecnologia, I.P, Portugal) under Projects No. UIDP/\-04564/\-2020, No. UIDB/\-04564/\-2020 and 2022.06460.PTDC and No. POCI-01-0145-FEDER-029912. 
%, one of the authors, would like to thank the FCT for its support through the Ph.D. grant number 2022.11685.BD. 
The authors acknowledge the Laboratory for Advanced Computing at the University of Coimbra for providing {HPC} resources that have contributed to the research results reported within this paper, URL: \url{https://www.uc.pt/lca}.

\appendix
\section{}
The posterior parameters for the three models DDB, NL and DDH,  obtained in Sec. \ref{sec:bayesian_inf}  are given in Table \ref{tab:para}.

\begin{table*}
\caption{Based on the posterior distribution for DDB, NL, and DDH restricted to nucleonic degrees of freedom, the median values and the 90\% confidence intervals (CI) for the parameters were calculated. Please refer to Section \ref{sec:bayesian_inf} for the specific terminology used for parameter names. The masses of the nucleon, $\omega$ meson, and $\rho$ meson in all models are 939 MeV, 783 MeV, and 763 MeV, respectively. However, the $\sigma$ meson mass is fixed to 500 MeV for NL and 550 MeV for DDB and DDH.}
\centering
\label{tab:para}
 \setlength{\tabcolsep}{6.5pt}
      \renewcommand{\arraystretch}{1.4}
\begin{tabular}{cccccccccccccc}
\hline \hline 
\multicolumn{4}{c}{DDB}                                                             &  & \multicolumn{4}{c}{NL}                                                              &  & \multicolumn{4}{c}{DDH}                                                            \\ \cline{1-4} \cline{6-9} \cline{11-14} 
\multirow{2}{*}{Parameters} & \multirow{2}{*}{median} & \multicolumn{2}{c}{90\% CI} &  & \multirow{2}{*}{Parameters} & \multirow{2}{*}{median} & \multicolumn{2}{c}{90\% CI} &  & \multirow{2}{*}{Parameters} & \multirow{2}{*}{median} & \multicolumn{2}{c}{90\% CI} \\ \cline{3-4} \cline{8-9} \cline{13-14} 
                            &                         & min          & max          &  &                             &                         & min          & max          &  &                             &                         & min          & max          \\ \hline
$g_\sigma$                  & 9.024                   & 8.170        & 10.059       &  & $g_\sigma$                  & 8.454                   & 8.010        & 9.691        &  & $g_\sigma$                  & 8.827                   & 8.146        & 9.322        \\
$g_\omega$                  & 10.761                  & 9.413        & 12.313       &  & $g_\omega$                  & 9.915                   & 9.084        & 12.167       &  & $g_\omega$                  & 10.475                  & 9.378        & 11.224       \\
$g_\rho$                    & 3.954                   & 3.485        & 4.389        &  & $g_\rho$                    & 12.193                  & 9.546        & 14.599       &  & $g_\rho$                    & 3.976                   & 3.475        & 4.434        \\
$a_\sigma$                  & 0.080                   & 0.054        & 0.113        &  & $B$                         & 4.586                   & 2.205        & 6.903        &  & $a_\sigma$                  & 1.247                   & 1.158        & 1.499        \\
$a_\omega$                  & 0.039                   & 0.004        & 0.105        &  & $C$                         & -1.985                  & -4.627       & 3.530        &  & $b_\sigma$                  & 1.392                   & 0.585        & 4.082        \\
$a_\rho$                    & 0.542                   & 0.318        & 0.703        &  & $\xi$                       & 0.004                   & 0.000        & 0.016        &  & $c_\sigma$                  & 1.867                   & 0.775        & 6.011        \\
                            &                         &              &              &  & $\Lambda$                   & 0.064                   & 0.036        & 0.103        &  & $d_\sigma$                  & 0.423                   & 0.235        & 0.656        \\
                            &                         &              &              &  &                             &                         &              &              &  & $a_\omega$                  & 1.215                   & 1.022        & 1.663        \\
                            &                         &              &              &  &                             &                         &              &              &  & $b_\omega$                  & 7.544                   & 1.876        & 14.074       \\
                            &                         &              &              &  &                             &                         &              &              &  & $c_\omega$                  & 9.546                   & 2.307        & 19.140       \\
                            &                         &              &              &  &                             &                         &              &              &  & $d_\omega$                  & 0.187                   & 0.132        & 0.380        \\
                            &                         &              &              &  &                             &                         &              &              &  &  $a_\rho$                           & 0.500                   & 0.304        & 0.720        \\ \hline
\end{tabular}
\end{table*}

%\bibliography{biblio}

\begin{thebibliography}{110}%
\makeatletter
\providecommand \@ifxundefined [1]{%
 \@ifx{#1\undefined}
}%
\providecommand \@ifnum [1]{%
 \ifnum #1\expandafter \@firstoftwo
 \else \expandafter \@secondoftwo
 \fi
}%
\providecommand \@ifx [1]{%
 \ifx #1\expandafter \@firstoftwo
 \else \expandafter \@secondoftwo
 \fi
}%
\providecommand \natexlab [1]{#1}%
\providecommand \enquote  [1]{``#1''}%
\providecommand \bibnamefont  [1]{#1}%
\providecommand \bibfnamefont [1]{#1}%
\providecommand \citenamefont [1]{#1}%
\providecommand \href@noop [0]{\@secondoftwo}%
\providecommand \href [0]{\begingroup \@sanitize@url \@href}%
\providecommand \@href[1]{\@@startlink{#1}\@@href}%
\providecommand \@@href[1]{\endgroup#1\@@endlink}%
\providecommand \@sanitize@url [0]{\catcode `\\12\catcode `\$12\catcode
  `\&12\catcode `\#12\catcode `\^12\catcode `\_12\catcode `\%12\relax}%
\providecommand \@@startlink[1]{}%
\providecommand \@@endlink[0]{}%
\providecommand \url  [0]{\begingroup\@sanitize@url \@url }%
\providecommand \@url [1]{\endgroup\@href {#1}{\urlprefix }}%
\providecommand \urlprefix  [0]{URL }%
\providecommand \Eprint [0]{\href }%
\providecommand \doibase [0]{https://doi.org/}%
\providecommand \selectlanguage [0]{\@gobble}%
\providecommand \bibinfo  [0]{\@secondoftwo}%
\providecommand \bibfield  [0]{\@secondoftwo}%
\providecommand \translation [1]{[#1]}%
\providecommand \BibitemOpen [0]{}%
\providecommand \bibitemStop [0]{}%
\providecommand \bibitemNoStop [0]{.\EOS\space}%
\providecommand \EOS [0]{\spacefactor3000\relax}%
\providecommand \BibitemShut  [1]{\csname bibitem#1\endcsname}%
\let\auto@bib@innerbib\@empty
%</preamble>
\bibitem [{\citenamefont {Abbott}\ \emph {et~al.}(2017)\citenamefont {Abbott}
  \emph {et~al.}}]{TheLIGOScientific:2017qsa}%
  \BibitemOpen
  \bibfield  {author} {\bibinfo {author} {\bibfnamefont {B.~P.}\ \bibnamefont
  {Abbott}} \emph {et~al.} (\bibinfo {collaboration} {LIGO Scientific,
  Virgo}),\ }\href {https://doi.org/10.1103/PhysRevLett.119.161101} {\bibfield
  {journal} {\bibinfo  {journal} {Phys. Rev. Lett.}\ }\textbf {\bibinfo
  {volume} {119}},\ \bibinfo {pages} {161101} (\bibinfo {year} {2017})},\
  \Eprint {https://arxiv.org/abs/1710.05832} {arXiv:1710.05832 [gr-qc]}
  \BibitemShut {NoStop}%
%%CITATION = ARXIV:1710.05832;%%
\bibitem [{\citenamefont {Abbott}\ \emph {et~al.}(2020)\citenamefont {Abbott}
  \emph {et~al.}}]{LIGOScientific:2020aai}%
  \BibitemOpen
  \bibfield  {author} {\bibinfo {author} {\bibfnamefont {B.~P.}\ \bibnamefont
  {Abbott}} \emph {et~al.} (\bibinfo {collaboration} {LIGO Scientific,
  Virgo}),\ }\href {https://doi.org/10.3847/2041-8213/ab75f5} {\bibfield
  {journal} {\bibinfo  {journal} {Astrophys. J. Lett.}\ }\textbf {\bibinfo
  {volume} {892}},\ \bibinfo {pages} {L3} (\bibinfo {year} {2020})},\ \Eprint
  {https://arxiv.org/abs/2001.01761} {arXiv:2001.01761 [astro-ph.HE]}
  \BibitemShut {NoStop}%
\bibitem [{\citenamefont {Demorest}\ \emph {et~al.}(2010)\citenamefont
  {Demorest}, \citenamefont {Pennucci}, \citenamefont {Ransom}, \citenamefont
  {Roberts},\ and\ \citenamefont {Hessels}}]{Demorest2010}%
  \BibitemOpen
  \bibfield  {author} {\bibinfo {author} {\bibfnamefont {P.}~\bibnamefont
  {Demorest}}, \bibinfo {author} {\bibfnamefont {T.}~\bibnamefont {Pennucci}},
  \bibinfo {author} {\bibfnamefont {S.}~\bibnamefont {Ransom}}, \bibinfo
  {author} {\bibfnamefont {M.}~\bibnamefont {Roberts}},\ and\ \bibinfo {author}
  {\bibfnamefont {J.}~\bibnamefont {Hessels}},\ }\href
  {https://doi.org/10.1038/nature09466} {\bibfield  {journal} {\bibinfo
  {journal} {Nature}\ }\textbf {\bibinfo {volume} {467}},\ \bibinfo {pages}
  {1081} (\bibinfo {year} {2010})}\BibitemShut {NoStop}%
%%CITATION = ARXIV:1010.5788;%%
\bibitem [{\citenamefont {{Antoniadis}}\ \emph {et~al.}(2013)\citenamefont
  {{Antoniadis}}, \citenamefont {{Freire}}, \citenamefont {{Wex}},
  \citenamefont {{Tauris}}, \citenamefont {{Lynch}}, \citenamefont {{van
  Kerkwijk}}, \citenamefont {{Kramer}}, \citenamefont {{Bassa}}, \citenamefont
  {{Dhillon}}, \citenamefont {{Driebe}}, \citenamefont {{Hessels}},
  \citenamefont {{Kaspi}}, \citenamefont {{Kondratiev}}, \citenamefont
  {{Langer}}, \citenamefont {{Marsh}}, \citenamefont {{McLaughlin}},
  \citenamefont {{Pennucci}}, \citenamefont {{Ransom}}, \citenamefont
  {{Stairs}}, \citenamefont {{van Leeuwen}}, \citenamefont {{Verbiest}},\ and\
  \citenamefont {{Whelan}}}]{Antoniadis2013}%
  \BibitemOpen
  \bibfield  {author} {\bibinfo {author} {\bibfnamefont {J.}~\bibnamefont
  {{Antoniadis}}}, \bibinfo {author} {\bibfnamefont {P.~C.~C.}\ \bibnamefont
  {{Freire}}}, \bibinfo {author} {\bibfnamefont {N.}~\bibnamefont {{Wex}}},
  \bibinfo {author} {\bibfnamefont {T.~M.}\ \bibnamefont {{Tauris}}}, \bibinfo
  {author} {\bibfnamefont {R.~S.}\ \bibnamefont {{Lynch}}}, \bibinfo {author}
  {\bibfnamefont {M.~H.}\ \bibnamefont {{van Kerkwijk}}}, \bibinfo {author}
  {\bibfnamefont {M.}~\bibnamefont {{Kramer}}}, \bibinfo {author}
  {\bibfnamefont {C.}~\bibnamefont {{Bassa}}}, \bibinfo {author} {\bibfnamefont
  {V.~S.}\ \bibnamefont {{Dhillon}}}, \bibinfo {author} {\bibfnamefont
  {T.}~\bibnamefont {{Driebe}}}, \bibinfo {author} {\bibfnamefont {J.~W.~T.}\
  \bibnamefont {{Hessels}}}, \bibinfo {author} {\bibfnamefont {V.~M.}\
  \bibnamefont {{Kaspi}}}, \bibinfo {author} {\bibfnamefont {V.~I.}\
  \bibnamefont {{Kondratiev}}}, \bibinfo {author} {\bibfnamefont
  {N.}~\bibnamefont {{Langer}}}, \bibinfo {author} {\bibfnamefont {T.~R.}\
  \bibnamefont {{Marsh}}}, \bibinfo {author} {\bibfnamefont {M.~A.}\
  \bibnamefont {{McLaughlin}}}, \bibinfo {author} {\bibfnamefont {T.~T.}\
  \bibnamefont {{Pennucci}}}, \bibinfo {author} {\bibfnamefont {S.~M.}\
  \bibnamefont {{Ransom}}}, \bibinfo {author} {\bibfnamefont {I.~H.}\
  \bibnamefont {{Stairs}}}, \bibinfo {author} {\bibfnamefont {J.}~\bibnamefont
  {{van Leeuwen}}}, \bibinfo {author} {\bibfnamefont {J.~P.~W.}\ \bibnamefont
  {{Verbiest}}},\ and\ \bibinfo {author} {\bibfnamefont {D.~G.}\ \bibnamefont
  {{Whelan}}},\ }\href {https://doi.org/10.1126/science.1233232} {\bibfield
  {journal} {\bibinfo  {journal} {Science}\ }\textbf {\bibinfo {volume}
  {340}},\ \bibinfo {pages} {448} (\bibinfo {year} {2013})}\BibitemShut
  {NoStop}%
\bibitem [{\citenamefont {Cromartie}\ \emph {et~al.}(2019)\citenamefont
  {Cromartie} \emph {et~al.}}]{NANOGrav:2019jur}%
  \BibitemOpen
  \bibfield  {author} {\bibinfo {author} {\bibfnamefont {H.~T.}\ \bibnamefont
  {Cromartie}} \emph {et~al.} (\bibinfo {collaboration} {NANOGrav}),\ }\href
  {https://doi.org/10.1038/s41550-019-0880-2} {\bibfield  {journal} {\bibinfo
  {journal} {Nature Astron.}\ }\textbf {\bibinfo {volume} {4}},\ \bibinfo
  {pages} {72} (\bibinfo {year} {2019})},\ \Eprint
  {https://arxiv.org/abs/1904.06759} {arXiv:1904.06759 [astro-ph.HE]}
  \BibitemShut {NoStop}%
\bibitem [{\citenamefont {Fonseca}\ \emph {et~al.}(2021)\citenamefont {Fonseca}
  \emph {et~al.}}]{Fonseca:2021wxt}%
  \BibitemOpen
  \bibfield  {author} {\bibinfo {author} {\bibfnamefont {E.}~\bibnamefont
  {Fonseca}} \emph {et~al.},\ }\href {https://doi.org/10.3847/2041-8213/ac03b8}
  {\bibfield  {journal} {\bibinfo  {journal} {Astrophys. J. Lett.}\ }\textbf
  {\bibinfo {volume} {915}},\ \bibinfo {pages} {L12} (\bibinfo {year}
  {2021})},\ \Eprint {https://arxiv.org/abs/2104.00880} {arXiv:2104.00880
  [astro-ph.HE]} \BibitemShut {NoStop}%
\bibitem [{\citenamefont {Romani}\ \emph {et~al.}(2021)\citenamefont {Romani},
  \citenamefont {Kandel}, \citenamefont {Filippenko}, \citenamefont {Brink},\
  and\ \citenamefont {Zheng}}]{Romani:2021xmb}%
  \BibitemOpen
  \bibfield  {author} {\bibinfo {author} {\bibfnamefont {R.~W.}\ \bibnamefont
  {Romani}}, \bibinfo {author} {\bibfnamefont {D.}~\bibnamefont {Kandel}},
  \bibinfo {author} {\bibfnamefont {A.~V.}\ \bibnamefont {Filippenko}},
  \bibinfo {author} {\bibfnamefont {T.~G.}\ \bibnamefont {Brink}},\ and\
  \bibinfo {author} {\bibfnamefont {W.}~\bibnamefont {Zheng}},\ }\href
  {https://doi.org/10.3847/2041-8213/abe2b4} {\bibfield  {journal} {\bibinfo
  {journal} {Astrophys. J. Lett.}\ }\textbf {\bibinfo {volume} {908}},\
  \bibinfo {pages} {L46} (\bibinfo {year} {2021})},\ \Eprint
  {https://arxiv.org/abs/2101.09822} {arXiv:2101.09822 [astro-ph.HE]}
  \BibitemShut {NoStop}%
\bibitem [{\citenamefont {Riley}\ \emph {et~al.}(2019)\citenamefont {Riley}
  \emph {et~al.}}]{Riley:2019yda}%
  \BibitemOpen
  \bibfield  {author} {\bibinfo {author} {\bibfnamefont {T.~E.}\ \bibnamefont
  {Riley}} \emph {et~al.},\ }\href {https://doi.org/10.3847/2041-8213/ab481c}
  {\bibfield  {journal} {\bibinfo  {journal} {Astrophys. J. Lett.}\ }\textbf
  {\bibinfo {volume} {887}},\ \bibinfo {pages} {L21} (\bibinfo {year}
  {2019})},\ \Eprint {https://arxiv.org/abs/1912.05702} {arXiv:1912.05702
  [astro-ph.HE]} \BibitemShut {NoStop}%
\bibitem [{\citenamefont {Miller}\ \emph {et~al.}(2019)\citenamefont {Miller}
  \emph {et~al.}}]{Miller:2019cac}%
  \BibitemOpen
  \bibfield  {author} {\bibinfo {author} {\bibfnamefont {M.~C.}\ \bibnamefont
  {Miller}} \emph {et~al.},\ }\href {https://doi.org/10.3847/2041-8213/ab50c5}
  {\bibfield  {journal} {\bibinfo  {journal} {Astrophys. J. Lett.}\ }\textbf
  {\bibinfo {volume} {887}},\ \bibinfo {pages} {L24} (\bibinfo {year}
  {2019})},\ \Eprint {https://arxiv.org/abs/1912.05705} {arXiv:1912.05705
  [astro-ph.HE]} \BibitemShut {NoStop}%
\bibitem [{\citenamefont {Riley}\ \emph {et~al.}(2021)\citenamefont {Riley}
  \emph {et~al.}}]{Riley:2021pdl}%
  \BibitemOpen
  \bibfield  {author} {\bibinfo {author} {\bibfnamefont {T.~E.}\ \bibnamefont
  {Riley}} \emph {et~al.},\ }\href {https://doi.org/10.3847/2041-8213/ac0a81}
  {\bibfield  {journal} {\bibinfo  {journal} {Astrophys. J. Lett.}\ }\textbf
  {\bibinfo {volume} {918}},\ \bibinfo {pages} {L27} (\bibinfo {year}
  {2021})},\ \Eprint {https://arxiv.org/abs/2105.06980} {arXiv:2105.06980
  [astro-ph.HE]} \BibitemShut {NoStop}%
\bibitem [{\citenamefont {Miller}\ \emph {et~al.}(2021)\citenamefont {Miller}
  \emph {et~al.}}]{Miller:2021qha}%
  \BibitemOpen
  \bibfield  {author} {\bibinfo {author} {\bibfnamefont {M.~C.}\ \bibnamefont
  {Miller}} \emph {et~al.},\ }\href {https://doi.org/10.3847/2041-8213/ac089b}
  {\bibfield  {journal} {\bibinfo  {journal} {Astrophys. J. Lett.}\ }\textbf
  {\bibinfo {volume} {918}},\ \bibinfo {pages} {L28} (\bibinfo {year}
  {2021})},\ \Eprint {https://arxiv.org/abs/2105.06979} {arXiv:2105.06979
  [astro-ph.HE]} \BibitemShut {NoStop}%
\bibitem [{\citenamefont {Raaijmakers}\ \emph {et~al.}(2021)\citenamefont
  {Raaijmakers}, \citenamefont {Greif}, \citenamefont {Hebeler}, \citenamefont
  {Hinderer}, \citenamefont {Nissanke}, \citenamefont {Schwenk}, \citenamefont
  {Riley}, \citenamefont {Watts}, \citenamefont {Lattimer},\ and\ \citenamefont
  {Ho}}]{Raaijmakers:2021uju}%
  \BibitemOpen
  \bibfield  {author} {\bibinfo {author} {\bibfnamefont {G.}~\bibnamefont
  {Raaijmakers}}, \bibinfo {author} {\bibfnamefont {S.~K.}\ \bibnamefont
  {Greif}}, \bibinfo {author} {\bibfnamefont {K.}~\bibnamefont {Hebeler}},
  \bibinfo {author} {\bibfnamefont {T.}~\bibnamefont {Hinderer}}, \bibinfo
  {author} {\bibfnamefont {S.}~\bibnamefont {Nissanke}}, \bibinfo {author}
  {\bibfnamefont {A.}~\bibnamefont {Schwenk}}, \bibinfo {author} {\bibfnamefont
  {T.~E.}\ \bibnamefont {Riley}}, \bibinfo {author} {\bibfnamefont {A.~L.}\
  \bibnamefont {Watts}}, \bibinfo {author} {\bibfnamefont {J.~M.}\ \bibnamefont
  {Lattimer}},\ and\ \bibinfo {author} {\bibfnamefont {W.~C.~G.}\ \bibnamefont
  {Ho}},\ }\href {https://doi.org/10.3847/2041-8213/ac089a} {\bibfield
  {journal} {\bibinfo  {journal} {Astrophys. J. Lett.}\ }\textbf {\bibinfo
  {volume} {918}},\ \bibinfo {pages} {L29} (\bibinfo {year} {2021})},\ \Eprint
  {https://arxiv.org/abs/2105.06981} {arXiv:2105.06981 [astro-ph.HE]}
  \BibitemShut {NoStop}%
\bibitem [{\citenamefont {Margueron}\ \emph
  {et~al.}(2018{\natexlab{a}})\citenamefont {Margueron}, \citenamefont
  {Hoffmann~Casali},\ and\ \citenamefont {Gulminelli}}]{Margueron2018a}%
  \BibitemOpen
  \bibfield  {author} {\bibinfo {author} {\bibfnamefont {J.}~\bibnamefont
  {Margueron}}, \bibinfo {author} {\bibfnamefont {R.}~\bibnamefont
  {Hoffmann~Casali}},\ and\ \bibinfo {author} {\bibfnamefont {F.}~\bibnamefont
  {Gulminelli}},\ }\href {https://doi.org/10.1103/PhysRevC.97.025805}
  {\bibfield  {journal} {\bibinfo  {journal} {Phys. Rev.}\ }\textbf {\bibinfo
  {volume} {C97}},\ \bibinfo {pages} {025805} (\bibinfo {year}
  {2018}{\natexlab{a}})},\ \Eprint {https://arxiv.org/abs/1708.06894}
  {arXiv:1708.06894 [nucl-th]} \BibitemShut {NoStop}%
%%CITATION = ARXIV:1708.06894;%%
\bibitem [{\citenamefont {Hebeler}\ \emph {et~al.}(2013)\citenamefont
  {Hebeler}, \citenamefont {Lattimer}, \citenamefont {Pethick},\ and\
  \citenamefont {Schwenk}}]{Hebeler2013}%
  \BibitemOpen
  \bibfield  {author} {\bibinfo {author} {\bibfnamefont {K.}~\bibnamefont
  {Hebeler}}, \bibinfo {author} {\bibfnamefont {J.~M.}\ \bibnamefont
  {Lattimer}}, \bibinfo {author} {\bibfnamefont {C.~J.}\ \bibnamefont
  {Pethick}},\ and\ \bibinfo {author} {\bibfnamefont {A.}~\bibnamefont
  {Schwenk}},\ }\href {https://doi.org/10.1088/0004-637X/773/1/11} {\bibfield
  {journal} {\bibinfo  {journal} {Astrophys. J.}\ }\textbf {\bibinfo {volume}
  {773}},\ \bibinfo {pages} {11} (\bibinfo {year} {2013})},\ \Eprint
  {https://arxiv.org/abs/1303.4662} {arXiv:1303.4662 [astro-ph.SR]}
  \BibitemShut {NoStop}%
\bibitem [{\citenamefont {Drischler}\ \emph {et~al.}(2019)\citenamefont
  {Drischler}, \citenamefont {Hebeler},\ and\ \citenamefont
  {Schwenk}}]{Drischler:2017wtt}%
  \BibitemOpen
  \bibfield  {author} {\bibinfo {author} {\bibfnamefont {C.}~\bibnamefont
  {Drischler}}, \bibinfo {author} {\bibfnamefont {K.}~\bibnamefont {Hebeler}},\
  and\ \bibinfo {author} {\bibfnamefont {A.}~\bibnamefont {Schwenk}},\ }\href
  {https://doi.org/10.1103/PhysRevLett.122.042501} {\bibfield  {journal}
  {\bibinfo  {journal} {Phys. Rev. Lett.}\ }\textbf {\bibinfo {volume} {122}},\
  \bibinfo {pages} {042501} (\bibinfo {year} {2019})},\ \Eprint
  {https://arxiv.org/abs/1710.08220} {arXiv:1710.08220 [nucl-th]} \BibitemShut
  {NoStop}%
\bibitem [{\citenamefont {Drischler}\ \emph {et~al.}(2020)\citenamefont
  {Drischler}, \citenamefont {Melendez}, \citenamefont {Furnstahl},\ and\
  \citenamefont {Phillips}}]{Drischler:2020yad}%
  \BibitemOpen
  \bibfield  {author} {\bibinfo {author} {\bibfnamefont {C.}~\bibnamefont
  {Drischler}}, \bibinfo {author} {\bibfnamefont {J.~A.}\ \bibnamefont
  {Melendez}}, \bibinfo {author} {\bibfnamefont {R.~J.}\ \bibnamefont
  {Furnstahl}},\ and\ \bibinfo {author} {\bibfnamefont {D.~R.}\ \bibnamefont
  {Phillips}},\ }\href {https://doi.org/10.1103/PhysRevC.102.054315} {\bibfield
   {journal} {\bibinfo  {journal} {Phys. Rev. C}\ }\textbf {\bibinfo {volume}
  {102}},\ \bibinfo {pages} {054315} (\bibinfo {year} {2020})},\ \Eprint
  {https://arxiv.org/abs/2004.07805} {arXiv:2004.07805 [nucl-th]} \BibitemShut
  {NoStop}%
\bibitem [{\citenamefont {Kurkela}\ \emph {et~al.}(2010)\citenamefont
  {Kurkela}, \citenamefont {Romatschke},\ and\ \citenamefont
  {Vuorinen}}]{Kurkela2009}%
  \BibitemOpen
  \bibfield  {author} {\bibinfo {author} {\bibfnamefont {A.}~\bibnamefont
  {Kurkela}}, \bibinfo {author} {\bibfnamefont {P.}~\bibnamefont
  {Romatschke}},\ and\ \bibinfo {author} {\bibfnamefont {A.}~\bibnamefont
  {Vuorinen}},\ }\href {https://doi.org/10.1103/PhysRevD.81.105021} {\bibfield
  {journal} {\bibinfo  {journal} {Phys. Rev. D}\ }\textbf {\bibinfo {volume}
  {81}},\ \bibinfo {pages} {105021} (\bibinfo {year} {2010})},\ \Eprint
  {https://arxiv.org/abs/0912.1856} {arXiv:0912.1856 [hep-ph]} \BibitemShut
  {NoStop}%
\bibitem [{\citenamefont {Kurkela}\ \emph {et~al.}(2014)\citenamefont
  {Kurkela}, \citenamefont {Fraga}, \citenamefont {Schaffner-Bielich},\ and\
  \citenamefont {Vuorinen}}]{Kurkela:2014vha}%
  \BibitemOpen
  \bibfield  {author} {\bibinfo {author} {\bibfnamefont {A.}~\bibnamefont
  {Kurkela}}, \bibinfo {author} {\bibfnamefont {E.~S.}\ \bibnamefont {Fraga}},
  \bibinfo {author} {\bibfnamefont {J.}~\bibnamefont {Schaffner-Bielich}},\
  and\ \bibinfo {author} {\bibfnamefont {A.}~\bibnamefont {Vuorinen}},\ }\href
  {https://doi.org/10.1088/0004-637X/789/2/127} {\bibfield  {journal} {\bibinfo
   {journal} {Astrophys. J.}\ }\textbf {\bibinfo {volume} {789}},\ \bibinfo
  {pages} {127} (\bibinfo {year} {2014})},\ \Eprint
  {https://arxiv.org/abs/1402.6618} {arXiv:1402.6618 [astro-ph.HE]}
  \BibitemShut {NoStop}%
\bibitem [{\citenamefont {Gorda}\ \emph {et~al.}(2022)\citenamefont {Gorda},
  \citenamefont {Komoltsev},\ and\ \citenamefont {Kurkela}}]{Gorda:2022jvk}%
  \BibitemOpen
  \bibfield  {author} {\bibinfo {author} {\bibfnamefont {T.}~\bibnamefont
  {Gorda}}, \bibinfo {author} {\bibfnamefont {O.}~\bibnamefont {Komoltsev}},\
  and\ \bibinfo {author} {\bibfnamefont {A.}~\bibnamefont {Kurkela}},\
  }\href@noop {} {\  (\bibinfo {year} {2022})},\ \Eprint
  {https://arxiv.org/abs/2204.11877} {arXiv:2204.11877 [nucl-th]} \BibitemShut
  {NoStop}%
\bibitem [{\citenamefont {{Lindblom}}(1992)}]{Lindblom:1992}%
  \BibitemOpen
  \bibfield  {author} {\bibinfo {author} {\bibfnamefont {L.}~\bibnamefont
  {{Lindblom}}},\ }\href {https://doi.org/10.1086/171882} {\bibfield  {journal}
  {\bibinfo  {journal} {\apj}\ }\textbf {\bibinfo {volume} {398}},\ \bibinfo
  {pages} {569} (\bibinfo {year} {1992})}\BibitemShut {NoStop}%
\bibitem [{\citenamefont {Steiner}\ \emph {et~al.}(2010)\citenamefont
  {Steiner}, \citenamefont {Lattimer},\ and\ \citenamefont
  {Brown}}]{Steiner:2010fz}%
  \BibitemOpen
  \bibfield  {author} {\bibinfo {author} {\bibfnamefont {A.~W.}\ \bibnamefont
  {Steiner}}, \bibinfo {author} {\bibfnamefont {J.~M.}\ \bibnamefont
  {Lattimer}},\ and\ \bibinfo {author} {\bibfnamefont {E.~F.}\ \bibnamefont
  {Brown}},\ }\href {https://doi.org/10.1088/0004-637X/722/1/33} {\bibfield
  {journal} {\bibinfo  {journal} {Astrophys. J.}\ }\textbf {\bibinfo {volume}
  {722}},\ \bibinfo {pages} {33} (\bibinfo {year} {2010})},\ \Eprint
  {https://arxiv.org/abs/1005.0811} {arXiv:1005.0811 [astro-ph.HE]}
  \BibitemShut {NoStop}%
\bibitem [{\citenamefont {Ozel}\ \emph {et~al.}(2010)\citenamefont {Ozel},
  \citenamefont {Baym},\ and\ \citenamefont {Guver}}]{Ozel:2010fw}%
  \BibitemOpen
  \bibfield  {author} {\bibinfo {author} {\bibfnamefont {F.}~\bibnamefont
  {Ozel}}, \bibinfo {author} {\bibfnamefont {G.}~\bibnamefont {Baym}},\ and\
  \bibinfo {author} {\bibfnamefont {T.}~\bibnamefont {Guver}},\ }\href
  {https://doi.org/10.1103/PhysRevD.82.101301} {\bibfield  {journal} {\bibinfo
  {journal} {Phys. Rev. D}\ }\textbf {\bibinfo {volume} {82}},\ \bibinfo
  {pages} {101301} (\bibinfo {year} {2010})},\ \Eprint
  {https://arxiv.org/abs/1002.3153} {arXiv:1002.3153 [astro-ph.HE]}
  \BibitemShut {NoStop}%
\bibitem [{\citenamefont {Steiner}\ \emph {et~al.}(2013)\citenamefont
  {Steiner}, \citenamefont {Lattimer},\ and\ \citenamefont
  {Brown}}]{Steiner:2012xt}%
  \BibitemOpen
  \bibfield  {author} {\bibinfo {author} {\bibfnamefont {A.~W.}\ \bibnamefont
  {Steiner}}, \bibinfo {author} {\bibfnamefont {J.~M.}\ \bibnamefont
  {Lattimer}},\ and\ \bibinfo {author} {\bibfnamefont {E.~F.}\ \bibnamefont
  {Brown}},\ }\href {https://doi.org/10.1088/2041-8205/765/1/L5} {\bibfield
  {journal} {\bibinfo  {journal} {Astrophys. J. Lett.}\ }\textbf {\bibinfo
  {volume} {765}},\ \bibinfo {pages} {L5} (\bibinfo {year} {2013})},\ \Eprint
  {https://arxiv.org/abs/1205.6871} {arXiv:1205.6871 [nucl-th]} \BibitemShut
  {NoStop}%
\bibitem [{\citenamefont {Raithel}\ \emph {et~al.}(2016)\citenamefont
  {Raithel}, \citenamefont {Ozel},\ and\ \citenamefont
  {Psaltis}}]{Raithel:2016bux}%
  \BibitemOpen
  \bibfield  {author} {\bibinfo {author} {\bibfnamefont {C.~A.}\ \bibnamefont
  {Raithel}}, \bibinfo {author} {\bibfnamefont {F.}~\bibnamefont {Ozel}},\ and\
  \bibinfo {author} {\bibfnamefont {D.}~\bibnamefont {Psaltis}},\ }\href
  {https://doi.org/10.3847/0004-637X/831/1/44} {\bibfield  {journal} {\bibinfo
  {journal} {Astrophys. J.}\ }\textbf {\bibinfo {volume} {831}},\ \bibinfo
  {pages} {44} (\bibinfo {year} {2016})},\ \Eprint
  {https://arxiv.org/abs/1605.03591} {arXiv:1605.03591 [astro-ph.HE]}
  \BibitemShut {NoStop}%
\bibitem [{\citenamefont {Fujimoto}\ \emph {et~al.}(2018)\citenamefont
  {Fujimoto}, \citenamefont {Fukushima},\ and\ \citenamefont
  {Murase}}]{Fujimoto_2018}%
  \BibitemOpen
  \bibfield  {author} {\bibinfo {author} {\bibfnamefont {Y.}~\bibnamefont
  {Fujimoto}}, \bibinfo {author} {\bibfnamefont {K.}~\bibnamefont
  {Fukushima}},\ and\ \bibinfo {author} {\bibfnamefont {K.}~\bibnamefont
  {Murase}},\ }\bibfield  {journal} {\bibinfo  {journal} {Physical Review D}\
  }\textbf {\bibinfo {volume} {98}},\ \href
  {https://doi.org/10.1103/physrevd.98.023019} {10.1103/physrevd.98.023019}
  (\bibinfo {year} {2018})\BibitemShut {NoStop}%
\bibitem [{\citenamefont {Fujimoto}\ \emph {et~al.}(2020)\citenamefont
  {Fujimoto}, \citenamefont {Fukushima},\ and\ \citenamefont
  {Murase}}]{Fujimoto_2020}%
  \BibitemOpen
  \bibfield  {author} {\bibinfo {author} {\bibfnamefont {Y.}~\bibnamefont
  {Fujimoto}}, \bibinfo {author} {\bibfnamefont {K.}~\bibnamefont
  {Fukushima}},\ and\ \bibinfo {author} {\bibfnamefont {K.}~\bibnamefont
  {Murase}},\ }\bibfield  {journal} {\bibinfo  {journal} {Physical Review D}\
  }\textbf {\bibinfo {volume} {101}},\ \href
  {https://doi.org/10.1103/physrevd.101.054016} {10.1103/physrevd.101.054016}
  (\bibinfo {year} {2020})\BibitemShut {NoStop}%
\bibitem [{\citenamefont {Fujimoto}\ \emph {et~al.}(2021)\citenamefont
  {Fujimoto}, \citenamefont {Fukushima},\ and\ \citenamefont
  {Murase}}]{fujimoto2021extensive}%
  \BibitemOpen
  \bibfield  {author} {\bibinfo {author} {\bibfnamefont {Y.}~\bibnamefont
  {Fujimoto}}, \bibinfo {author} {\bibfnamefont {K.}~\bibnamefont
  {Fukushima}},\ and\ \bibinfo {author} {\bibfnamefont {K.}~\bibnamefont
  {Murase}},\ }\href@noop {} {\bibfield  {journal} {\bibinfo  {journal}
  {Journal of High Energy Physics}\ }\textbf {\bibinfo {volume} {2021}},\
  \bibinfo {pages} {1} (\bibinfo {year} {2021})}\BibitemShut {NoStop}%
\bibitem [{\citenamefont {Ferreira}\ and\ \citenamefont
  {Provid\^encia}(2021{\natexlab{a}})}]{Ferreira:2019bny}%
  \BibitemOpen
  \bibfield  {author} {\bibinfo {author} {\bibfnamefont {M.~a.}\ \bibnamefont
  {Ferreira}}\ and\ \bibinfo {author} {\bibfnamefont {C.}~\bibnamefont
  {Provid\^encia}},\ }\href {https://doi.org/10.1088/1475-7516/2021/07/011}
  {\bibfield  {journal} {\bibinfo  {journal} {JCAP}\ }\textbf {\bibinfo
  {volume} {07}},\ \bibinfo {pages} {011}},\ \Eprint
  {https://arxiv.org/abs/1910.05554} {arXiv:1910.05554 [nucl-th]} \BibitemShut
  {NoStop}%
\bibitem [{\citenamefont {Morawski}\ and\ \citenamefont
  {Bejger}(2020)}]{morawski2020neural}%
  \BibitemOpen
  \bibfield  {author} {\bibinfo {author} {\bibfnamefont {F.}~\bibnamefont
  {Morawski}}\ and\ \bibinfo {author} {\bibfnamefont {M.}~\bibnamefont
  {Bejger}},\ }\href@noop {} {\bibfield  {journal} {\bibinfo  {journal}
  {Astronomy \& Astrophysics}\ }\textbf {\bibinfo {volume} {642}},\ \bibinfo
  {pages} {A78} (\bibinfo {year} {2020})}\BibitemShut {NoStop}%
\bibitem [{\citenamefont {Carvalho}\ \emph {et~al.}(2023)\citenamefont
  {Carvalho}, \citenamefont {Ferreira}, \citenamefont {Malik},\ and\
  \citenamefont {Provid\^encia}}]{Carvalho:2023ele}%
  \BibitemOpen
  \bibfield  {author} {\bibinfo {author} {\bibfnamefont {V.}~\bibnamefont
  {Carvalho}}, \bibinfo {author} {\bibfnamefont {M.}~\bibnamefont {Ferreira}},
  \bibinfo {author} {\bibfnamefont {T.}~\bibnamefont {Malik}},\ and\ \bibinfo
  {author} {\bibfnamefont {C.}~\bibnamefont {Provid\^encia}},\ }\href@noop {}
  {\  (\bibinfo {year} {2023})},\ \Eprint {https://arxiv.org/abs/2306.06929}
  {arXiv:2306.06929 [nucl-th]} \BibitemShut {NoStop}%
\bibitem [{\citenamefont {Zhou}\ \emph
  {et~al.}(2023{\natexlab{a}})\citenamefont {Zhou}, \citenamefont {Wang},
  \citenamefont {Pang},\ and\ \citenamefont {Shi}}]{Zhou:2023pti}%
  \BibitemOpen
  \bibfield  {author} {\bibinfo {author} {\bibfnamefont {K.}~\bibnamefont
  {Zhou}}, \bibinfo {author} {\bibfnamefont {L.}~\bibnamefont {Wang}}, \bibinfo
  {author} {\bibfnamefont {L.-G.}\ \bibnamefont {Pang}},\ and\ \bibinfo
  {author} {\bibfnamefont {S.}~\bibnamefont {Shi}},\ }\href@noop {} {\
  (\bibinfo {year} {2023}{\natexlab{a}})},\ \Eprint
  {https://arxiv.org/abs/2303.15136} {arXiv:2303.15136 [hep-ph]} \BibitemShut
  {NoStop}%
\bibitem [{\citenamefont {Mondal}\ and\ \citenamefont
  {Gulminelli}(2021)}]{Mondal2021}%
  \BibitemOpen
  \bibfield  {author} {\bibinfo {author} {\bibfnamefont {C.}~\bibnamefont
  {Mondal}}\ and\ \bibinfo {author} {\bibfnamefont {F.}~\bibnamefont
  {Gulminelli}},\ }\href@noop {} {\  (\bibinfo {year} {2021})},\ \Eprint
  {https://arxiv.org/abs/2111.04520} {arXiv:2111.04520 [nucl-th]} \BibitemShut
  {NoStop}%
\bibitem [{\citenamefont {Imam}\ \emph {et~al.}(2022)\citenamefont {Imam},
  \citenamefont {Patra}, \citenamefont {Mondal}, \citenamefont {Malik},\ and\
  \citenamefont {Agrawal}}]{Imam:2021dbe}%
  \BibitemOpen
  \bibfield  {author} {\bibinfo {author} {\bibfnamefont {S.~M.~A.}\
  \bibnamefont {Imam}}, \bibinfo {author} {\bibfnamefont {N.~K.}\ \bibnamefont
  {Patra}}, \bibinfo {author} {\bibfnamefont {C.}~\bibnamefont {Mondal}},
  \bibinfo {author} {\bibfnamefont {T.}~\bibnamefont {Malik}},\ and\ \bibinfo
  {author} {\bibfnamefont {B.~K.}\ \bibnamefont {Agrawal}},\ }\href
  {https://doi.org/10.1103/PhysRevC.105.015806} {\bibfield  {journal} {\bibinfo
   {journal} {Phys. Rev. C}\ }\textbf {\bibinfo {volume} {105}},\ \bibinfo
  {pages} {015806} (\bibinfo {year} {2022})},\ \Eprint
  {https://arxiv.org/abs/2110.15776} {arXiv:2110.15776 [nucl-th]} \BibitemShut
  {NoStop}%
\bibitem [{\citenamefont {de~Tovar}\ \emph {et~al.}(2021)\citenamefont
  {de~Tovar}, \citenamefont {Ferreira},\ and\ \citenamefont
  {Provid\^encia}}]{Tovar2021}%
  \BibitemOpen
  \bibfield  {author} {\bibinfo {author} {\bibfnamefont {P.~B.}\ \bibnamefont
  {de~Tovar}}, \bibinfo {author} {\bibfnamefont {M.}~\bibnamefont {Ferreira}},\
  and\ \bibinfo {author} {\bibfnamefont {C.}~\bibnamefont {Provid\^encia}},\
  }\href {https://doi.org/10.1103/PhysRevD.104.123036} {\bibfield  {journal}
  {\bibinfo  {journal} {Phys. Rev. D}\ }\textbf {\bibinfo {volume} {104}},\
  \bibinfo {pages} {123036} (\bibinfo {year} {2021})},\ \Eprint
  {https://arxiv.org/abs/2112.05551} {arXiv:2112.05551 [nucl-th]} \BibitemShut
  {NoStop}%
\bibitem [{\citenamefont {Annala}\ \emph {et~al.}(2021)\citenamefont {Annala},
  \citenamefont {Gorda}, \citenamefont {Katerini}, \citenamefont {Kurkela},
  \citenamefont {N\"attil\"a}, \citenamefont {Paschalidis},\ and\ \citenamefont
  {Vuorinen}}]{Annala:2021gom}%
  \BibitemOpen
  \bibfield  {author} {\bibinfo {author} {\bibfnamefont {E.}~\bibnamefont
  {Annala}}, \bibinfo {author} {\bibfnamefont {T.}~\bibnamefont {Gorda}},
  \bibinfo {author} {\bibfnamefont {E.}~\bibnamefont {Katerini}}, \bibinfo
  {author} {\bibfnamefont {A.}~\bibnamefont {Kurkela}}, \bibinfo {author}
  {\bibfnamefont {J.}~\bibnamefont {N\"attil\"a}}, \bibinfo {author}
  {\bibfnamefont {V.}~\bibnamefont {Paschalidis}},\ and\ \bibinfo {author}
  {\bibfnamefont {A.}~\bibnamefont {Vuorinen}},\ }\href@noop {} {\  (\bibinfo
  {year} {2021})},\ \Eprint {https://arxiv.org/abs/2105.05132}
  {arXiv:2105.05132 [astro-ph.HE]} \BibitemShut {NoStop}%
\bibitem [{\citenamefont {Lindblom}(2010)}]{Lindblom:2010bb}%
  \BibitemOpen
  \bibfield  {author} {\bibinfo {author} {\bibfnamefont {L.}~\bibnamefont
  {Lindblom}},\ }\href {https://doi.org/10.1103/PhysRevD.82.103011} {\bibfield
  {journal} {\bibinfo  {journal} {Phys. Rev. D}\ }\textbf {\bibinfo {volume}
  {82}},\ \bibinfo {pages} {103011} (\bibinfo {year} {2010})},\ \Eprint
  {https://arxiv.org/abs/1009.0738} {arXiv:1009.0738 [astro-ph.HE]}
  \BibitemShut {NoStop}%
\bibitem [{\citenamefont {Annala}\ \emph {et~al.}(2020)\citenamefont {Annala},
  \citenamefont {Gorda}, \citenamefont {Kurkela}, \citenamefont {N\"attil\"a},\
  and\ \citenamefont {Vuorinen}}]{Annala2019}%
  \BibitemOpen
  \bibfield  {author} {\bibinfo {author} {\bibfnamefont {E.}~\bibnamefont
  {Annala}}, \bibinfo {author} {\bibfnamefont {T.}~\bibnamefont {Gorda}},
  \bibinfo {author} {\bibfnamefont {A.}~\bibnamefont {Kurkela}}, \bibinfo
  {author} {\bibfnamefont {J.}~\bibnamefont {N\"attil\"a}},\ and\ \bibinfo
  {author} {\bibfnamefont {A.}~\bibnamefont {Vuorinen}},\ }\href
  {https://doi.org/10.1038/s41567-020-0914-9} {\bibfield  {journal} {\bibinfo
  {journal} {Nature Phys.}\ }\textbf {\bibinfo {volume} {16}},\ \bibinfo
  {pages} {907} (\bibinfo {year} {2020})},\ \Eprint
  {https://arxiv.org/abs/1903.09121} {arXiv:1903.09121 [astro-ph.HE]}
  \BibitemShut {NoStop}%
\bibitem [{\citenamefont {Altiparmak}\ \emph {et~al.}(2022)\citenamefont
  {Altiparmak}, \citenamefont {Ecker},\ and\ \citenamefont
  {Rezzolla}}]{Altiparmak:2022bke}%
  \BibitemOpen
  \bibfield  {author} {\bibinfo {author} {\bibfnamefont {S.}~\bibnamefont
  {Altiparmak}}, \bibinfo {author} {\bibfnamefont {C.}~\bibnamefont {Ecker}},\
  and\ \bibinfo {author} {\bibfnamefont {L.}~\bibnamefont {Rezzolla}},\ }\href
  {https://doi.org/10.3847/2041-8213/ac9b2a} {\bibfield  {journal} {\bibinfo
  {journal} {Astrophys. J. Lett.}\ }\textbf {\bibinfo {volume} {939}},\
  \bibinfo {pages} {L34} (\bibinfo {year} {2022})},\ \Eprint
  {https://arxiv.org/abs/2203.14974} {arXiv:2203.14974 [astro-ph.HE]}
  \BibitemShut {NoStop}%
\bibitem [{\citenamefont {Somasundaram}\ \emph {et~al.}(2023)\citenamefont
  {Somasundaram}, \citenamefont {Tews},\ and\ \citenamefont
  {Margueron}}]{Somasundaram:2022ztm}%
  \BibitemOpen
  \bibfield  {author} {\bibinfo {author} {\bibfnamefont {R.}~\bibnamefont
  {Somasundaram}}, \bibinfo {author} {\bibfnamefont {I.}~\bibnamefont {Tews}},\
  and\ \bibinfo {author} {\bibfnamefont {J.}~\bibnamefont {Margueron}},\ }\href
  {https://doi.org/10.1103/PhysRevC.107.L052801} {\bibfield  {journal}
  {\bibinfo  {journal} {Phys. Rev. C}\ }\textbf {\bibinfo {volume} {107}},\
  \bibinfo {pages} {L052801} (\bibinfo {year} {2023})},\ \Eprint
  {https://arxiv.org/abs/2204.14039} {arXiv:2204.14039 [nucl-th]} \BibitemShut
  {NoStop}%
\bibitem [{\citenamefont {Margueron}\ \emph
  {et~al.}(2018{\natexlab{b}})\citenamefont {Margueron}, \citenamefont
  {Hoffmann~Casali},\ and\ \citenamefont {Gulminelli}}]{Margueron:2017eqc}%
  \BibitemOpen
  \bibfield  {author} {\bibinfo {author} {\bibfnamefont {J.}~\bibnamefont
  {Margueron}}, \bibinfo {author} {\bibfnamefont {R.}~\bibnamefont
  {Hoffmann~Casali}},\ and\ \bibinfo {author} {\bibfnamefont {F.}~\bibnamefont
  {Gulminelli}},\ }\href {https://doi.org/10.1103/PhysRevC.97.025805}
  {\bibfield  {journal} {\bibinfo  {journal} {Phys. Rev. C}\ }\textbf {\bibinfo
  {volume} {97}},\ \bibinfo {pages} {025805} (\bibinfo {year}
  {2018}{\natexlab{b}})},\ \Eprint {https://arxiv.org/abs/1708.06894}
  {arXiv:1708.06894 [nucl-th]} \BibitemShut {NoStop}%
\bibitem [{\citenamefont {Margueron}\ \emph
  {et~al.}(2018{\natexlab{c}})\citenamefont {Margueron}, \citenamefont
  {Hoffmann~Casali},\ and\ \citenamefont {Gulminelli}}]{Margueron:2017lup}%
  \BibitemOpen
  \bibfield  {author} {\bibinfo {author} {\bibfnamefont {J.}~\bibnamefont
  {Margueron}}, \bibinfo {author} {\bibfnamefont {R.}~\bibnamefont
  {Hoffmann~Casali}},\ and\ \bibinfo {author} {\bibfnamefont {F.}~\bibnamefont
  {Gulminelli}},\ }\href {https://doi.org/10.1103/PhysRevC.97.025806}
  {\bibfield  {journal} {\bibinfo  {journal} {Phys. Rev. C}\ }\textbf {\bibinfo
  {volume} {97}},\ \bibinfo {pages} {025806} (\bibinfo {year}
  {2018}{\natexlab{c}})},\ \Eprint {https://arxiv.org/abs/1708.06895}
  {arXiv:1708.06895 [nucl-th]} \BibitemShut {NoStop}%
\bibitem [{\citenamefont {Ferreira}\ \emph {et~al.}(2020)\citenamefont
  {Ferreira}, \citenamefont {Fortin}, \citenamefont {Malik}, \citenamefont
  {Agrawal},\ and\ \citenamefont {Provid\^encia}}]{Ferreira:2019bgy}%
  \BibitemOpen
  \bibfield  {author} {\bibinfo {author} {\bibfnamefont {M.}~\bibnamefont
  {Ferreira}}, \bibinfo {author} {\bibfnamefont {M.}~\bibnamefont {Fortin}},
  \bibinfo {author} {\bibfnamefont {T.}~\bibnamefont {Malik}}, \bibinfo
  {author} {\bibfnamefont {B.~K.}\ \bibnamefont {Agrawal}},\ and\ \bibinfo
  {author} {\bibfnamefont {C.}~\bibnamefont {Provid\^encia}},\ }\href
  {https://doi.org/10.1103/PhysRevD.101.043021} {\bibfield  {journal} {\bibinfo
   {journal} {Phys. Rev. D}\ }\textbf {\bibinfo {volume} {101}},\ \bibinfo
  {pages} {043021} (\bibinfo {year} {2020})},\ \Eprint
  {https://arxiv.org/abs/1912.11131} {arXiv:1912.11131 [nucl-th]} \BibitemShut
  {NoStop}%
\bibitem [{\citenamefont {Xie}\ and\ \citenamefont {Li}(2019)}]{Xie:2019sqb}%
  \BibitemOpen
  \bibfield  {author} {\bibinfo {author} {\bibfnamefont {W.-J.}\ \bibnamefont
  {Xie}}\ and\ \bibinfo {author} {\bibfnamefont {B.-A.}\ \bibnamefont {Li}},\
  }\href {https://doi.org/10.3847/1538-4357/ab3f37} {\bibfield  {journal}
  {\bibinfo  {journal} {Astrophys. J.}\ }\textbf {\bibinfo {volume} {883}},\
  \bibinfo {pages} {174} (\bibinfo {year} {2019})},\ \Eprint
  {https://arxiv.org/abs/1907.10741} {arXiv:1907.10741 [astro-ph.HE]}
  \BibitemShut {NoStop}%
\bibitem [{\citenamefont {Xie}\ and\ \citenamefont {Li}(2020)}]{Xie:2020tdo}%
  \BibitemOpen
  \bibfield  {author} {\bibinfo {author} {\bibfnamefont {W.-J.}\ \bibnamefont
  {Xie}}\ and\ \bibinfo {author} {\bibfnamefont {B.-A.}\ \bibnamefont {Li}},\
  }\href {https://doi.org/10.3847/1538-4357/aba271} {\bibfield  {journal}
  {\bibinfo  {journal} {Astrophys. J.}\ }\textbf {\bibinfo {volume} {899}},\
  \bibinfo {pages} {4} (\bibinfo {year} {2020})},\ \Eprint
  {https://arxiv.org/abs/2005.07216} {arXiv:2005.07216 [astro-ph.HE]}
  \BibitemShut {NoStop}%
\bibitem [{\citenamefont {Ferreira}\ and\ \citenamefont
  {Provid\^encia}(2021{\natexlab{b}})}]{Ferreira:2021pni}%
  \BibitemOpen
  \bibfield  {author} {\bibinfo {author} {\bibfnamefont {M.}~\bibnamefont
  {Ferreira}}\ and\ \bibinfo {author} {\bibfnamefont {C.}~\bibnamefont
  {Provid\^encia}},\ }\href {https://doi.org/10.1103/PhysRevD.104.063006}
  {\bibfield  {journal} {\bibinfo  {journal} {Phys. Rev. D}\ }\textbf {\bibinfo
  {volume} {104}},\ \bibinfo {pages} {063006} (\bibinfo {year}
  {2021}{\natexlab{b}})},\ \Eprint {https://arxiv.org/abs/2110.00305}
  {arXiv:2110.00305 [nucl-th]} \BibitemShut {NoStop}%
\bibitem [{\citenamefont {Thi}\ \emph {et~al.}(2021)\citenamefont {Thi},
  \citenamefont {Mondal},\ and\ \citenamefont {Gulminelli}}]{Thi:2021jhz}%
  \BibitemOpen
  \bibfield  {author} {\bibinfo {author} {\bibfnamefont {H.~D.}\ \bibnamefont
  {Thi}}, \bibinfo {author} {\bibfnamefont {C.}~\bibnamefont {Mondal}},\ and\
  \bibinfo {author} {\bibfnamefont {F.}~\bibnamefont {Gulminelli}},\ }\href
  {https://doi.org/10.3390/universe7100373} {\bibfield  {journal} {\bibinfo
  {journal} {Universe}\ }\textbf {\bibinfo {volume} {7}},\ \bibinfo {pages}
  {373} (\bibinfo {year} {2021})},\ \Eprint {https://arxiv.org/abs/2109.09675}
  {arXiv:2109.09675 [astro-ph.HE]} \BibitemShut {NoStop}%
\bibitem [{\citenamefont {Landry}\ and\ \citenamefont
  {Essick}(2019)}]{Landry:2018prl}%
  \BibitemOpen
  \bibfield  {author} {\bibinfo {author} {\bibfnamefont {P.}~\bibnamefont
  {Landry}}\ and\ \bibinfo {author} {\bibfnamefont {R.}~\bibnamefont
  {Essick}},\ }\href {https://doi.org/10.1103/PhysRevD.99.084049} {\bibfield
  {journal} {\bibinfo  {journal} {Phys. Rev. D}\ }\textbf {\bibinfo {volume}
  {99}},\ \bibinfo {pages} {084049} (\bibinfo {year} {2019})},\ \Eprint
  {https://arxiv.org/abs/1811.12529} {arXiv:1811.12529 [gr-qc]} \BibitemShut
  {NoStop}%
\bibitem [{\citenamefont {Essick}\ \emph {et~al.}(2020)\citenamefont {Essick},
  \citenamefont {Landry},\ and\ \citenamefont {Holz}}]{Essick:2019ldf}%
  \BibitemOpen
  \bibfield  {author} {\bibinfo {author} {\bibfnamefont {R.}~\bibnamefont
  {Essick}}, \bibinfo {author} {\bibfnamefont {P.}~\bibnamefont {Landry}},\
  and\ \bibinfo {author} {\bibfnamefont {D.~E.}\ \bibnamefont {Holz}},\ }\href
  {https://doi.org/10.1103/PhysRevD.101.063007} {\bibfield  {journal} {\bibinfo
   {journal} {Phys. Rev. D}\ }\textbf {\bibinfo {volume} {101}},\ \bibinfo
  {pages} {063007} (\bibinfo {year} {2020})},\ \Eprint
  {https://arxiv.org/abs/1910.09740} {arXiv:1910.09740 [astro-ph.HE]}
  \BibitemShut {NoStop}%
\bibitem [{\citenamefont {Zhou}\ \emph
  {et~al.}(2023{\natexlab{b}})\citenamefont {Zhou}, \citenamefont {Hu},
  \citenamefont {Zhang},\ and\ \citenamefont {Shen}}]{Zhou:2023cfs}%
  \BibitemOpen
  \bibfield  {author} {\bibinfo {author} {\bibfnamefont {W.}~\bibnamefont
  {Zhou}}, \bibinfo {author} {\bibfnamefont {J.}~\bibnamefont {Hu}}, \bibinfo
  {author} {\bibfnamefont {Y.}~\bibnamefont {Zhang}},\ and\ \bibinfo {author}
  {\bibfnamefont {H.}~\bibnamefont {Shen}},\ }\href
  {https://doi.org/10.3847/1538-4357/acd335} {\bibfield  {journal} {\bibinfo
  {journal} {Astrophys. J.}\ }\textbf {\bibinfo {volume} {950}},\ \bibinfo
  {pages} {186} (\bibinfo {year} {2023}{\natexlab{b}})},\ \Eprint
  {https://arxiv.org/abs/2305.03323} {arXiv:2305.03323 [nucl-th]} \BibitemShut
  {NoStop}%
\bibitem [{\citenamefont {Fujimoto}\ \emph {et~al.}(2022)\citenamefont
  {Fujimoto}, \citenamefont {Fukushima}, \citenamefont {McLerran},\ and\
  \citenamefont {Praszalowicz}}]{Fujimoto:2022ohj}%
  \BibitemOpen
  \bibfield  {author} {\bibinfo {author} {\bibfnamefont {Y.}~\bibnamefont
  {Fujimoto}}, \bibinfo {author} {\bibfnamefont {K.}~\bibnamefont {Fukushima}},
  \bibinfo {author} {\bibfnamefont {L.~D.}\ \bibnamefont {McLerran}},\ and\
  \bibinfo {author} {\bibfnamefont {M.}~\bibnamefont {Praszalowicz}},\ }\href
  {https://doi.org/10.1103/PhysRevLett.129.252702} {\bibfield  {journal}
  {\bibinfo  {journal} {Phys. Rev. Lett.}\ }\textbf {\bibinfo {volume} {129}},\
  \bibinfo {pages} {252702} (\bibinfo {year} {2022})},\ \Eprint
  {https://arxiv.org/abs/2207.06753} {arXiv:2207.06753 [nucl-th]} \BibitemShut
  {NoStop}%
\bibitem [{\citenamefont {Annala}\ \emph {et~al.}(2023)\citenamefont {Annala},
  \citenamefont {Gorda}, \citenamefont {Hirvonen}, \citenamefont {Komoltsev},
  \citenamefont {Kurkela}, \citenamefont {N\"attil\"a},\ and\ \citenamefont
  {Vuorinen}}]{Annala:2023cwx}%
  \BibitemOpen
  \bibfield  {author} {\bibinfo {author} {\bibfnamefont {E.}~\bibnamefont
  {Annala}}, \bibinfo {author} {\bibfnamefont {T.}~\bibnamefont {Gorda}},
  \bibinfo {author} {\bibfnamefont {J.}~\bibnamefont {Hirvonen}}, \bibinfo
  {author} {\bibfnamefont {O.}~\bibnamefont {Komoltsev}}, \bibinfo {author}
  {\bibfnamefont {A.}~\bibnamefont {Kurkela}}, \bibinfo {author} {\bibfnamefont
  {J.}~\bibnamefont {N\"attil\"a}},\ and\ \bibinfo {author} {\bibfnamefont
  {A.}~\bibnamefont {Vuorinen}},\ }\href@noop {} {\  (\bibinfo {year}
  {2023})},\ \Eprint {https://arxiv.org/abs/2303.11356} {arXiv:2303.11356
  [astro-ph.HE]} \BibitemShut {NoStop}%
\bibitem [{\citenamefont {Malik}\ \emph
  {et~al.}(2022{\natexlab{a}})\citenamefont {Malik}, \citenamefont {Agrawal},\
  and\ \citenamefont {Provid\^encia}}]{Malik:2022ilb}%
  \BibitemOpen
  \bibfield  {author} {\bibinfo {author} {\bibfnamefont {T.}~\bibnamefont
  {Malik}}, \bibinfo {author} {\bibfnamefont {B.~K.}\ \bibnamefont {Agrawal}},\
  and\ \bibinfo {author} {\bibfnamefont {C.}~\bibnamefont {Provid\^encia}},\
  }\href {https://doi.org/10.1103/PhysRevC.106.L042801} {\bibfield  {journal}
  {\bibinfo  {journal} {Phys. Rev. C}\ }\textbf {\bibinfo {volume} {106}},\
  \bibinfo {pages} {L042801} (\bibinfo {year} {2022}{\natexlab{a}})},\ \Eprint
  {https://arxiv.org/abs/2206.15404} {arXiv:2206.15404 [nucl-th]} \BibitemShut
  {NoStop}%
\bibitem [{\citenamefont {Malik}\ and\ \citenamefont
  {Provid\^encia}(2022)}]{Malik:2022jqc}%
  \BibitemOpen
  \bibfield  {author} {\bibinfo {author} {\bibfnamefont {T.}~\bibnamefont
  {Malik}}\ and\ \bibinfo {author} {\bibfnamefont {C.}~\bibnamefont
  {Provid\^encia}},\ }\href@noop {} {\  (\bibinfo {year} {2022})},\ \Eprint
  {https://arxiv.org/abs/2205.15843} {arXiv:2205.15843 [nucl-th]} \BibitemShut
  {NoStop}%
\bibitem [{\citenamefont {Malik}\ \emph
  {et~al.}(2022{\natexlab{b}})\citenamefont {Malik}, \citenamefont {Ferreira},
  \citenamefont {Agrawal},\ and\ \citenamefont
  {Provid\^encia}}]{Malik:2022zol}%
  \BibitemOpen
  \bibfield  {author} {\bibinfo {author} {\bibfnamefont {T.}~\bibnamefont
  {Malik}}, \bibinfo {author} {\bibfnamefont {M.}~\bibnamefont {Ferreira}},
  \bibinfo {author} {\bibfnamefont {B.~K.}\ \bibnamefont {Agrawal}},\ and\
  \bibinfo {author} {\bibfnamefont {C.}~\bibnamefont {Provid\^encia}},\ }\href
  {https://doi.org/10.3847/1538-4357/ac5d3c} {\bibfield  {journal} {\bibinfo
  {journal} {Astrophys. J.}\ }\textbf {\bibinfo {volume} {930}},\ \bibinfo
  {pages} {17} (\bibinfo {year} {2022}{\natexlab{b}})},\ \Eprint
  {https://arxiv.org/abs/2201.12552} {arXiv:2201.12552 [nucl-th]} \BibitemShut
  {NoStop}%
\bibitem [{\citenamefont {Malik}\ \emph {et~al.}(2023)\citenamefont {Malik},
  \citenamefont {Ferreira}, \citenamefont {Albino},\ and\ \citenamefont
  {Provid\^encia}}]{Malik:2023mnx}%
  \BibitemOpen
  \bibfield  {author} {\bibinfo {author} {\bibfnamefont {T.}~\bibnamefont
  {Malik}}, \bibinfo {author} {\bibfnamefont {M.}~\bibnamefont {Ferreira}},
  \bibinfo {author} {\bibfnamefont {M.~B.}\ \bibnamefont {Albino}},\ and\
  \bibinfo {author} {\bibfnamefont {C.}~\bibnamefont {Provid\^encia}},\
  }\href@noop {} {\  (\bibinfo {year} {2023})},\ \Eprint
  {https://arxiv.org/abs/2301.08169} {arXiv:2301.08169 [nucl-th]} \BibitemShut
  {NoStop}%
\bibitem [{\citenamefont {Boguta}\ and\ \citenamefont
  {Bodmer}(1977)}]{Boguta1977}%
  \BibitemOpen
  \bibfield  {author} {\bibinfo {author} {\bibfnamefont {J.}~\bibnamefont
  {Boguta}}\ and\ \bibinfo {author} {\bibfnamefont {A.~R.}\ \bibnamefont
  {Bodmer}},\ }\href {https://doi.org/10.1016/0375-9474(77)90626-1} {\bibfield
  {journal} {\bibinfo  {journal} {Nucl. Phys. A}\ }\textbf {\bibinfo {volume}
  {292}},\ \bibinfo {pages} {413} (\bibinfo {year} {1977})}\BibitemShut
  {NoStop}%
\bibitem [{\citenamefont {Mueller}\ and\ \citenamefont
  {Serot}(1996{\natexlab{a}})}]{Mueller1996}%
  \BibitemOpen
  \bibfield  {author} {\bibinfo {author} {\bibfnamefont {H.}~\bibnamefont
  {Mueller}}\ and\ \bibinfo {author} {\bibfnamefont {B.~D.}\ \bibnamefont
  {Serot}},\ }\href {https://doi.org/10.1016/0375-9474(96)00187-X} {\bibfield
  {journal} {\bibinfo  {journal} {Nucl. Phys. A}\ }\textbf {\bibinfo {volume}
  {606}},\ \bibinfo {pages} {508} (\bibinfo {year} {1996}{\natexlab{a}})},\
  \Eprint {https://arxiv.org/abs/nucl-th/9603037} {arXiv:nucl-th/9603037}
  \BibitemShut {NoStop}%
\bibitem [{\citenamefont {Typel}\ and\ \citenamefont
  {Wolter}(1999)}]{Typel1999}%
  \BibitemOpen
  \bibfield  {author} {\bibinfo {author} {\bibfnamefont {S.}~\bibnamefont
  {Typel}}\ and\ \bibinfo {author} {\bibfnamefont {H.~H.}\ \bibnamefont
  {Wolter}},\ }\href {https://doi.org/10.1016/S0375-9474(99)00310-3} {\bibfield
   {journal} {\bibinfo  {journal} {Nucl. Phys. A}\ }\textbf {\bibinfo {volume}
  {656}},\ \bibinfo {pages} {331} (\bibinfo {year} {1999})}\BibitemShut
  {NoStop}%
\bibitem [{\citenamefont {Typel}\ \emph {et~al.}(2010)\citenamefont {Typel},
  \citenamefont {Ropke}, \citenamefont {Klahn}, \citenamefont {Blaschke},\ and\
  \citenamefont {Wolter}}]{Typel2009}%
  \BibitemOpen
  \bibfield  {author} {\bibinfo {author} {\bibfnamefont {S.}~\bibnamefont
  {Typel}}, \bibinfo {author} {\bibfnamefont {G.}~\bibnamefont {Ropke}},
  \bibinfo {author} {\bibfnamefont {T.}~\bibnamefont {Klahn}}, \bibinfo
  {author} {\bibfnamefont {D.}~\bibnamefont {Blaschke}},\ and\ \bibinfo
  {author} {\bibfnamefont {H.~H.}\ \bibnamefont {Wolter}},\ }\href
  {https://doi.org/10.1103/PhysRevC.81.015803} {\bibfield  {journal} {\bibinfo
  {journal} {Phys. Rev. C}\ }\textbf {\bibinfo {volume} {81}},\ \bibinfo
  {pages} {015803} (\bibinfo {year} {2010})},\ \Eprint
  {https://arxiv.org/abs/0908.2344} {arXiv:0908.2344 [nucl-th]} \BibitemShut
  {NoStop}%
\bibitem [{\citenamefont {Traversi}\ \emph {et~al.}(2020)\citenamefont
  {Traversi}, \citenamefont {Char},\ and\ \citenamefont
  {Pagliara}}]{Traversi:2020aaa}%
  \BibitemOpen
  \bibfield  {author} {\bibinfo {author} {\bibfnamefont {S.}~\bibnamefont
  {Traversi}}, \bibinfo {author} {\bibfnamefont {P.}~\bibnamefont {Char}},\
  and\ \bibinfo {author} {\bibfnamefont {G.}~\bibnamefont {Pagliara}},\ }\href
  {https://doi.org/10.3847/1538-4357/ab99c1} {\bibfield  {journal} {\bibinfo
  {journal} {Astrophys. J.}\ }\textbf {\bibinfo {volume} {897}},\ \bibinfo
  {pages} {165} (\bibinfo {year} {2020})},\ \Eprint
  {https://arxiv.org/abs/2002.08951} {arXiv:2002.08951 [astro-ph.HE]}
  \BibitemShut {NoStop}%
\bibitem [{\citenamefont {Sun}\ \emph {et~al.}(2023)\citenamefont {Sun},
  \citenamefont {Miao}, \citenamefont {Sun},\ and\ \citenamefont
  {Li}}]{Sun:2022yor}%
  \BibitemOpen
  \bibfield  {author} {\bibinfo {author} {\bibfnamefont {X.}~\bibnamefont
  {Sun}}, \bibinfo {author} {\bibfnamefont {Z.}~\bibnamefont {Miao}}, \bibinfo
  {author} {\bibfnamefont {B.}~\bibnamefont {Sun}},\ and\ \bibinfo {author}
  {\bibfnamefont {A.}~\bibnamefont {Li}},\ }\href
  {https://doi.org/10.3847/1538-4357/ac9d9a} {\bibfield  {journal} {\bibinfo
  {journal} {Astrophys. J.}\ }\textbf {\bibinfo {volume} {942}},\ \bibinfo
  {pages} {55} (\bibinfo {year} {2023})},\ \Eprint
  {https://arxiv.org/abs/2205.10631} {arXiv:2205.10631 [astro-ph.HE]}
  \BibitemShut {NoStop}%
\bibitem [{\citenamefont {Beznogov}\ and\ \citenamefont
  {Raduta}(2023)}]{Beznogov:2022rri}%
  \BibitemOpen
  \bibfield  {author} {\bibinfo {author} {\bibfnamefont {M.~V.}\ \bibnamefont
  {Beznogov}}\ and\ \bibinfo {author} {\bibfnamefont {A.~R.}\ \bibnamefont
  {Raduta}},\ }\href {https://doi.org/10.1103/PhysRevC.107.045803} {\bibfield
  {journal} {\bibinfo  {journal} {Phys. Rev. C}\ }\textbf {\bibinfo {volume}
  {107}},\ \bibinfo {pages} {045803} (\bibinfo {year} {2023})},\ \Eprint
  {https://arxiv.org/abs/2212.07168} {arXiv:2212.07168 [nucl-th]} \BibitemShut
  {NoStop}%
\bibitem [{\citenamefont {Huang}\ \emph {et~al.}(2023)\citenamefont {Huang},
  \citenamefont {Raaijmakers}, \citenamefont {Watts}, \citenamefont {Tolos},\
  and\ \citenamefont {Provid\^encia}}]{Huang:2023grj}%
  \BibitemOpen
  \bibfield  {author} {\bibinfo {author} {\bibfnamefont {C.}~\bibnamefont
  {Huang}}, \bibinfo {author} {\bibfnamefont {G.}~\bibnamefont {Raaijmakers}},
  \bibinfo {author} {\bibfnamefont {A.~L.}\ \bibnamefont {Watts}}, \bibinfo
  {author} {\bibfnamefont {L.}~\bibnamefont {Tolos}},\ and\ \bibinfo {author}
  {\bibfnamefont {C.}~\bibnamefont {Provid\^encia}},\ }\href@noop {} {\
  (\bibinfo {year} {2023})},\ \Eprint {https://arxiv.org/abs/2303.17518}
  {arXiv:2303.17518 [astro-ph.HE]} \BibitemShut {NoStop}%
\bibitem [{\citenamefont {{Watts}}(2019)}]{extp_watts}%
  \BibitemOpen
  \bibfield  {author} {\bibinfo {author} {\bibfnamefont {A.~L.~{\it et al.}.}\
  \bibnamefont {{Watts}}},\ }\href {https://doi.org/10.1007/s11433-017-9188-4}
  {\bibfield  {journal} {\bibinfo  {journal} {Science China Physics, Mechanics,
  and Astronomy}\ }\textbf {\bibinfo {volume} {62}},\ \bibinfo {eid} {29503}
  (\bibinfo {year} {2019})},\ \Eprint {https://arxiv.org/abs/1812.04021}
  {arXiv:1812.04021 [astro-ph.HE]} \BibitemShut {NoStop}%
\bibitem [{\citenamefont {{Ray}}(2019)}]{strobex}%
  \BibitemOpen
  \bibfield  {author} {\bibinfo {author} {\bibfnamefont {P.~S.~{\it et al.}.}\
  \bibnamefont {{Ray}}},\ }\href@noop {} {\bibfield  {journal} {\bibinfo
  {journal} {arXiv e-prints}\ ,\ \bibinfo {eid} {arXiv:1903.03035}} (\bibinfo
  {year} {2019})},\ \Eprint {https://arxiv.org/abs/1903.03035}
  {arXiv:1903.03035 [astro-ph.IM]} \BibitemShut {NoStop}%
\bibitem [{\citenamefont {Mueller}\ and\ \citenamefont
  {Serot}(1996{\natexlab{b}})}]{Mueller:1996pm}%
  \BibitemOpen
  \bibfield  {author} {\bibinfo {author} {\bibfnamefont {H.}~\bibnamefont
  {Mueller}}\ and\ \bibinfo {author} {\bibfnamefont {B.~D.}\ \bibnamefont
  {Serot}},\ }\href {https://doi.org/10.1016/0375-9474(96)00187-X} {\bibfield
  {journal} {\bibinfo  {journal} {Nucl. Phys. A}\ }\textbf {\bibinfo {volume}
  {606}},\ \bibinfo {pages} {508} (\bibinfo {year} {1996}{\natexlab{b}})},\
  \Eprint {https://arxiv.org/abs/nucl-th/9603037} {arXiv:nucl-th/9603037}
  \BibitemShut {NoStop}%
\bibitem [{\citenamefont {Sugahara}\ and\ \citenamefont
  {Toki}(1994)}]{Sugahara:1993wz}%
  \BibitemOpen
  \bibfield  {author} {\bibinfo {author} {\bibfnamefont {Y.}~\bibnamefont
  {Sugahara}}\ and\ \bibinfo {author} {\bibfnamefont {H.}~\bibnamefont
  {Toki}},\ }\href {https://doi.org/10.1016/0375-9474(94)90923-7} {\bibfield
  {journal} {\bibinfo  {journal} {Nucl. Phys. A}\ }\textbf {\bibinfo {volume}
  {579}},\ \bibinfo {pages} {557} (\bibinfo {year} {1994})}\BibitemShut
  {NoStop}%
\bibitem [{\citenamefont {Cavagnoli}\ \emph {et~al.}(2011)\citenamefont
  {Cavagnoli}, \citenamefont {Menezes},\ and\ \citenamefont
  {Providencia}}]{Cavagnoli:2011ft}%
  \BibitemOpen
  \bibfield  {author} {\bibinfo {author} {\bibfnamefont {R.}~\bibnamefont
  {Cavagnoli}}, \bibinfo {author} {\bibfnamefont {D.~P.}\ \bibnamefont
  {Menezes}},\ and\ \bibinfo {author} {\bibfnamefont {C.}~\bibnamefont
  {Providencia}},\ }\href {https://doi.org/10.1103/PhysRevC.84.065810}
  {\bibfield  {journal} {\bibinfo  {journal} {Phys. Rev. C}\ }\textbf {\bibinfo
  {volume} {84}},\ \bibinfo {pages} {065810} (\bibinfo {year} {2011})},\
  \Eprint {https://arxiv.org/abs/1108.1733} {arXiv:1108.1733 [hep-ph]}
  \BibitemShut {NoStop}%
\bibitem [{\citenamefont {Gelman}\ \emph {et~al.}(2013)\citenamefont {Gelman},
  \citenamefont {Carlin}, \citenamefont {Stern}, \citenamefont {Dunson},
  \citenamefont {Vehtari}, \citenamefont {Rubin}, \citenamefont {Carlin},
  \citenamefont {Stern}, \citenamefont {Rubin},\ and\ \citenamefont
  {Dunson}}]{Gelman2013}%
  \BibitemOpen
  \bibfield  {author} {\bibinfo {author} {\bibfnamefont {A.}~\bibnamefont
  {Gelman}}, \bibinfo {author} {\bibfnamefont {J.~B.}\ \bibnamefont {Carlin}},
  \bibinfo {author} {\bibfnamefont {H.~S.}\ \bibnamefont {Stern}}, \bibinfo
  {author} {\bibfnamefont {D.~B.}\ \bibnamefont {Dunson}}, \bibinfo {author}
  {\bibfnamefont {A.}~\bibnamefont {Vehtari}}, \bibinfo {author} {\bibfnamefont
  {D.~B.}\ \bibnamefont {Rubin}}, \bibinfo {author} {\bibfnamefont
  {J.}~\bibnamefont {Carlin}}, \bibinfo {author} {\bibfnamefont
  {H.}~\bibnamefont {Stern}}, \bibinfo {author} {\bibfnamefont
  {D.}~\bibnamefont {Rubin}},\ and\ \bibinfo {author} {\bibfnamefont
  {D.}~\bibnamefont {Dunson}},\ }\href@noop {} {\emph {\bibinfo {title}
  {{Bayesian Data Analysis Third edition}}}}\ (\bibinfo  {publisher} {CRC
  Press, Boca Raton, Florida},\ \bibinfo {year} {2013})\BibitemShut {NoStop}%
\bibitem [{\citenamefont {{Skilling}}(2004)}]{Skilling2004}%
  \BibitemOpen
  \bibfield  {author} {\bibinfo {author} {\bibfnamefont {J.}~\bibnamefont
  {{Skilling}}},\ }in\ \href {https://doi.org/10.1063/1.1835238} {\emph
  {\bibinfo {booktitle} {Bayesian Inference and Maximum Entropy Methods in
  Science and Engineering: 24th International Workshop on Bayesian Inference
  and Maximum Entropy Methods in Science and Engineering}}},\ \bibinfo {series}
  {American Institute of Physics Conference Series}, Vol.\ \bibinfo {volume}
  {735},\ \bibinfo {editor} {edited by\ \bibinfo {editor} {\bibfnamefont
  {R.}~\bibnamefont {{Fischer}}}, \bibinfo {editor} {\bibfnamefont
  {R.}~\bibnamefont {{Preuss}}},\ and\ \bibinfo {editor} {\bibfnamefont
  {U.~V.}\ \bibnamefont {{Toussaint}}}}\ (\bibinfo {year} {2004})\ pp.\
  \bibinfo {pages} {395--405}\BibitemShut {NoStop}%
\bibitem [{\citenamefont {Buchner}\ \emph {et~al.}(2014)\citenamefont
  {Buchner}, \citenamefont {Georgakakis}, \citenamefont {Nandra}, \citenamefont
  {Hsu}, \citenamefont {Rangel}, \citenamefont {Brightman}, \citenamefont
  {Merloni}, \citenamefont {Salvato}, \citenamefont {Donley},\ and\
  \citenamefont {Kocevski}}]{Buchner:2014nha}%
  \BibitemOpen
  \bibfield  {author} {\bibinfo {author} {\bibfnamefont {J.}~\bibnamefont
  {Buchner}}, \bibinfo {author} {\bibfnamefont {A.}~\bibnamefont
  {Georgakakis}}, \bibinfo {author} {\bibfnamefont {K.}~\bibnamefont {Nandra}},
  \bibinfo {author} {\bibfnamefont {L.}~\bibnamefont {Hsu}}, \bibinfo {author}
  {\bibfnamefont {C.}~\bibnamefont {Rangel}}, \bibinfo {author} {\bibfnamefont
  {M.}~\bibnamefont {Brightman}}, \bibinfo {author} {\bibfnamefont
  {A.}~\bibnamefont {Merloni}}, \bibinfo {author} {\bibfnamefont
  {M.}~\bibnamefont {Salvato}}, \bibinfo {author} {\bibfnamefont
  {J.}~\bibnamefont {Donley}},\ and\ \bibinfo {author} {\bibfnamefont
  {D.}~\bibnamefont {Kocevski}},\ }\href
  {https://doi.org/10.1051/0004-6361/201322971} {\bibfield  {journal} {\bibinfo
   {journal} {Astron. Astrophys.}\ }\textbf {\bibinfo {volume} {564}},\
  \bibinfo {pages} {A125} (\bibinfo {year} {2014})},\ \Eprint
  {https://arxiv.org/abs/1402.0004} {arXiv:1402.0004 [astro-ph.HE]}
  \BibitemShut {NoStop}%
\bibitem [{\citenamefont {Buchner}(2021)}]{buchner2021nested}%
  \BibitemOpen
  \bibfield  {author} {\bibinfo {author} {\bibfnamefont {J.}~\bibnamefont
  {Buchner}},\ }\href@noop {} {\bibinfo {title} {Nested sampling methods}}
  (\bibinfo {year} {2021}),\ \Eprint {https://arxiv.org/abs/2101.09675}
  {arXiv:2101.09675 [stat.CO]} \BibitemShut {NoStop}%
\bibitem [{\citenamefont {Dutra}\ \emph {et~al.}(2014)\citenamefont {Dutra},
  \citenamefont {Louren\c{c}o}, \citenamefont {Avancini}, \citenamefont
  {Carlson}, \citenamefont {Delfino}, \citenamefont {Menezes}, \citenamefont
  {Provid\^encia}, \citenamefont {Typel},\ and\ \citenamefont
  {Stone}}]{Dutra:2014qga}%
  \BibitemOpen
  \bibfield  {author} {\bibinfo {author} {\bibfnamefont {M.}~\bibnamefont
  {Dutra}}, \bibinfo {author} {\bibfnamefont {O.}~\bibnamefont {Louren\c{c}o}},
  \bibinfo {author} {\bibfnamefont {S.~S.}\ \bibnamefont {Avancini}}, \bibinfo
  {author} {\bibfnamefont {B.~V.}\ \bibnamefont {Carlson}}, \bibinfo {author}
  {\bibfnamefont {A.}~\bibnamefont {Delfino}}, \bibinfo {author} {\bibfnamefont
  {D.~P.}\ \bibnamefont {Menezes}}, \bibinfo {author} {\bibfnamefont
  {C.}~\bibnamefont {Provid\^encia}}, \bibinfo {author} {\bibfnamefont
  {S.}~\bibnamefont {Typel}},\ and\ \bibinfo {author} {\bibfnamefont {J.~R.}\
  \bibnamefont {Stone}},\ }\href {https://doi.org/10.1103/PhysRevC.90.055203}
  {\bibfield  {journal} {\bibinfo  {journal} {Phys. Rev. C}\ }\textbf {\bibinfo
  {volume} {90}},\ \bibinfo {pages} {055203} (\bibinfo {year} {2014})},\
  \Eprint {https://arxiv.org/abs/1405.3633} {arXiv:1405.3633 [nucl-th]}
  \BibitemShut {NoStop}%
\bibitem [{\citenamefont {{Shlomo, S.}}\ \emph {et~al.}(2006)\citenamefont
  {{Shlomo, S.}}, \citenamefont {{Kolomietz, V. M.}},\ and\ \citenamefont
  {{Col\`o, G.}}}]{Shlomo2006}%
  \BibitemOpen
  \bibfield  {author} {\bibinfo {author} {\bibnamefont {{Shlomo, S.}}},
  \bibinfo {author} {\bibnamefont {{Kolomietz, V. M.}}},\ and\ \bibinfo
  {author} {\bibnamefont {{Col\`o, G.}}},\ }\href
  {https://doi.org/10.1140/epja/i2006-10100-3} {\bibfield  {journal} {\bibinfo
  {journal} {Eur. Phys. J. A}\ }\textbf {\bibinfo {volume} {30}},\ \bibinfo
  {pages} {23} (\bibinfo {year} {2006})}\BibitemShut {NoStop}%
\bibitem [{\citenamefont {Todd-Rutel}(2005)}]{Todd-Rutel2005}%
  \BibitemOpen
  \bibfield  {author} {\bibinfo {author} {\bibfnamefont {e.~a.}\ \bibnamefont
  {Todd-Rutel}},\ }\href {https://doi.org/10.1103/PhysRevLett.95.122501}
  {\bibfield  {journal} {\bibinfo  {journal} {Phys. Rev. Lett.}\ }\textbf
  {\bibinfo {volume} {95}},\ \bibinfo {pages} {122501} (\bibinfo {year}
  {2005})},\ \Eprint {https://arxiv.org/abs/nucl-th/0504034}
  {arXiv:nucl-th/0504034} \BibitemShut {NoStop}%
\bibitem [{\citenamefont {Essick}\ \emph {et~al.}(2021)\citenamefont {Essick},
  \citenamefont {Landry}, \citenamefont {Schwenk},\ and\ \citenamefont
  {Tews}}]{Essick:2021ezp}%
  \BibitemOpen
  \bibfield  {author} {\bibinfo {author} {\bibfnamefont {R.}~\bibnamefont
  {Essick}}, \bibinfo {author} {\bibfnamefont {P.}~\bibnamefont {Landry}},
  \bibinfo {author} {\bibfnamefont {A.}~\bibnamefont {Schwenk}},\ and\ \bibinfo
  {author} {\bibfnamefont {I.}~\bibnamefont {Tews}},\ }\href
  {https://doi.org/10.1103/PhysRevC.104.065804} {\bibfield  {journal} {\bibinfo
   {journal} {Phys. Rev. C}\ }\textbf {\bibinfo {volume} {104}},\ \bibinfo
  {pages} {065804} (\bibinfo {year} {2021})},\ \Eprint
  {https://arxiv.org/abs/2107.05528} {arXiv:2107.05528 [nucl-th]} \BibitemShut
  {NoStop}%
\bibitem [{\citenamefont {Abbott}\ \emph {et~al.}(2019)\citenamefont {Abbott}
  \emph {et~al.}}]{LIGOScientific:2018hze}%
  \BibitemOpen
  \bibfield  {author} {\bibinfo {author} {\bibfnamefont {B.~P.}\ \bibnamefont
  {Abbott}} \emph {et~al.} (\bibinfo {collaboration} {LIGO Scientific,
  Virgo}),\ }\href {https://doi.org/10.1103/PhysRevX.9.011001} {\bibfield
  {journal} {\bibinfo  {journal} {Phys. Rev. X}\ }\textbf {\bibinfo {volume}
  {9}},\ \bibinfo {pages} {011001} (\bibinfo {year} {2019})},\ \Eprint
  {https://arxiv.org/abs/1805.11579} {arXiv:1805.11579 [gr-qc]} \BibitemShut
  {NoStop}%
\bibitem [{\citenamefont {Tolman}(1939)}]{TOV1}%
  \BibitemOpen
  \bibfield  {author} {\bibinfo {author} {\bibfnamefont {R.~C.}\ \bibnamefont
  {Tolman}},\ }\href {https://doi.org/10.1103/PhysRev.55.364} {\bibfield
  {journal} {\bibinfo  {journal} {Phys. Rev.}\ }\textbf {\bibinfo {volume}
  {55}},\ \bibinfo {pages} {364} (\bibinfo {year} {1939})}\BibitemShut
  {NoStop}%
%%CITATION = PHRVA,55,364;%%
\bibitem [{\citenamefont {Oppenheimer}\ and\ \citenamefont
  {Volkoff}(1939)}]{TOV2}%
  \BibitemOpen
  \bibfield  {author} {\bibinfo {author} {\bibfnamefont {J.~R.}\ \bibnamefont
  {Oppenheimer}}\ and\ \bibinfo {author} {\bibfnamefont {G.~M.}\ \bibnamefont
  {Volkoff}},\ }\href {https://doi.org/10.1103/PhysRev.55.374} {\bibfield
  {journal} {\bibinfo  {journal} {Phys. Rev.}\ }\textbf {\bibinfo {volume}
  {55}},\ \bibinfo {pages} {374} (\bibinfo {year} {1939})}\BibitemShut
  {NoStop}%
%%CITATION = PHRVA,55,374;%%
\bibitem [{\citenamefont {{Glendenning}}(1996)}]{book.Glendenning1996}%
  \BibitemOpen
  \bibfield  {author} {\bibinfo {author} {\bibfnamefont {N.~K.}\ \bibnamefont
  {{Glendenning}}},\ }\href@noop {} {\emph {\bibinfo {title} {{Compact
  Stars}}}}\ (\bibinfo {year} {1996})\BibitemShut {NoStop}%
\bibitem [{\citenamefont {Abbott}\ \emph {et~al.}()\citenamefont {Abbott} \emph
  {et~al.}}]{LIGOScientific:2017ync}%
  \BibitemOpen
  \bibfield  {author} {\bibinfo {author} {\bibfnamefont {B.~P.}\ \bibnamefont
  {Abbott}} \emph {et~al.},\ }\href {https://doi.org/10.3847/2041-8213/aa91c9}
  {\ \textbf {\bibinfo {volume} {848}},\ \bibinfo {pages} {L12}},\ \Eprint
  {https://arxiv.org/abs/1710.05833} {arXiv:1710.05833} \BibitemShut {NoStop}%
\bibitem [{\citenamefont {Hinderer}(2008)}]{Hinderer2008}%
  \BibitemOpen
  \bibfield  {author} {\bibinfo {author} {\bibfnamefont {T.}~\bibnamefont
  {Hinderer}},\ }\href {https://doi.org/10.1086/533487} {\bibfield  {journal}
  {\bibinfo  {journal} {Astrophys. J.}\ }\textbf {\bibinfo {volume} {677}},\
  \bibinfo {pages} {1216} (\bibinfo {year} {2008})},\ \Eprint
  {https://arxiv.org/abs/0711.2420} {arXiv:0711.2420 [astro-ph]} \BibitemShut
  {NoStop}%
%%CITATION = ARXIV:0711.2420;%%
\bibitem [{\citenamefont {Abbott}\ \emph {et~al.}(2018)\citenamefont {Abbott}
  \emph {et~al.}}]{LIGOScientific:2018cki}%
  \BibitemOpen
  \bibfield  {author} {\bibinfo {author} {\bibfnamefont {B.~P.}\ \bibnamefont
  {Abbott}} \emph {et~al.} (\bibinfo {collaboration} {LIGO Scientific,
  Virgo}),\ }\href {https://doi.org/10.1103/PhysRevLett.121.161101} {\bibfield
  {journal} {\bibinfo  {journal} {Phys. Rev. Lett.}\ }\textbf {\bibinfo
  {volume} {121}},\ \bibinfo {pages} {161101} (\bibinfo {year} {2018})},\
  \Eprint {https://arxiv.org/abs/1805.11581} {arXiv:1805.11581 [gr-qc]}
  \BibitemShut {NoStop}%
\bibitem [{\citenamefont {Fortin}\ \emph {et~al.}(2017)\citenamefont {Fortin},
  \citenamefont {Avancini}, \citenamefont {Provid\^encia},\ and\ \citenamefont
  {Vida\~na}}]{Fortin:2017cvt}%
  \BibitemOpen
  \bibfield  {author} {\bibinfo {author} {\bibfnamefont {M.}~\bibnamefont
  {Fortin}}, \bibinfo {author} {\bibfnamefont {S.~S.}\ \bibnamefont
  {Avancini}}, \bibinfo {author} {\bibfnamefont {C.}~\bibnamefont
  {Provid\^encia}},\ and\ \bibinfo {author} {\bibfnamefont {I.}~\bibnamefont
  {Vida\~na}},\ }\href {https://doi.org/10.1103/PhysRevC.95.065803} {\bibfield
  {journal} {\bibinfo  {journal} {Phys. Rev. C}\ }\textbf {\bibinfo {volume}
  {95}},\ \bibinfo {pages} {065803} (\bibinfo {year} {2017})},\ \Eprint
  {https://arxiv.org/abs/1701.06373} {arXiv:1701.06373 [nucl-th]} \BibitemShut
  {NoStop}%
\bibitem [{\citenamefont {Provid\^encia}\ \emph {et~al.}(2018)\citenamefont
  {Provid\^encia}, \citenamefont {Fortin}, \citenamefont {Pais},\ and\
  \citenamefont {Rabhi}}]{Providencia:2018ywl}%
  \BibitemOpen
  \bibfield  {author} {\bibinfo {author} {\bibfnamefont {C.}~\bibnamefont
  {Provid\^encia}}, \bibinfo {author} {\bibfnamefont {M.}~\bibnamefont
  {Fortin}}, \bibinfo {author} {\bibfnamefont {H.}~\bibnamefont {Pais}},\ and\
  \bibinfo {author} {\bibfnamefont {A.}~\bibnamefont {Rabhi}}\ }\href
  {https://doi.org/10.3389/fspas.2019.00013} {10.3389/fspas.2019.00013}
  (\bibinfo {year} {2018}),\ \Eprint {https://arxiv.org/abs/1811.00786}
  {arXiv:1811.00786 [astro-ph.HE]} \BibitemShut {NoStop}%
\bibitem [{\citenamefont {Gal}\ \emph {et~al.}(2016)\citenamefont {Gal},
  \citenamefont {Hungerford},\ and\ \citenamefont {Millener}}]{Gal:2016boi}%
  \BibitemOpen
  \bibfield  {author} {\bibinfo {author} {\bibfnamefont {A.}~\bibnamefont
  {Gal}}, \bibinfo {author} {\bibfnamefont {E.~V.}\ \bibnamefont
  {Hungerford}},\ and\ \bibinfo {author} {\bibfnamefont {D.~J.}\ \bibnamefont
  {Millener}},\ }\href {https://doi.org/10.1103/RevModPhys.88.035004}
  {\bibfield  {journal} {\bibinfo  {journal} {Rev. Mod. Phys.}\ }\textbf
  {\bibinfo {volume} {88}},\ \bibinfo {pages} {035004} (\bibinfo {year}
  {2016})},\ \Eprint {https://arxiv.org/abs/1605.00557} {arXiv:1605.00557
  [nucl-th]} \BibitemShut {NoStop}%
\bibitem [{\citenamefont {Weissenborn}\ \emph {et~al.}(2012)\citenamefont
  {Weissenborn}, \citenamefont {Chatterjee},\ and\ \citenamefont
  {Schaffner-Bielich}}]{Weissenborn:2011kb}%
  \BibitemOpen
  \bibfield  {author} {\bibinfo {author} {\bibfnamefont {S.}~\bibnamefont
  {Weissenborn}}, \bibinfo {author} {\bibfnamefont {D.}~\bibnamefont
  {Chatterjee}},\ and\ \bibinfo {author} {\bibfnamefont {J.}~\bibnamefont
  {Schaffner-Bielich}},\ }\href
  {https://doi.org/10.1016/j.nuclphysa.2012.02.012} {\bibfield  {journal}
  {\bibinfo  {journal} {Nucl. Phys. A}\ }\textbf {\bibinfo {volume} {881}},\
  \bibinfo {pages} {62} (\bibinfo {year} {2012})},\ \Eprint
  {https://arxiv.org/abs/1111.6049} {arXiv:1111.6049 [astro-ph.HE]}
  \BibitemShut {NoStop}%
\bibitem [{\citenamefont {Fortin}\ \emph
  {et~al.}(2016{\natexlab{a}})\citenamefont {Fortin}, \citenamefont
  {Providencia}, \citenamefont {Raduta}, \citenamefont {Gulminelli},
  \citenamefont {Zdunik}, \citenamefont {Haensel},\ and\ \citenamefont
  {Bejger}}]{Fortin2016}%
  \BibitemOpen
  \bibfield  {author} {\bibinfo {author} {\bibfnamefont {M.}~\bibnamefont
  {Fortin}}, \bibinfo {author} {\bibfnamefont {C.}~\bibnamefont {Providencia}},
  \bibinfo {author} {\bibfnamefont {A.~R.}\ \bibnamefont {Raduta}}, \bibinfo
  {author} {\bibfnamefont {F.}~\bibnamefont {Gulminelli}}, \bibinfo {author}
  {\bibfnamefont {J.~L.}\ \bibnamefont {Zdunik}}, \bibinfo {author}
  {\bibfnamefont {P.}~\bibnamefont {Haensel}},\ and\ \bibinfo {author}
  {\bibfnamefont {M.}~\bibnamefont {Bejger}},\ }\href
  {https://doi.org/10.1103/PhysRevC.94.035804} {\bibfield  {journal} {\bibinfo
  {journal} {Phys. Rev. C}\ }\textbf {\bibinfo {volume} {94}},\ \bibinfo
  {pages} {035804} (\bibinfo {year} {2016}{\natexlab{a}})},\ \Eprint
  {https://arxiv.org/abs/1604.01944} {arXiv:1604.01944 [astro-ph.SR]}
  \BibitemShut {NoStop}%
\bibitem [{\citenamefont {Fortin}\ \emph {et~al.}(2020)\citenamefont {Fortin},
  \citenamefont {Raduta}, \citenamefont {Avancini},\ and\ \citenamefont
  {Provid\^encia}}]{Fortin:2020qin}%
  \BibitemOpen
  \bibfield  {author} {\bibinfo {author} {\bibfnamefont {M.}~\bibnamefont
  {Fortin}}, \bibinfo {author} {\bibfnamefont {A.~R.}\ \bibnamefont {Raduta}},
  \bibinfo {author} {\bibfnamefont {S.}~\bibnamefont {Avancini}},\ and\
  \bibinfo {author} {\bibfnamefont {C.}~\bibnamefont {Provid\^encia}},\ }\href
  {https://doi.org/10.1103/PhysRevD.101.034017} {\bibfield  {journal} {\bibinfo
   {journal} {Phys. Rev. D}\ }\textbf {\bibinfo {volume} {101}},\ \bibinfo
  {pages} {034017} (\bibinfo {year} {2020})},\ \Eprint
  {https://arxiv.org/abs/2001.08036} {arXiv:2001.08036 [hep-ph]} \BibitemShut
  {NoStop}%
\bibitem [{\citenamefont {Stone}\ \emph {et~al.}(2021)\citenamefont {Stone},
  \citenamefont {Dexheimer}, \citenamefont {Guichon}, \citenamefont {Thomas},\
  and\ \citenamefont {Typel}}]{Stone:2019blq}%
  \BibitemOpen
  \bibfield  {author} {\bibinfo {author} {\bibfnamefont {J.~R.}\ \bibnamefont
  {Stone}}, \bibinfo {author} {\bibfnamefont {V.}~\bibnamefont {Dexheimer}},
  \bibinfo {author} {\bibfnamefont {P.~A.~M.}\ \bibnamefont {Guichon}},
  \bibinfo {author} {\bibfnamefont {A.~W.}\ \bibnamefont {Thomas}},\ and\
  \bibinfo {author} {\bibfnamefont {S.}~\bibnamefont {Typel}},\ }\href
  {https://doi.org/10.1093/mnras/staa4006} {\bibfield  {journal} {\bibinfo
  {journal} {Mon. Not. Roy. Astron. Soc.}\ }\textbf {\bibinfo {volume} {502}},\
  \bibinfo {pages} {3476} (\bibinfo {year} {2021})},\ \Eprint
  {https://arxiv.org/abs/1906.11100} {arXiv:1906.11100 [nucl-th]} \BibitemShut
  {NoStop}%
\bibitem [{\citenamefont {Fortin}\ \emph {et~al.}(2018)\citenamefont {Fortin},
  \citenamefont {Oertel},\ and\ \citenamefont
  {Provid\^encia}}]{Fortin:2017dsj}%
  \BibitemOpen
  \bibfield  {author} {\bibinfo {author} {\bibfnamefont {M.}~\bibnamefont
  {Fortin}}, \bibinfo {author} {\bibfnamefont {M.}~\bibnamefont {Oertel}},\
  and\ \bibinfo {author} {\bibfnamefont {C.}~\bibnamefont {Provid\^encia}},\
  }\href {https://doi.org/10.1017/pasa.2018.32} {\bibfield  {journal} {\bibinfo
   {journal} {Publ. Astron. Soc. Austral.}\ }\textbf {\bibinfo {volume} {35}},\
  \bibinfo {pages} {44} (\bibinfo {year} {2018})},\ \Eprint
  {https://arxiv.org/abs/1711.09427} {arXiv:1711.09427 [astro-ph.HE]}
  \BibitemShut {NoStop}%
\bibitem [{\citenamefont {Yakovlev}\ \emph {et~al.}(2001)\citenamefont
  {Yakovlev}, \citenamefont {Kaminker}, \citenamefont {Gnedin},\ and\
  \citenamefont {Haensel}}]{Yakovlev:2000jp}%
  \BibitemOpen
  \bibfield  {author} {\bibinfo {author} {\bibfnamefont {D.~G.}\ \bibnamefont
  {Yakovlev}}, \bibinfo {author} {\bibfnamefont {A.~D.}\ \bibnamefont
  {Kaminker}}, \bibinfo {author} {\bibfnamefont {O.~Y.}\ \bibnamefont
  {Gnedin}},\ and\ \bibinfo {author} {\bibfnamefont {P.}~\bibnamefont
  {Haensel}},\ }\href {https://doi.org/10.1016/S0370-1573(00)00131-9}
  {\bibfield  {journal} {\bibinfo  {journal} {Phys. Rept.}\ }\textbf {\bibinfo
  {volume} {354}},\ \bibinfo {pages} {1} (\bibinfo {year} {2001})},\ \Eprint
  {https://arxiv.org/abs/astro-ph/0012122} {arXiv:astro-ph/0012122}
  \BibitemShut {NoStop}%
\bibitem [{\citenamefont {Yakovlev}\ and\ \citenamefont
  {Pethick}(2004)}]{Yakovlev:2004iq}%
  \BibitemOpen
  \bibfield  {author} {\bibinfo {author} {\bibfnamefont {D.~G.}\ \bibnamefont
  {Yakovlev}}\ and\ \bibinfo {author} {\bibfnamefont {C.~J.}\ \bibnamefont
  {Pethick}},\ }\href {https://doi.org/10.1146/annurev.astro.42.053102.134013}
  {\bibfield  {journal} {\bibinfo  {journal} {Ann. Rev. Astron. Astrophys.}\
  }\textbf {\bibinfo {volume} {42}},\ \bibinfo {pages} {169} (\bibinfo {year}
  {2004})},\ \Eprint {https://arxiv.org/abs/astro-ph/0402143}
  {arXiv:astro-ph/0402143} \BibitemShut {NoStop}%
\bibitem [{\citenamefont {Fortin}\ \emph
  {et~al.}(2016{\natexlab{b}})\citenamefont {Fortin}, \citenamefont
  {Providencia}, \citenamefont {Raduta}, \citenamefont {Gulminelli},
  \citenamefont {Zdunik}, \citenamefont {Haensel},\ and\ \citenamefont
  {Bejger}}]{Fortin:2016hny}%
  \BibitemOpen
  \bibfield  {author} {\bibinfo {author} {\bibfnamefont {M.}~\bibnamefont
  {Fortin}}, \bibinfo {author} {\bibfnamefont {C.}~\bibnamefont {Providencia}},
  \bibinfo {author} {\bibfnamefont {A.~R.}\ \bibnamefont {Raduta}}, \bibinfo
  {author} {\bibfnamefont {F.}~\bibnamefont {Gulminelli}}, \bibinfo {author}
  {\bibfnamefont {J.~L.}\ \bibnamefont {Zdunik}}, \bibinfo {author}
  {\bibfnamefont {P.}~\bibnamefont {Haensel}},\ and\ \bibinfo {author}
  {\bibfnamefont {M.}~\bibnamefont {Bejger}},\ }\href
  {https://doi.org/10.1103/PhysRevC.94.035804} {\bibfield  {journal} {\bibinfo
  {journal} {Phys. Rev. C}\ }\textbf {\bibinfo {volume} {94}},\ \bibinfo
  {pages} {035804} (\bibinfo {year} {2016}{\natexlab{b}})},\ \Eprint
  {https://arxiv.org/abs/1604.01944} {arXiv:1604.01944 [astro-ph.SR]}
  \BibitemShut {NoStop}%
\bibitem [{\citenamefont {Fortin}\ \emph {et~al.}(2021)\citenamefont {Fortin},
  \citenamefont {Raduta}, \citenamefont {Avancini},\ and\ \citenamefont
  {Provid\^encia}}]{Fortin:2021umb}%
  \BibitemOpen
  \bibfield  {author} {\bibinfo {author} {\bibfnamefont {M.}~\bibnamefont
  {Fortin}}, \bibinfo {author} {\bibfnamefont {A.~R.}\ \bibnamefont {Raduta}},
  \bibinfo {author} {\bibfnamefont {S.}~\bibnamefont {Avancini}},\ and\
  \bibinfo {author} {\bibfnamefont {C.}~\bibnamefont {Provid\^encia}},\ }\href
  {https://doi.org/10.1103/PhysRevD.103.083004} {\bibfield  {journal} {\bibinfo
   {journal} {Phys. Rev. D}\ }\textbf {\bibinfo {volume} {103}},\ \bibinfo
  {pages} {083004} (\bibinfo {year} {2021})},\ \Eprint
  {https://arxiv.org/abs/2102.07565} {arXiv:2102.07565 [nucl-th]} \BibitemShut
  {NoStop}%
\bibitem [{\citenamefont {Beznogov}\ and\ \citenamefont
  {Yakovlev}(2015)}]{Beznogov:2015ewa}%
  \BibitemOpen
  \bibfield  {author} {\bibinfo {author} {\bibfnamefont {M.~V.}\ \bibnamefont
  {Beznogov}}\ and\ \bibinfo {author} {\bibfnamefont {D.~G.}\ \bibnamefont
  {Yakovlev}},\ }\href {https://doi.org/10.1093/mnras/stv1293} {\bibfield
  {journal} {\bibinfo  {journal} {Mon. Not. Roy. Astron. Soc.}\ }\textbf
  {\bibinfo {volume} {452}},\ \bibinfo {pages} {540} (\bibinfo {year}
  {2015})},\ \Eprint {https://arxiv.org/abs/1507.04206} {arXiv:1507.04206
  [astro-ph.SR]} \BibitemShut {NoStop}%
\bibitem [{\citenamefont {Kurkela}(2022)}]{Kurkela:2022elj}%
  \BibitemOpen
  \bibfield  {author} {\bibinfo {author} {\bibfnamefont {A.}~\bibnamefont
  {Kurkela}},\ }in\ \href@noop {} {\emph {\bibinfo {booktitle} {{15th
  Conference on Quark Confinement and the Hadron Spectrum}}}}\ (\bibinfo {year}
  {2022})\ \Eprint {https://arxiv.org/abs/2211.11414} {arXiv:2211.11414
  [hep-ph]} \BibitemShut {NoStop}%
\bibitem [{\citenamefont {Kojo}(2021)}]{Kojo:2020krb}%
  \BibitemOpen
  \bibfield  {author} {\bibinfo {author} {\bibfnamefont {T.}~\bibnamefont
  {Kojo}},\ }\href {https://doi.org/10.1007/s43673-021-00011-6} {\bibfield
  {journal} {\bibinfo  {journal} {AAPPS Bull.}\ }\textbf {\bibinfo {volume}
  {31}},\ \bibinfo {pages} {11} (\bibinfo {year} {2021})},\ \Eprint
  {https://arxiv.org/abs/2011.10940} {arXiv:2011.10940 [nucl-th]} \BibitemShut
  {NoStop}%
\bibitem [{\citenamefont {Al-Mamun}\ \emph {et~al.}(2021)\citenamefont
  {Al-Mamun}, \citenamefont {Steiner}, \citenamefont {N\"attil\"a},
  \citenamefont {Lange}, \citenamefont {O'Shaughnessy}, \citenamefont {Tews},
  \citenamefont {Gandolfi}, \citenamefont {Heinke},\ and\ \citenamefont
  {Han}}]{Al-Mamun:2020vzu}%
  \BibitemOpen
  \bibfield  {author} {\bibinfo {author} {\bibfnamefont {M.}~\bibnamefont
  {Al-Mamun}}, \bibinfo {author} {\bibfnamefont {A.~W.}\ \bibnamefont
  {Steiner}}, \bibinfo {author} {\bibfnamefont {J.}~\bibnamefont
  {N\"attil\"a}}, \bibinfo {author} {\bibfnamefont {J.}~\bibnamefont {Lange}},
  \bibinfo {author} {\bibfnamefont {R.}~\bibnamefont {O'Shaughnessy}}, \bibinfo
  {author} {\bibfnamefont {I.}~\bibnamefont {Tews}}, \bibinfo {author}
  {\bibfnamefont {S.}~\bibnamefont {Gandolfi}}, \bibinfo {author}
  {\bibfnamefont {C.}~\bibnamefont {Heinke}},\ and\ \bibinfo {author}
  {\bibfnamefont {S.}~\bibnamefont {Han}},\ }\href
  {https://doi.org/10.1103/PhysRevLett.126.061101} {\bibfield  {journal}
  {\bibinfo  {journal} {Phys. Rev. Lett.}\ }\textbf {\bibinfo {volume} {126}},\
  \bibinfo {pages} {061101} (\bibinfo {year} {2021})},\ \Eprint
  {https://arxiv.org/abs/2008.12817} {arXiv:2008.12817 [astro-ph.HE]}
  \BibitemShut {NoStop}%
\bibitem [{\citenamefont {Drischler}\ \emph {et~al.}(2022)\citenamefont
  {Drischler}, \citenamefont {Han},\ and\ \citenamefont
  {Reddy}}]{Drischler:2021bup}%
  \BibitemOpen
  \bibfield  {author} {\bibinfo {author} {\bibfnamefont {C.}~\bibnamefont
  {Drischler}}, \bibinfo {author} {\bibfnamefont {S.}~\bibnamefont {Han}},\
  and\ \bibinfo {author} {\bibfnamefont {S.}~\bibnamefont {Reddy}},\ }\href
  {https://doi.org/10.1103/PhysRevC.105.035808} {\bibfield  {journal} {\bibinfo
   {journal} {Phys. Rev. C}\ }\textbf {\bibinfo {volume} {105}},\ \bibinfo
  {pages} {035808} (\bibinfo {year} {2022})},\ \Eprint
  {https://arxiv.org/abs/2110.14896} {arXiv:2110.14896 [nucl-th]} \BibitemShut
  {NoStop}%
\bibitem [{\citenamefont {Komoltsev}\ and\ \citenamefont
  {Kurkela}(2022)}]{Komoltsev:2021jzg}%
  \BibitemOpen
  \bibfield  {author} {\bibinfo {author} {\bibfnamefont {O.}~\bibnamefont
  {Komoltsev}}\ and\ \bibinfo {author} {\bibfnamefont {A.}~\bibnamefont
  {Kurkela}},\ }\href {https://doi.org/10.1103/PhysRevLett.128.202701}
  {\bibfield  {journal} {\bibinfo  {journal} {Phys. Rev. Lett.}\ }\textbf
  {\bibinfo {volume} {128}},\ \bibinfo {pages} {202701} (\bibinfo {year}
  {2022})},\ \Eprint {https://arxiv.org/abs/2111.05350} {arXiv:2111.05350
  [nucl-th]} \BibitemShut {NoStop}%
\bibitem [{\citenamefont {Horowitz}\ and\ \citenamefont
  {Piekarewicz}(2001)}]{Horowitz:2000xj}%
  \BibitemOpen
  \bibfield  {author} {\bibinfo {author} {\bibfnamefont {C.~J.}\ \bibnamefont
  {Horowitz}}\ and\ \bibinfo {author} {\bibfnamefont {J.}~\bibnamefont
  {Piekarewicz}},\ }\href {https://doi.org/10.1103/PhysRevLett.86.5647}
  {\bibfield  {journal} {\bibinfo  {journal} {Phys. Rev. Lett.}\ }\textbf
  {\bibinfo {volume} {86}},\ \bibinfo {pages} {5647} (\bibinfo {year}
  {2001})},\ \Eprint {https://arxiv.org/abs/astro-ph/0010227}
  {arXiv:astro-ph/0010227} \BibitemShut {NoStop}%
\bibitem [{\citenamefont {Lalazissis}\ \emph {et~al.}(2005)\citenamefont
  {Lalazissis}, \citenamefont {Niksic}, \citenamefont {Vretenar},\ and\
  \citenamefont {Ring}}]{Lalazissis2005}%
  \BibitemOpen
  \bibfield  {author} {\bibinfo {author} {\bibfnamefont {G.~A.}\ \bibnamefont
  {Lalazissis}}, \bibinfo {author} {\bibfnamefont {T.}~\bibnamefont {Niksic}},
  \bibinfo {author} {\bibfnamefont {D.}~\bibnamefont {Vretenar}},\ and\
  \bibinfo {author} {\bibfnamefont {P.}~\bibnamefont {Ring}},\ }\href
  {https://doi.org/10.1103/PhysRevC.71.024312} {\bibfield  {journal} {\bibinfo
  {journal} {Phys. Rev. C}\ }\textbf {\bibinfo {volume} {71}},\ \bibinfo
  {pages} {024312} (\bibinfo {year} {2005})}\BibitemShut {NoStop}%
\bibitem [{\citenamefont {Koch}\ \emph {et~al.}(1987)\citenamefont {Koch},
  \citenamefont {Biro}, \citenamefont {Kunz},\ and\ \citenamefont
  {Mosel}}]{Koch:1987py}%
  \BibitemOpen
  \bibfield  {author} {\bibinfo {author} {\bibfnamefont {V.}~\bibnamefont
  {Koch}}, \bibinfo {author} {\bibfnamefont {T.~S.}\ \bibnamefont {Biro}},
  \bibinfo {author} {\bibfnamefont {J.}~\bibnamefont {Kunz}},\ and\ \bibinfo
  {author} {\bibfnamefont {U.}~\bibnamefont {Mosel}},\ }\href
  {https://doi.org/10.1016/0370-2693(87)91517-6} {\bibfield  {journal}
  {\bibinfo  {journal} {Phys. Lett. B}\ }\textbf {\bibinfo {volume} {185}},\
  \bibinfo {pages} {1} (\bibinfo {year} {1987})}\BibitemShut {NoStop}%
\bibitem [{\citenamefont {da~Providencia}\ \emph {et~al.}(2002)\citenamefont
  {da~Providencia}, \citenamefont {da~Providencia},\ and\ \citenamefont
  {Moszkowski}}]{daProvidencia:2002pcg}%
  \BibitemOpen
  \bibfield  {author} {\bibinfo {author} {\bibfnamefont {C.}~\bibnamefont
  {da~Providencia}}, \bibinfo {author} {\bibfnamefont {J.}~\bibnamefont
  {da~Providencia}},\ and\ \bibinfo {author} {\bibfnamefont {S.~A.}\
  \bibnamefont {Moszkowski}},\ }\href
  {https://doi.org/10.1142/S021797920302034X} {\bibfield  {journal} {\bibinfo
  {journal} {Ser. Adv. Quant. Many Body Theor.}\ }\textbf {\bibinfo {volume}
  {6}},\ \bibinfo {pages} {242} (\bibinfo {year} {2002})}\BibitemShut {NoStop}%
\bibitem [{\citenamefont {Mishustin}\ \emph {et~al.}(2004)\citenamefont
  {Mishustin}, \citenamefont {Satarov},\ and\ \citenamefont
  {Greiner}}]{Mishustin:2003wq}%
  \BibitemOpen
  \bibfield  {author} {\bibinfo {author} {\bibfnamefont {I.~N.}\ \bibnamefont
  {Mishustin}}, \bibinfo {author} {\bibfnamefont {L.~M.}\ \bibnamefont
  {Satarov}},\ and\ \bibinfo {author} {\bibfnamefont {W.}~\bibnamefont
  {Greiner}},\ }\href {https://doi.org/10.1016/j.physrep.2003.10.010}
  {\bibfield  {journal} {\bibinfo  {journal} {Phys. Rept.}\ }\textbf {\bibinfo
  {volume} {391}},\ \bibinfo {pages} {363} (\bibinfo {year} {2004})},\ \Eprint
  {https://arxiv.org/abs/hep-ph/0304296} {arXiv:hep-ph/0304296} \BibitemShut
  {NoStop}%
\bibitem [{\citenamefont {Pais}\ \emph {et~al.}(2016)\citenamefont {Pais},
  \citenamefont {Menezes},\ and\ \citenamefont {Provid\^encia}}]{Pais:2016dng}%
  \BibitemOpen
  \bibfield  {author} {\bibinfo {author} {\bibfnamefont {H.}~\bibnamefont
  {Pais}}, \bibinfo {author} {\bibfnamefont {D.~P.}\ \bibnamefont {Menezes}},\
  and\ \bibinfo {author} {\bibfnamefont {C.}~\bibnamefont {Provid\^encia}},\
  }\href {https://doi.org/10.1103/PhysRevC.93.065805} {\bibfield  {journal}
  {\bibinfo  {journal} {Phys. Rev. C}\ }\textbf {\bibinfo {volume} {93}},\
  \bibinfo {pages} {065805} (\bibinfo {year} {2016})},\ \Eprint
  {https://arxiv.org/abs/1603.01239} {arXiv:1603.01239 [nucl-th]} \BibitemShut
  {NoStop}%
\bibitem [{\citenamefont {Wei}\ \emph {et~al.}(2016)\citenamefont {Wei},
  \citenamefont {Jiang}, \citenamefont {Yang},\ and\ \citenamefont
  {Zhang}}]{Wei:2015aep}%
  \BibitemOpen
  \bibfield  {author} {\bibinfo {author} {\bibfnamefont {S.-N.}\ \bibnamefont
  {Wei}}, \bibinfo {author} {\bibfnamefont {W.-Z.}\ \bibnamefont {Jiang}},
  \bibinfo {author} {\bibfnamefont {R.-Y.}\ \bibnamefont {Yang}},\ and\
  \bibinfo {author} {\bibfnamefont {D.-R.}\ \bibnamefont {Zhang}},\ }\href
  {https://doi.org/10.1016/j.physletb.2016.10.019} {\bibfield  {journal}
  {\bibinfo  {journal} {Phys. Lett. B}\ }\textbf {\bibinfo {volume} {763}},\
  \bibinfo {pages} {145} (\bibinfo {year} {2016})},\ \Eprint
  {https://arxiv.org/abs/1511.07144} {arXiv:1511.07144 [nucl-th]} \BibitemShut
  {NoStop}%
\bibitem [{\citenamefont {Liu}\ \emph {et~al.}(2002)\citenamefont {Liu},
  \citenamefont {Greco}, \citenamefont {Baran}, \citenamefont {Colonna},\ and\
  \citenamefont {Di~Toro}}]{Liu:2001iz}%
  \BibitemOpen
  \bibfield  {author} {\bibinfo {author} {\bibfnamefont {B.}~\bibnamefont
  {Liu}}, \bibinfo {author} {\bibfnamefont {V.}~\bibnamefont {Greco}}, \bibinfo
  {author} {\bibfnamefont {V.}~\bibnamefont {Baran}}, \bibinfo {author}
  {\bibfnamefont {M.}~\bibnamefont {Colonna}},\ and\ \bibinfo {author}
  {\bibfnamefont {M.}~\bibnamefont {Di~Toro}},\ }\href
  {https://doi.org/10.1103/PhysRevC.65.045201} {\bibfield  {journal} {\bibinfo
  {journal} {Phys. Rev. C}\ }\textbf {\bibinfo {volume} {65}},\ \bibinfo
  {pages} {045201} (\bibinfo {year} {2002})},\ \Eprint
  {https://arxiv.org/abs/nucl-th/0112034} {arXiv:nucl-th/0112034} \BibitemShut
  {NoStop}%
\bibitem [{\citenamefont {Fortin}\ \emph {et~al.}(2015)\citenamefont {Fortin},
  \citenamefont {Zdunik}, \citenamefont {Haensel},\ and\ \citenamefont
  {Bejger}}]{Fortin:2014mya}%
  \BibitemOpen
  \bibfield  {author} {\bibinfo {author} {\bibfnamefont {M.}~\bibnamefont
  {Fortin}}, \bibinfo {author} {\bibfnamefont {J.~L.}\ \bibnamefont {Zdunik}},
  \bibinfo {author} {\bibfnamefont {P.}~\bibnamefont {Haensel}},\ and\ \bibinfo
  {author} {\bibfnamefont {M.}~\bibnamefont {Bejger}},\ }\href
  {https://doi.org/10.1051/0004-6361/201424800} {\bibfield  {journal} {\bibinfo
   {journal} {Astron. Astrophys.}\ }\textbf {\bibinfo {volume} {576}},\
  \bibinfo {pages} {A68} (\bibinfo {year} {2015})},\ \Eprint
  {https://arxiv.org/abs/1408.3052} {arXiv:1408.3052 [astro-ph.SR]}
  \BibitemShut {NoStop}%
\end{thebibliography}

\end{document}